\documentclass[a4paper,fleqn,usenatbib]{mnras}

\usepackage{newtxtext,newtxmath}

\UseRawInputEncoding

\usepackage[T1]{fontenc}
\usepackage{ae,aecompl}
\usepackage{deluxetable}
\usepackage{adjustbox}
\usepackage{threeparttable}

\usepackage{graphicx}	
\usepackage{amsmath}	
\usepackage{amssymb}	

\title[Complex structure of a proto-brown dwarf]{Complex structure of a proto-brown dwarf}

\author[Riaz \& Machida]{
B. Riaz,$^{1}$\thanks{E-mail: briaz@usm.lmu.de}
M. N. Machida$^{2}$
\\
$^{1}$Universit\"ats-Sternwarte M\"unchen, Ludwig-Maximilians-Universit\"at, Scheinerstr.~1, 81679 M\"unchen, Germany	\\
$^{2}$Department of Earth and Planetary Sciences, Faculty of Sciences, Kyushu University, Fukuoka, Japan
}

\date{Accepted 2021 February 17. Received 2021 February 15; in original form 2020 December 29.}

\pubyear{2021}

\begin{document}
\label{firstpage}
\pagerange{\pageref{firstpage}--\pageref{lastpage}}
\maketitle

\begin{abstract}

We present ALMA $^{12}$CO (2-1), $^{13}$CO (2-1), C$^{18}$O (2-1) molecular line observations of a very young proto-brown dwarf system, ISO-OPH 200. We have conducted physical+chemical modelling of the complex internal structure for this system using the core collapse simulations for brown dwarf formation. The model at an age of $\sim$6000 yr can provide a good fit to the observed kinematics, spectra, and reproduce the complex structures seen in the moment maps. Results from modelling indicate that $^{12}$CO emission is tracing an extended ($\sim$1000 au) molecular outflow and a bright shock knot, $^{13}$CO is tracing the outer ($\sim$1000 au) envelope/pseudo-disc, and C$^{18}$O is tracing the inner ($\sim$500 au) pseudo-disc. The source size of $\sim$8.6 au measured in the 873$\mu$m image is comparable to the inner Keplerian disc size predicted by the model. A 3D model structure of ISO-OPH 200 suggests that this system is viewed partially through a wide outflow cavity resulting in a direct view of the outflow and a partial view of the envelope/pseudo-disc. We have argued that ISO-OPH 200 has been mis-classified as a Class Flat object due to the unusual orientation. The various signatures of this system, notably, the young $\sim$616 yr outflow dynamical age and high outflow rate ($\sim$1$\times$10$^{-7}$ M$_{\sun}$ yr$^{-1}$), silicate absorption in the 10$\micron$ mid-infrared spectrum, pristine ISM-like dust in the envelope/disc, comparable sizes of the extended envelope and outflow, indicate that ISO-OPH 200 is an early Class 0 stage system formed in a star-like mechanism via gravitational collapse of a very low-mass core.

\end{abstract}




\section{Introduction} 
\label{intro}

The early stages of star formation are characterized by several processes that unravel through multiple physical components. A low-mass star forms by accreting gas from its surroundings while at the same time drives a powerful bipolar outflow that carries away the excess angular momentum (e.g., Banerjee \& Pudritz 2006; Machida \& Basu 2019). The jets/outflows are launched close to the forming star, and can cause turbulence by injecting energy and momentum into the dense circumstellar envelope, resulting in its dispersal over time and halting the infalling of material onto the nascent star (e.g., Machida \& Basu 2019; Dapp et al. 2012; Wurster et al. 2018; Tsukamoto et al. 2015). Protostellar outflows therefore play an important role during the early accretion phase and can determine the star formation efficiency and the final mass of the star.

Since the gaseous environment of an early-stage Class 0/I object exhibits complicated morphologies and kinematics, a thorough study of such regions requires observations of different molecular species that can trace different density and kinematic regimes. The most commonly observed tracers are CO and the $^{13}$CO and C$^{18}$O isotopologues. The highly abundant $^{12}$CO can probe the low-density, high-velocity outflow gas, while $^{13}$CO can trace denser outflow material than $^{12}$CO and is a good probe of the low-velocity, high-density gas in the walls of the outflow cavities and the outer circumstellar envelope. The C$^{18}$O isotope can trace the kinematics of the dense gas in the inner envelope/pseudo-disc region. The combination of these lines can provide information on the structure and kinematics of the outflow/envelope/disc components and on the outflow-envelope interaction in a protostellar system (e.g., Redman et al. 2002; Hatchell et al. 1999; Thi et al. 2004; Carolan et al. 2008; Arce et al. 2001; 2005). High angular resolution observations with ALMA in the $^{12}$CO, $^{13}$CO, and C$^{18}$O molecular lines have allowed disentangling the infall, outflow, rotational kinematics in protostellar systems and to understand the interaction among them (e.g., Aso et al. 2017ab; Ohashi et al. 2014; Sakai et al. 2014; Lefloch et al. 2015; van der Marel et al. 2013; Lee et al. 2017; Hirota et al. 2017; Bjerkeli et al. 2016; Plunkett et al. 2015; Yen et al. 2013; 2015; 2017).

A recurring question in star formation is whether the formation and evolutionary stages for sub-stellar mass  ($\leq$0.08 M$_{\sun}$) objects are similar to low-mass star formation. Numerical simulation studies have provided an understanding of the various phenomenon of the brown dwarf formation processes, whether it is a star-like formation via gravitational core collapse (e.g., Machida et al. 2009), or alternative mechanisms of formation via disc fragmentation and as ejected embryos (Stamatellos \& Whitworth 2008; Goodwin \& Whitworth 2007; Bate 2012). The physical scales of early-stage Class 0/I proto-brown dwarfs (proto-BDs) are expected to be at least ten times smaller than a low-mass protostar (e.g., Riaz et al. 2019a; Machida et al. 2009), which makes it difficult to resolve the inner few tens of au scales where various processes are expected to originate from. In our recent work on ALMA CO line and continuum observations of a Class I proto-BD in Orion, we showed on the basis of physical+chemical modelling that the observed CO line emission is tracing the infall+rotational kinematics originating from a pseudo-disc structure with a total (gas+dust) mass of $\sim$0.02 M$_{\sun}$ and a size of 178$\pm$16 au in this proto-BD system (Riaz et al. 2019a). By modelling the observed position and velocity offsets in the CO line position-velocity diagram, we were able to constrain the kinematical age of this system to be 0.03$\pm$0.01 Myr (Riaz et al. 2019a).

Here, we present an interpretation of the complex internal structure for the much younger proto-BD system, ISO-Oph 200 (L$_{bol}\sim$0.08 L$_{\sun}$; L$_{int}\sim$0.06 L$_{\sun}$), based on physical+chemical modelling of ALMA molecular line observations. ISO-Oph 200 is located in the Ophiuchus region at a distance of 144.2$\pm$1.3 pc (Ortiz-Le\'{o}n et al. 2018). Riaz et al. (2018; 2019) first reported the detection of several high-density molecular lines along with CO and isotopologues in IRAM 30m observations of ISO-Oph 200. They identified this young stellar object (YSO) as a Stage 0/I proto-BD based on the molecular line intensity in the high-density tracers including the HCO$^{+}$ (3-2) line, and the total (dust+gas) mass below the sub-stellar limit. ISO-OPH 200 shows strong accretion and outflow activity in near-infrared observations (Whelan et al. 2018). ISO-OPH 200 shows CN but no HCN emission, which is indicative of photo-dissociation of HCN by the UV radiation field originating from an active accretion zone. This object shows an unusually high H$_{2}$CO ortho/para ratio of $\sim$4, which indicates gas-phase formation for H$_{2}$CO, and it is an interesting case for HNC being formed at very cold temperatures independent of the formation/destruction of HCN (Riaz et al. 2018; 2019b). As we show in the present work, the various observational and model-derived signatures of ISO-Oph 200 indicate it to be a very young proto-BD system in the early stages of formation.

Section~\ref{obs} presents the ALMA CO line (and isotopologues) and continuum observations, the analysis of which is described in Section~\ref{results}. Section~\ref{modelling} presents the results from the physical+chemical modelling of the molecular line observations. Section~\ref{discuss} presents a description of the 3D structure and morphology of the proto-BD system, the physical parameters of the molecular outflow, the various indicators of the extreme youth and classification of ISO-OPH 200, and a star-like formation mechanism of this proto-BD system. Section~\ref{alma-iram} presents a comparison of the ALMA interferometric observations with the IRAM, JCMT, and APEX single-dish data.

\section{Observations}
\label{obs}

The data presented in this paper are from the ALMA programmes 2015.1.00741.S, 2016.1.00545.S, and 2017.1.00107.S. The observations were carried out between March, 2016, and May, 2018. The continuum data from these datasets are in Band 4 (125-163 GHz), Band 6 (211-275 GHz), and Band 7 (275-373 GHz). The Band 4 continuum data was taken in four spectral windows with central wavelength, $\lambda_{c}$, of 1.85 mm, 1.87 mm, 1.99 mm, and 2.02 mm. The Band 6 continuum data was taken in one spectral window with central wavelength of 1.33 mm. The Band 7 continuum data was also taken in four spectral windows with $\lambda_{c}$ of 855.26 $\mu$m, 860.15 $\mu$m, 885.81 $\mu$m, and 890.91 $\mu$m. The angular resolution of the continuum data ranges between 0.15$\arcsec$--1.15$\arcsec$. The molecular line data is only in Band 6; the spectral setup covers the $^{12}$CO (2-1), $^{13}$CO (2-1), C$^{18}$O (2-1) lines. The angular resolution of the molecular line data is $\sim$1.4$\arcsec$, and the velocity resolution is $\sim$0.08 km s$^{-1}$ at 217 GHz. The largest angular structure that can be observed in these configurations is $\sim$12$\arcsec$ in Band 4 and 6 and $\sim$2$\arcsec$ in Band 7. We employed the calibrated data delivered by the EU-ARC. ISO-Oph 200 is detected in all continuum and line images. Data analysis was performed using the CASA software. We have used the convention of measuring the position angle (PA) East of North.

\section{Data Analysis and Results}
\label{results}

\subsection{Continuum Emission}
\label{continuum}

\begin{figure*}
  \centering              
     \includegraphics[width=2.3in]{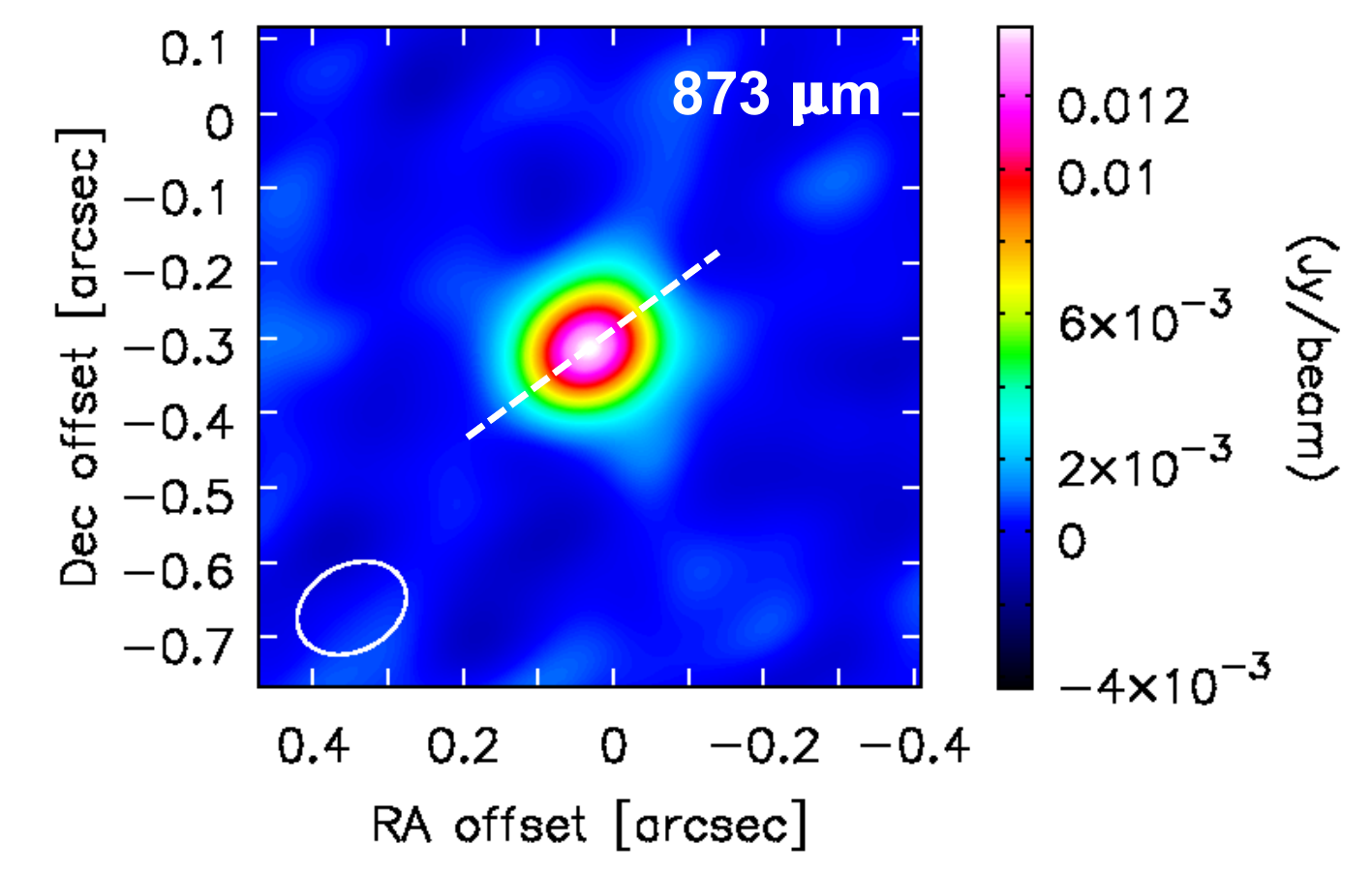} 
     \includegraphics[width=2.3in]{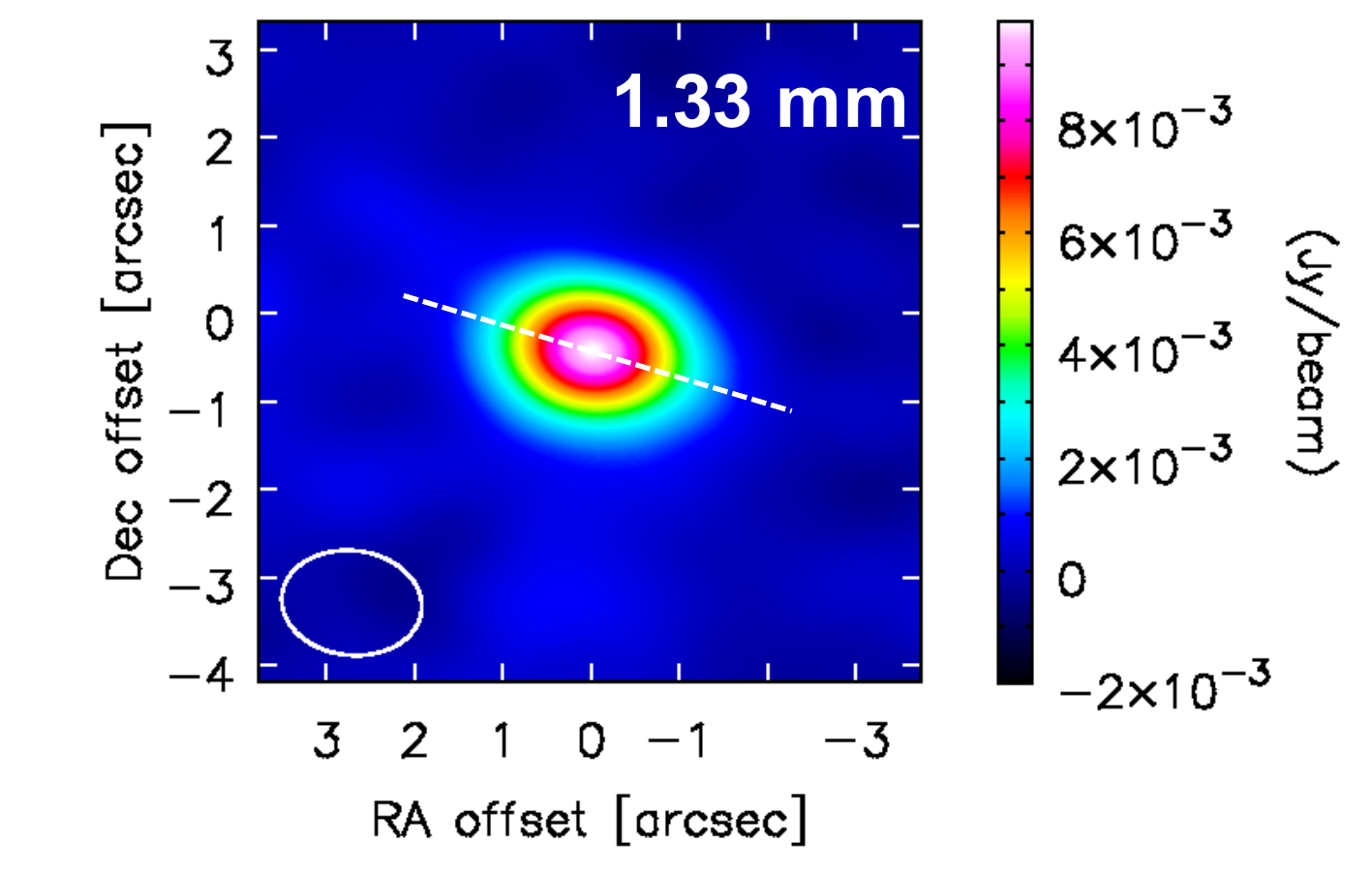}     
     \includegraphics[width=2.29in]{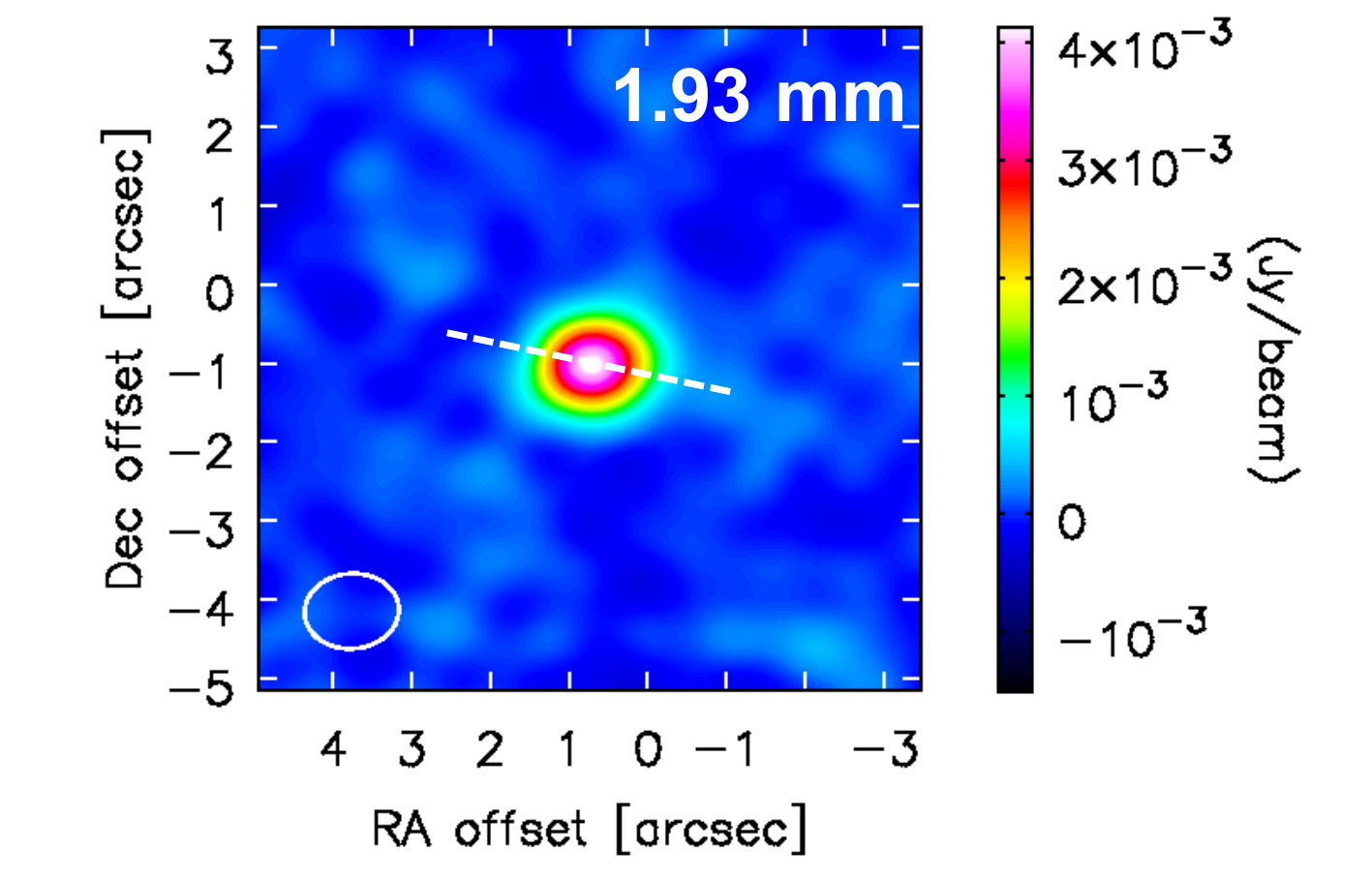}     
     \caption{The Band 7 (left), Band 6 (middle), and Band 4 (right) continuum images for ISO-Oph 200. The white dashed line marks the PA. The colour bar shows the flux scale in units of Jy beam$^{-1}$. The 1-$\sigma$ rms is listed in Table~\ref{cont-data}. The beam size is shown at the bottom, left corner. The xy-axes show the position offset relative to the ISO-Oph 200 position. North is up, east is to the left.   }
     \label{cont-img}
  \end{figure*}

Figure~\ref{cont-img} shows the Band 7 (873 $\mu$m), Band 6 (1.33 mm), and Band 4 (1.93 mm) continuum images for ISO-Oph 200. For the Band 4 and Band 7 data, we have produced a continuum image by combining all four spectral windows. We have used the CASA {\it uvmodelfit} and {\it imfit} tasks to measure the physical parameters and fluxes in the continuum images. The results are listed in Table~\ref{cont-data}. The proto-BD system is marginally resolved in all images. The projected size deconvolved from the beam is 0.4$\arcsec$$\times$0.3$\arcsec$ ($\sim$58$\times$43 au) in Band 4, 0.8$\arcsec$$\times$0.5$\arcsec$ ($\sim$115$\times$72 au) in Band 6, and 0.07$\arcsec$$\times$0.04$\arcsec$ ($\sim$8.6$\times$8.6 au) in Band 7. The differences in the PA and sizes are likely due to the beam convolution effect. The peak flux does not precisely coincide with (RA, Dec) = (0,0) position offset but shows a slight shift of $\sim$0.3$\arcsec$--0.5$\arcsec$. However, the object position measured by {\it imfit} is consistent with the 2MASS position within the position uncertainty (Table~\ref{cont-data}). The position offset is likely due to the difference in the target position between the near-infrared and millimeter observations and the beam convolution effect.

Table~\ref{cont-data} lists the peak and the integrated fluxes in the continuum images for ISO-Oph 200. Assuming that the dust emission at sub-millimeter/millimeter (sub-mm/mm) wavelengths is optically thin, the dust continuum mass can be calculated from the sub-mm/mm flux density as

\begin{equation}
M_{dust} = \frac{F_{\nu}~d^{2}}{B_{\nu}(T_{dust})~ \kappa_{\nu}}~,
\end{equation}

\noindent where $F_{\nu}$ is the flux density, $d$ is the distance to the source, $B_{\nu}$ is the Planck function at the dust temperature $T_{dust}$, and $\kappa_{\nu}$ is the dust opacity. 

The frequency dependence of the dust mass opacity is assumed to be

\begin{equation}
\kappa_{\nu} = 0.1 \times \left( \frac{\nu}{10^{12}} \right)^{\beta} [cm^{-2} g^{-1}] 
\end{equation}

\noindent for optically thin emission (Beckwith et al. 1990). Here $\beta$ is the slope of the opacity law and is related to the sub-mm to mm spectral slope of the SED, $\alpha$, as $\alpha \approx \beta$ + 2 in the Rayleigh-Jeans limit (e.g., Beckwith \& Sargent 1991). We have used all of the nine spectral windows in Bands 4, 6 and 7 and measured the slope between $\sim$873$\mu$m to $\sim$1.93 mm. The fit to these data points has a slope $\alpha$ = 3.2$\pm$0.3, which implies $\beta$ = 1.2$\pm$0.3 for ISO-Oph 200. The integrated fluxes were convolved over the beam size to measure the indices. These estimates are consistent with optically thin envelope emission, which would have $\alpha \sim$ 3.5-4 for typical dust opacities and $\beta \sim$ 1.7 similar to interstellar dust (e.g., Lin et al. 2016; Li et al. 2017; Joergensen et al. 2007). The $\alpha$ and $\beta$ values for this proto-BD are within the range estimated for low-mass Class 0/I protostars (e.g., Li et al. 2017; Joergensen et al. 2007). 

For $\beta$ = 1.2, the dust opacity, $\kappa_{\nu}$, is 0.030111 cm$^{2}$ g$^{-1}$ at 872.7 $\mu$m (343.517 GHz), 0.01876 cm$^{2}$ g$^{-1}$ at 1.33 mm (225.496 GHz), and 0.01231 cm$^{2}$ g$^{-1}$ at 1.93 mm (154.985 GHz). The kinetic temperature is expected to be relatively low ($\sim$10 K) throughout the outer and inner envelope layers in proto-brown dwarfs (Machida et al. 2009). We have assumed a dust and gas temperature of 10 K, and a gas to dust mass ratio of 100. The resulting measurements on the total (dust+gas) mass, M$_{d+g}$, are listed in Table~\ref{cont-data}, and range between 4--7 M$_{Jup}$ for ISO-Oph 200. The uncertainties in the mass estimates due to the assumption on $T_{dust}$ and the gas to dust mass ratio could result in $\sim$2-10 times lower M$_{d+g}$ than estimated here, for e.g., using a higher dust temperature of 20 K will result in a factor of $\sim$2 lower mass (Riaz et al. 2019a).

The M$_{d+g}$ derived from ALMA observations with standard parameter sets is lower than the mass M$_{d+g}$ = 10$\pm$2 M$_{Jup}$ derived from the single-dish JCMT/SCUBA-2 850 $\micron$ continuum flux. The SCUBA-2 image is shown in Appendix~\ref{alma-iram}. A similar difference of a factor of $\sim$2 was noted in the M$_{d+g}$ derived from ALMA and SCUBA-2 observations for the proto-BD M1701117 (Riaz et al. 2019a). The difference can be explained by the source size measured from the ALMA continuum image and the $\sim$14.5$\arcsec$ ($\sim$2088 au) beam size of the SCUBA-2 bolocam. The SCUBA-2 image (Appendix~\ref{alma-iram}) shows a slightly elongated structure instead of a point-like source seen in the ALMA continuum images. However, we do not see any complex extended structures in the SCUBA-2 or ALMA continuum images as seen in the ALMA molecular line images. The largest structure that can be observed in the present ALMA configuration is $\sim$10$\arcsec$ or $\sim$1440 au. The 1-$\sigma$ rms sensitivity is $\sim$1 mJy beam$^{-1}$ in the ALMA continuum image and $\sim$4 mJy beam$^{-1}$ in the SCUBA-2 image. The small difference in the masses suggests that ALMA has not resolved out any dense circumstellar material but rather the diffuse emission in the outer envelope regions. It may also be the case that the much larger beam size of SCUBA-2 has resulted in contamination from the unrelated surrounding cloud material that can enhance the observed emission.

\begin{table*}
\centering
\caption{Parameters derived from the continuum images}
\label{cont-data}
\begin{adjustbox}{scale=0.9,center}
\begin{threeparttable}
\begin{tabular}{lcccccccccc} 
\hline

Band & $\lambda_{c}$ & Position (J2000)\tnote{a} & 1-$\sigma$ rms & Beam Size  & PA & Peak Flux & Integrated Flux & M$_{d+g}$\tnote{b} \\
 &   &  & [mJy beam$^{-1}$] &   &  & [mJy beam$^{-1}$] & [mJy] & [M$_{Jup}$] \\
\hline 		

7 & 872.7 $\mu$m & 16:31:43.75 -24:55:24.92 & 2.6  & 0.15$\arcsec$$\times$0.11$\arcsec$ & 127$\degr \pm$10$\degr$ & 13.93$\pm$0.38  & 18.08$\pm$0.002  & 4.2$\pm$0.5  \\
6 & 1.33 mm & 16:31:43.75 -24:55:25.04 & 1.5  & 1.59$\arcsec$$\times$1.19$\arcsec$ & 73$\degr \pm$10$\degr$ & 9.51$\pm$0.32 & 10.57$\pm$0.00015  & 6.4$\pm$0.5  \\
4 & 1.93 mm & 16:31:43.75 -24:55:25.01 & 0.8  & 1.20$\arcsec$$\times$0.95$\arcsec$ & 80$\degr \pm$5$\degr$ & 4.21$\pm$0.10  & 4.36$\pm$0.18  & 7.1$\pm$0.7  \\	 

\hline
\end{tabular}
\begin{tablenotes}
  \item[a] Position uncertainty is $\pm$0.002$\arcsec$. 
  \item[b] The uncertainty on M$_{d+g}$ was calculated from the flux error. 
\end{tablenotes}
\end{threeparttable}
\end{adjustbox}
\end{table*}

\subsection{C$^{18}$O (2-1) Emission}
\label{C18O}

\begin{figure*}
  \centering              
     \includegraphics[width=2.2in]{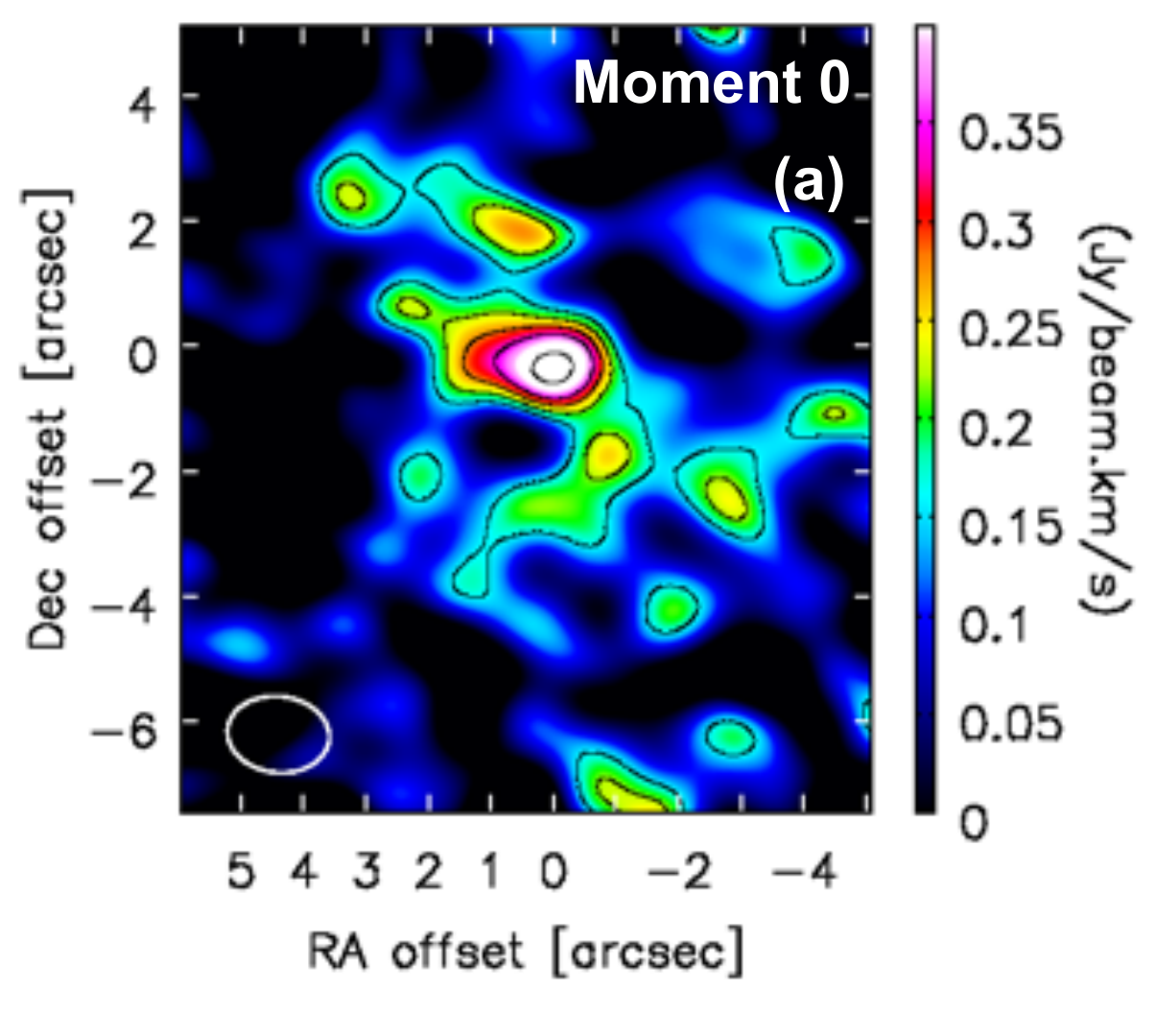}  
     \includegraphics[width=2.2in]{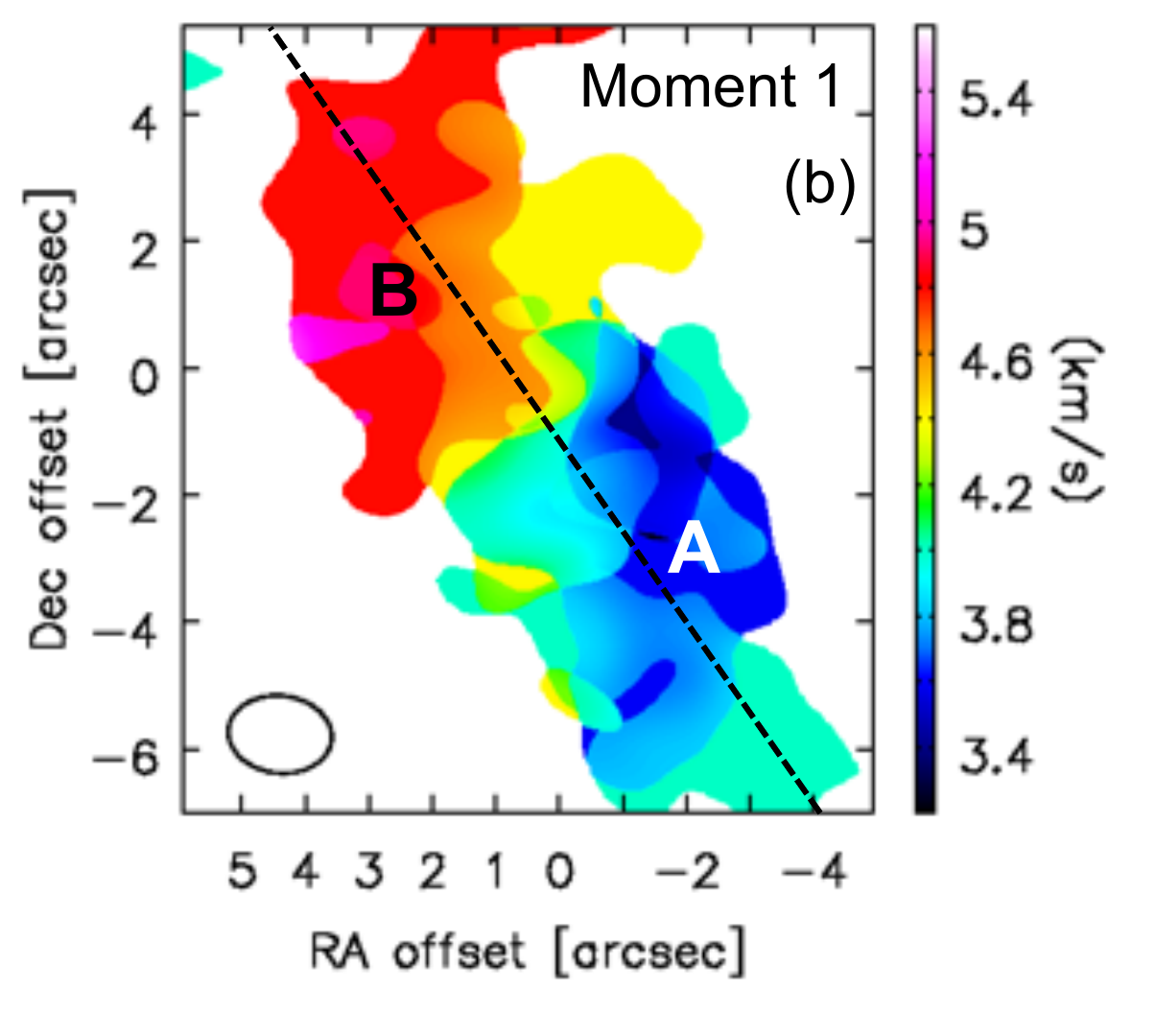}  
     \includegraphics[width=2.2in]{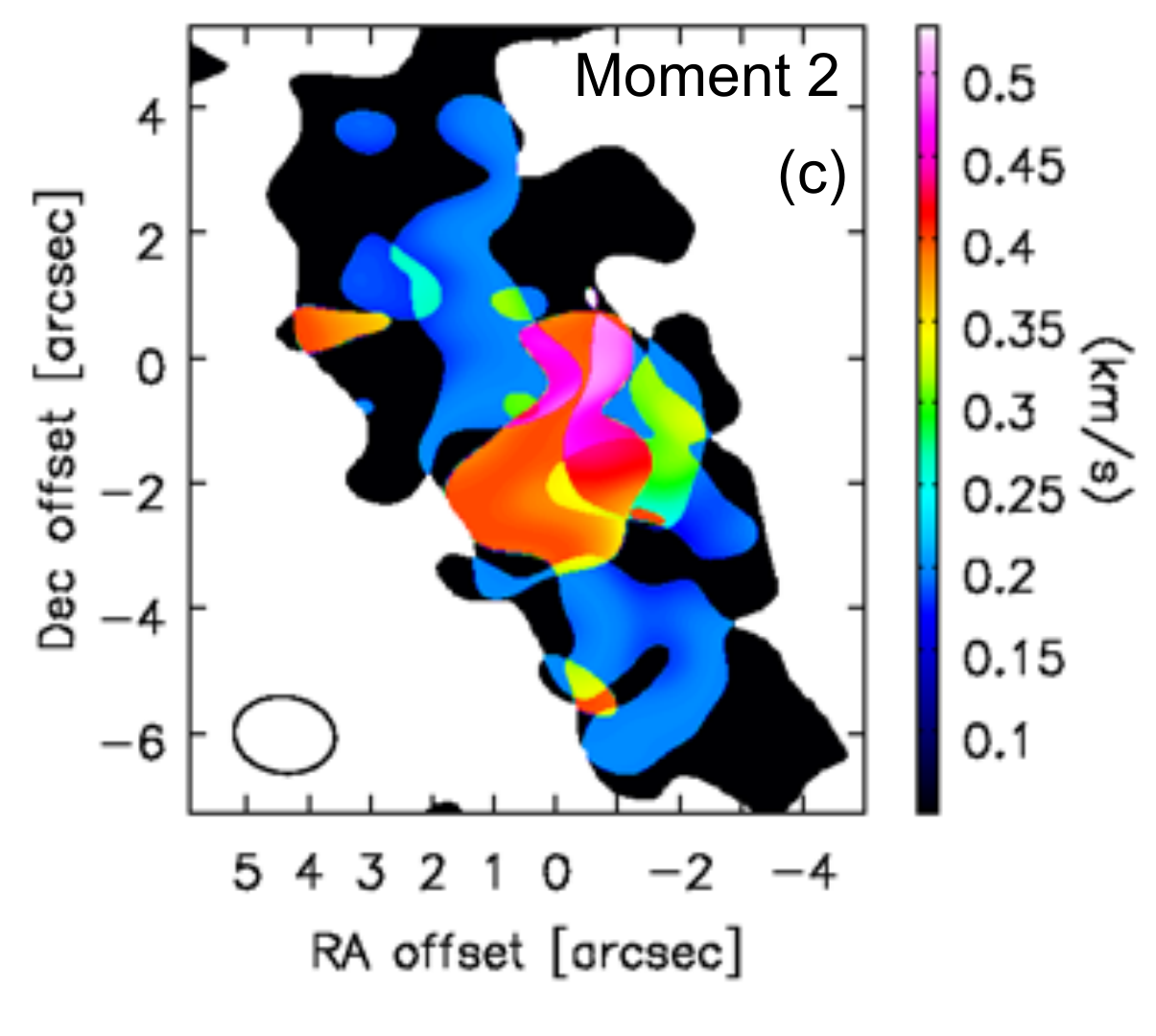}      
     \caption{The moment maps in the C$^{18}$O (2-1) line for ISO-OPH 200. {\bf (a)} Moment 0 map. The colour bar shows the flux scale in units of Jy beam$^{-1}$.km s$^{-1}$. The contour levels are from 2-$\sigma$ to 10-$\sigma$ in steps of 2-$\sigma$. The 1-$\sigma$ rms is $\sim$0.05 Jy beam$^{-1}$.km s$^{-1}$. {\bf (b)} Moment 1 map and {\bf (c)} moment 2 map. Dashed lines mark the PAs. The xy-axes show the position offset relative to the position of ISO-OPH 200. North is up, east is to the left. }
     \label{C18O-moment}
  \end{figure*}

\begin{figure}
  \centering              
     \includegraphics[width=3in]{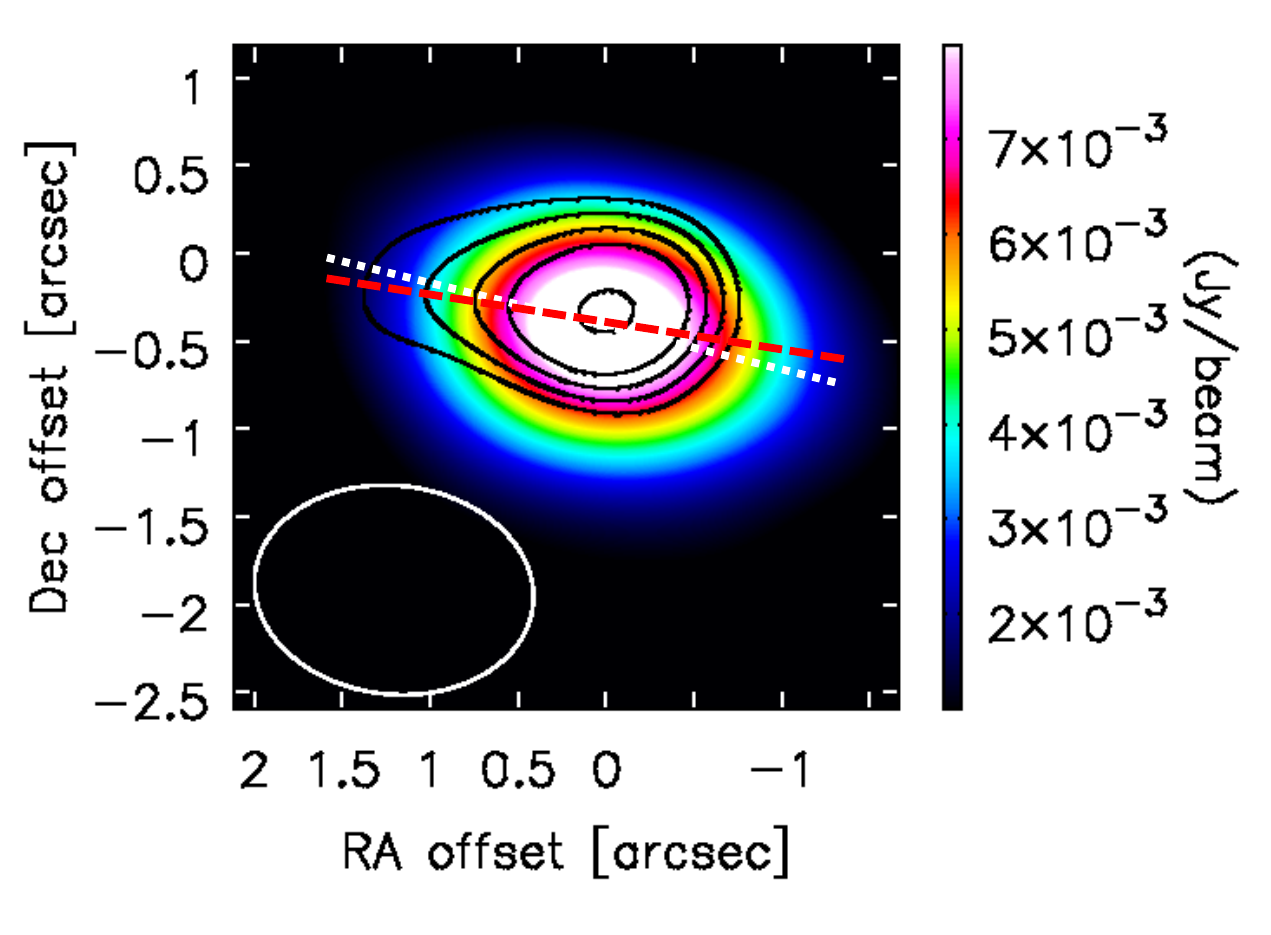} 
     \caption{The 1.3 mm continuum image (raster map) overplotted on the C$^{18}$O moment 0 map (black contours). The colour bar for the continuum image shows the flux scale in units of Jy beam$^{-1}$. The 1-$\sigma$ rms is 1.5 mJy beam$^{-1}$. The contour levels in the C$^{18}$O map are from 6-$\sigma$ to 14-$\sigma$ in steps of 2-$\sigma$; the 1-$\sigma$ rms is $\sim$0.05 Jy beam$^{-1}$.km s$^{-1}$. The beam size is shown in the bottom, left corner. The red and white dashed lines mark the PA for the C$^{18}$O and continuum images, respectively. }
     \label{cont-c18o}
  \end{figure}

Figure~\ref{C18O-moment} shows the moment maps in the C$^{18}$O line emission. The individual velocity channel maps are discussed in Appendix~\ref{channel}. The C$^{18}$O moment 0 map (Fig.~\ref{C18O-moment}a) shows a compact structure centered at the proto-BD position, with a major axis PA = 83$\degr \pm$5$\degr$. The extended emission in the C$^{18}$O moment 0 map at $>$3$\arcsec$ is at a $\leq$3-$\sigma$ level. The $>$5-$\sigma$ emission is only seen in a compact shape very close to the source position. This can also be seen in the velocity channel maps (Fig.~\ref{C18O-chmaps}). There is a nice overlap between the C$^{18}$O moment 0 and the 1.3 mm continuum image (Fig.~\ref{cont-c18o}), both of which were observed at the same angular resolution of $\sim$1.4$\arcsec$. Figure~\ref{cont-c18o} shows the C$^{18}$O emission detected at a $>$5-$\sigma$ level. The PA measured for the continuum image (73$\degr \pm$10$\degr$) is similar to the PA for the C$^{18}$O emission. Note that the PA may not be the real orientation of these images as it seems strongly affected by the beam shape. While the peak emission in C$^{18}$O is seen within $\sim$2$\arcsec$ of the source position, we used a beam size of approximately 5$\arcsec \times$6$\arcsec$ to extract the spectrum so as to cover the full spatial scale over which the C$^{18}$O emission is seen in the velocity channels (Fig.~\ref{C18O-chmaps}).

\begin{figure}
\center 
     \includegraphics[width=2.3in]{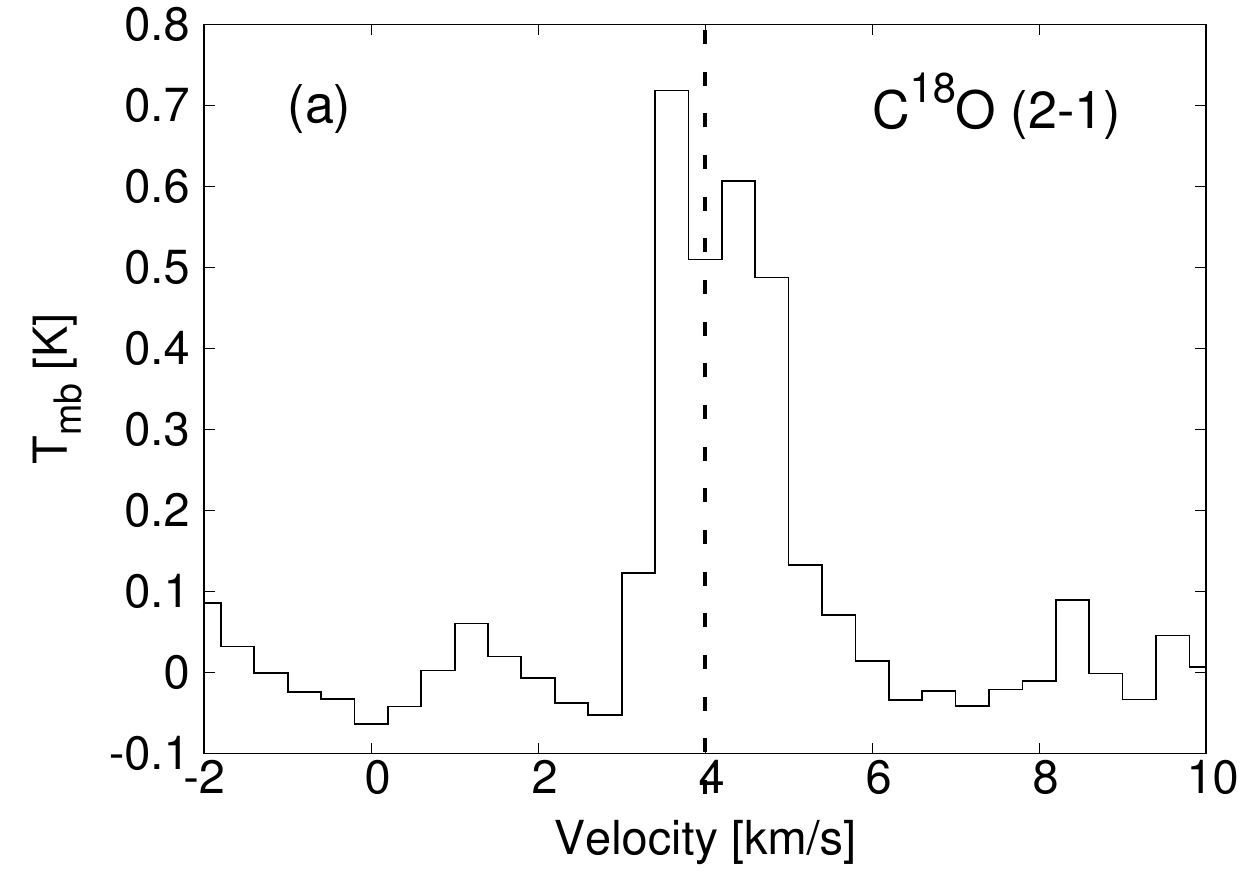}  
     \includegraphics[width=2.3in]{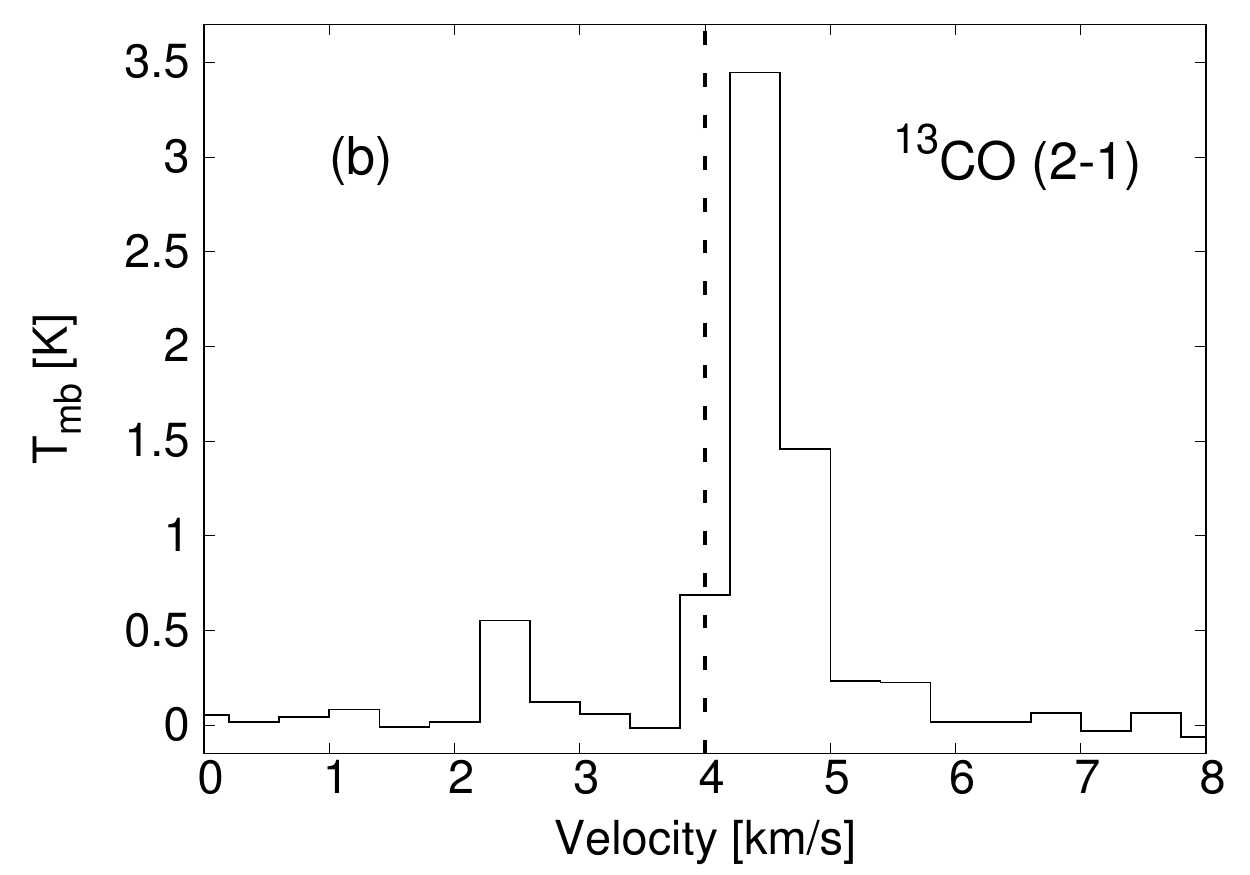} 
     \includegraphics[width=2.3in]{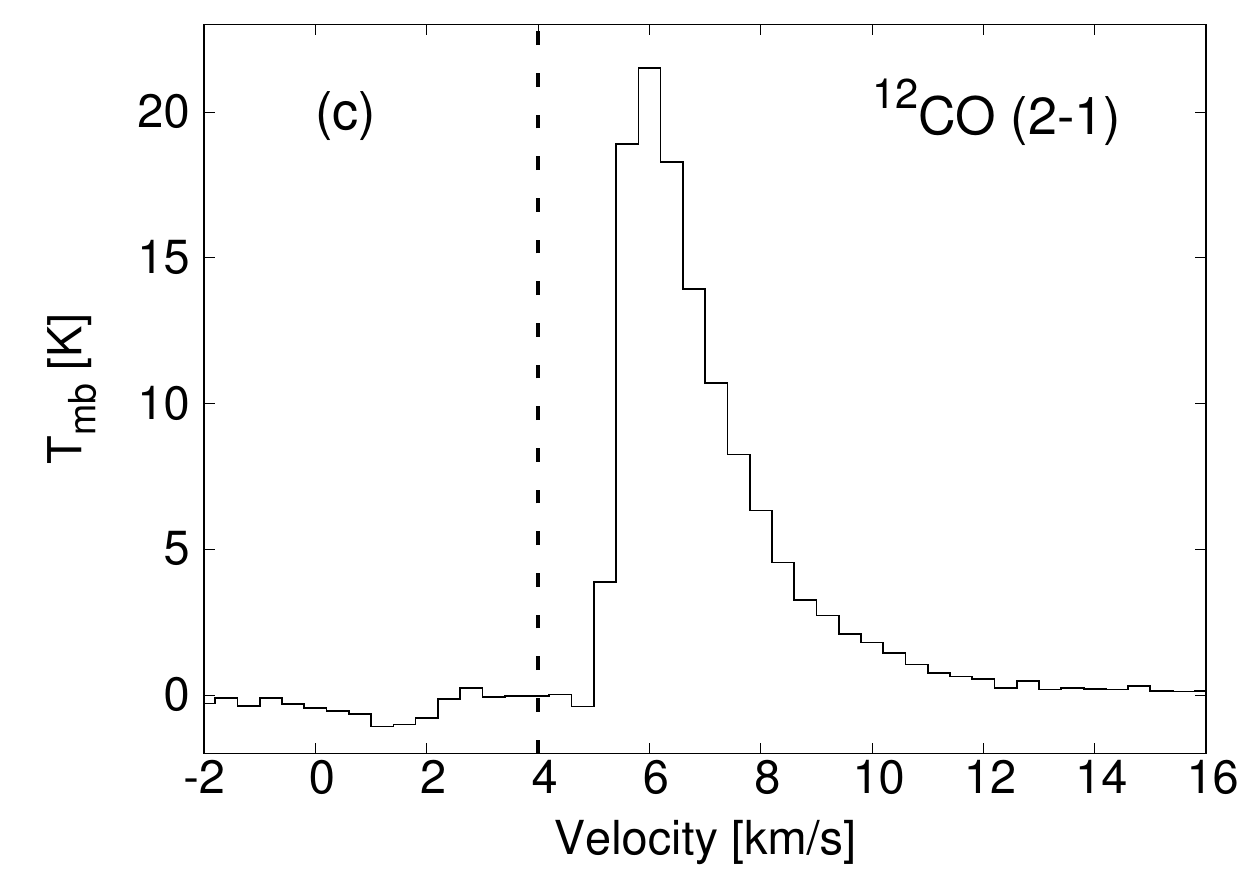}     
    \caption{The C$^{18}$O (2-1) (left), $^{13}$CO (2-1) (middle), and $^{12}$CO (2-1) (right) spectra for ISO-OPH 200. Dashed line marks the source V$_{LSR}$. } 
    \label{spectra} 
\end{figure}

\begin{figure}
  \centering              
     \includegraphics[width=2.8in]{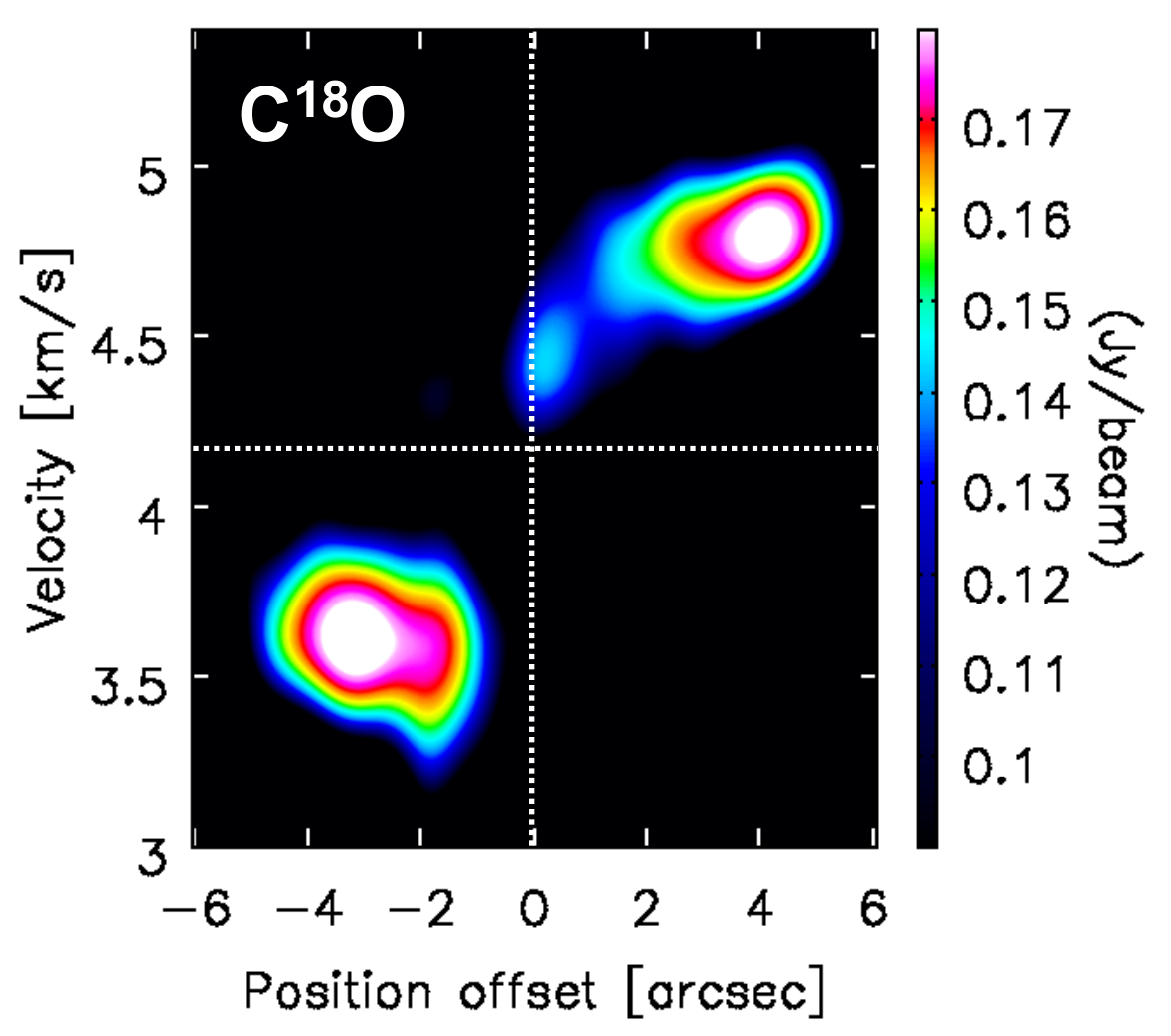} 
     \caption{The PV diagram in the C$^{18}$O line emission. Dashed lines mark the source position and V$_{LSR}$. The colour bar shows the flux scale in units of Jy beam$^{-1}$. The 1-$\sigma$ rms is $\sim$0.05 Jy beam$^{-1}$. The red-shifted lobe is located towards the north-east of the source. }
     \label{pvd}
  \end{figure}

The C$^{18}$O moment 1 map (Fig.~\ref{C18O-moment}b) shows that the velocity is $\sim$4.0 -- 4.2 km s$^{-1}$ at the proto-BD position or the (0,0) position offset. The C$^{18}$O (2-1) spectrum in Fig.~\ref{spectra}a shows a double-peaked line profile, centered at $\sim$4 km s$^{-1}$. We have therefore set the source V$_{LSR} \sim$4.0$\pm$0.2 km s$^{-1}$ for ISO-OPH 200. The moment 1 map (Fig.~\ref{C18O-moment}b) in C$^{18}$O shows two bright structures, labelled {\bf A} and {\bf B}, at blue- and red-shifted velocities with respect to the source V$_{LSR}$, respectively, and the velocity spread is seen along a major axis PA = 37$\degr \pm$10$\degr$. We have constructed a position-velocity diagram (PVD) in the C$^{18}$O line for a cut along this PA (Fig.~\ref{pvd}). The PVD shows two bright lobes with peak emission at position offsets of approximately +3.8$\arcsec$ and -3.3$\arcsec$, and velocity offset of approximately $\pm$0.6 km s$^{-1}$ with respect to the source position and V$_{LSR}$. The C$^{18}$O moment 2 map (Fig.~\ref{C18O-moment}c) shows that the velocity dispersion increases towards the central source position.

\subsection{$^{13}$CO Line Emission}
\label{13CO}

\begin{figure*}
  \centering              
     \includegraphics[width=2in]{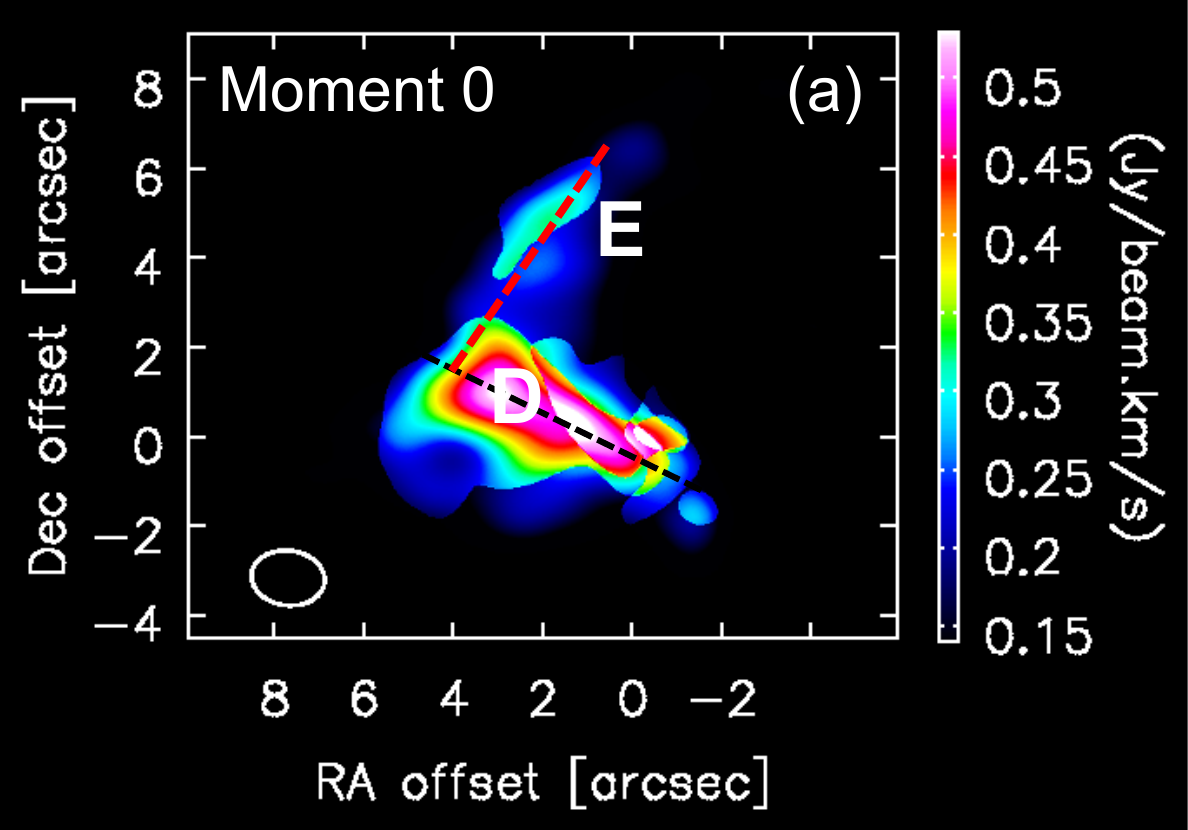}  \hspace{0.17in}
     \includegraphics[width=2in]{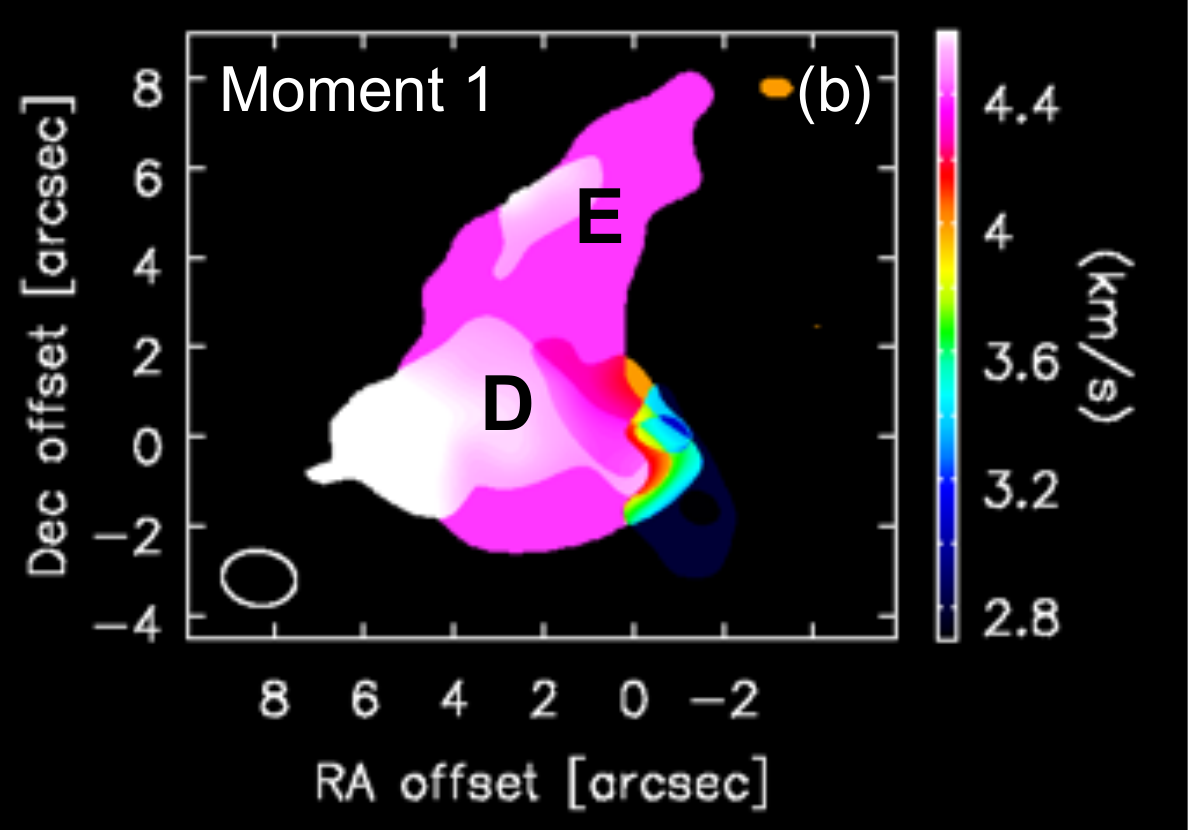}  \hspace{0.17in}
     \includegraphics[width=2in]{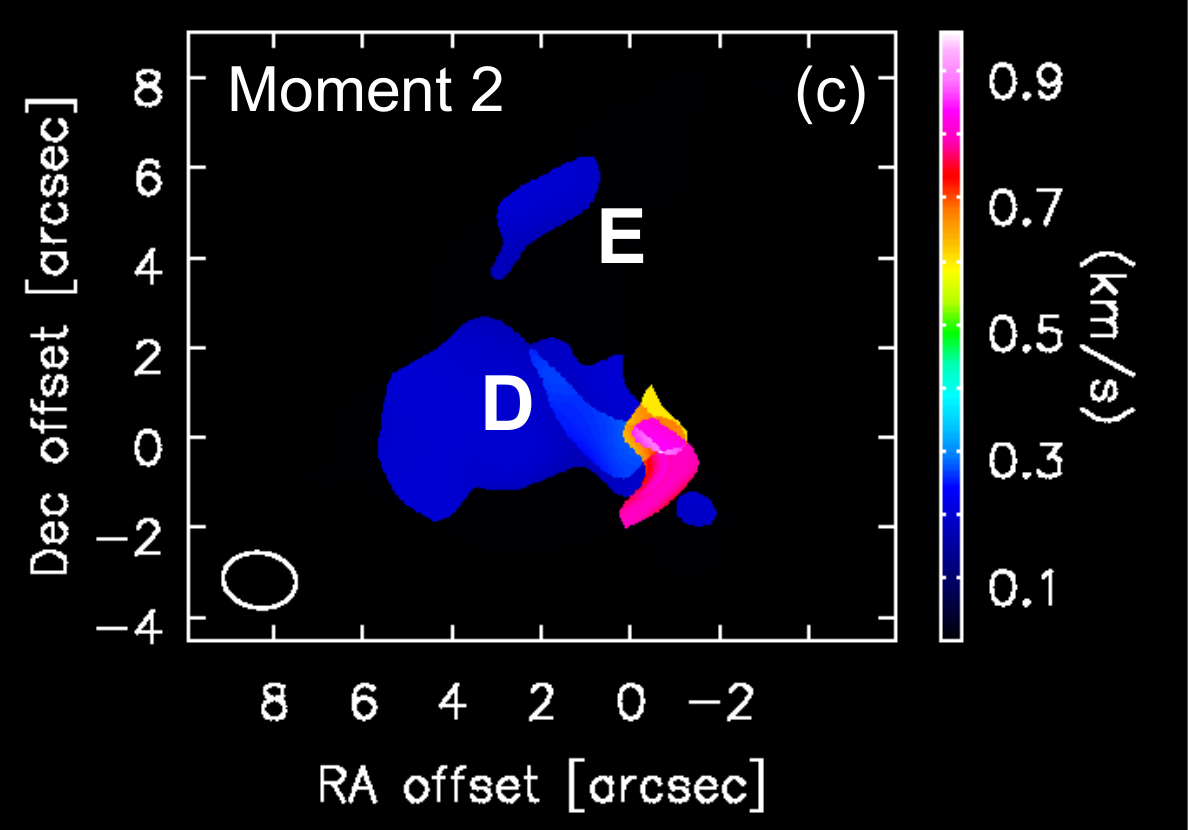}            
     \caption{The moment maps in the $^{13}$CO (2-1) line for ISO-OPH 200. {\bf (a)} Moment 0 map. Dashed lines are the PAs measured for the {\bf D E} structures. The colour bar shows the flux scale in units of Jy beam$^{-1}$.km s$^{-1}$. The 1-$\sigma$ rms is 0.1 Jy beam$^{-1}$.km s$^{-1}$. {\bf (b)} Moment 1 map and {\bf (c)} moment 2 map. The xy-axes show the position offset relative to the position of ISO-OPH 200. North is up, east is to the left. }
     \label{13CO-moment}
  \end{figure*}

Figure~\ref{13CO-moment} shows the moment maps in the $^{13}$CO (2-1) line. The individual velocity channel maps are discussed in Appendix~\ref{channel}. The brightest emission is seen in the extended, complex {\bf D} structure (Fig.~\ref{13CO-moment}a). We have measured a PA 63$\degr$$\pm$5$\degr$ for this structure. A much fainter extended structure {\bf E} is seen at a different PA of 326$\degr \pm$5$\degr$. These extended structures are nearly perpendicular to each other (Fig.~\ref{13CO-moment}a). As seen in the moment 1 map in Fig.~\ref{13CO-moment}b, the emission in the {\bf D} and {\bf E} structures is red-shifted with respect to the source V$_{LSR} \sim$4.0 km s$^{-1}$, while much fainter blue-shifted emission is seen at a position offset of about (-1,-1) and velocities of $\sim$2.8-3.4 km s$^{-1}$. The moment 2 map (Fig.~\ref{13CO-moment}c) shows that the highest velocity dispersion is towards the central source position at (0,0) offset. In comparison, the velocity gradient is nearly constant along the {\bf D} and {\bf E} structures, and shows very little dispersion of $\leq$0.1 km s$^{-1}$. The $^{13}$CO (2-1) spectrum (Fig.~\ref{spectra}b) shows a red-shifted peak that is much stronger than the weak blue-shifted lobe. The spectrum was extracted using a beam size of approximately 10$\arcsec \times$10$\arcsec$ to cover the full spatial scale of the observed emission in the velocity channel maps (Fig.~\ref{13CO-chmaps}).

\subsection{$^{12}$CO Line Emission}
\label{CO}

\begin{figure*}
  \centering              
     \includegraphics[width=2in]{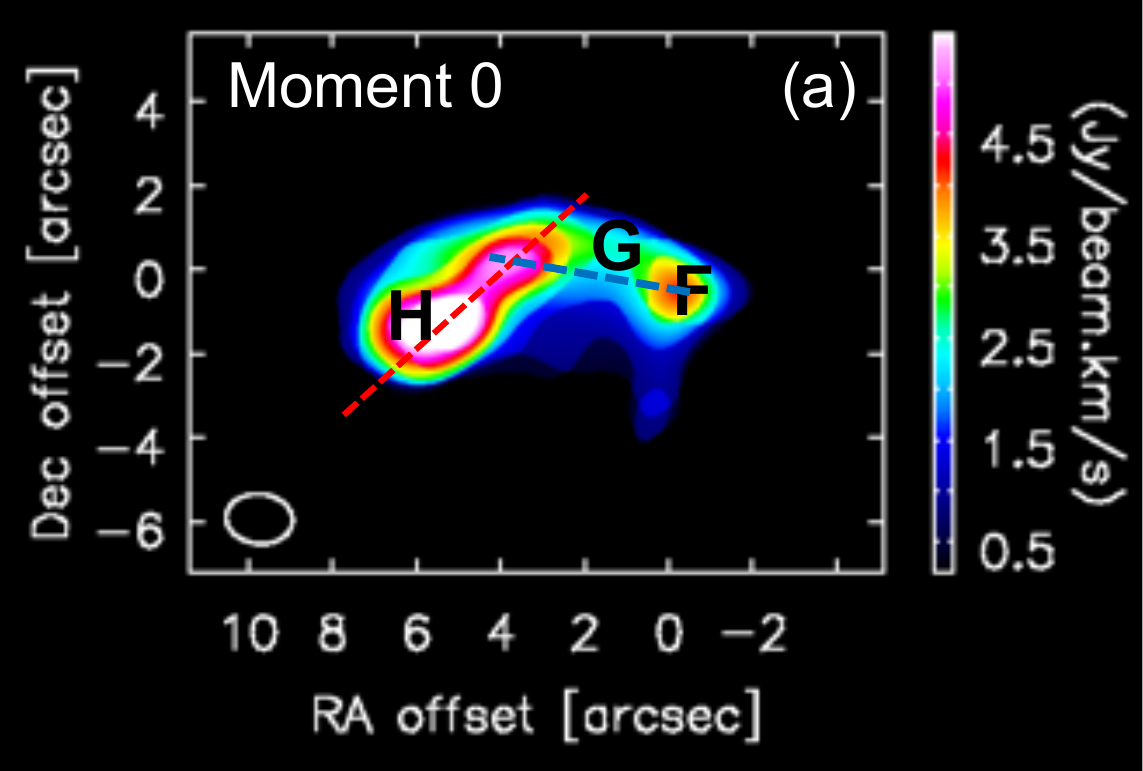}   \hspace{0.17in}
     \includegraphics[width=2in]{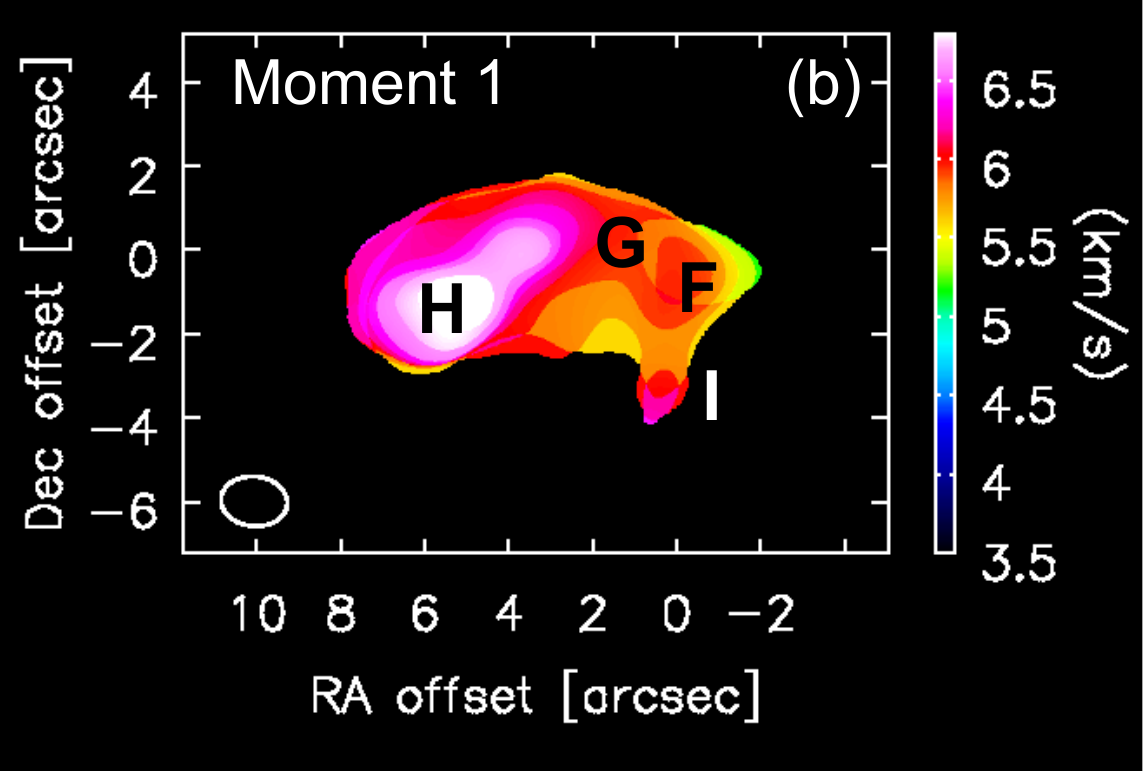}   \hspace{0.17in}
     \includegraphics[width=2in]{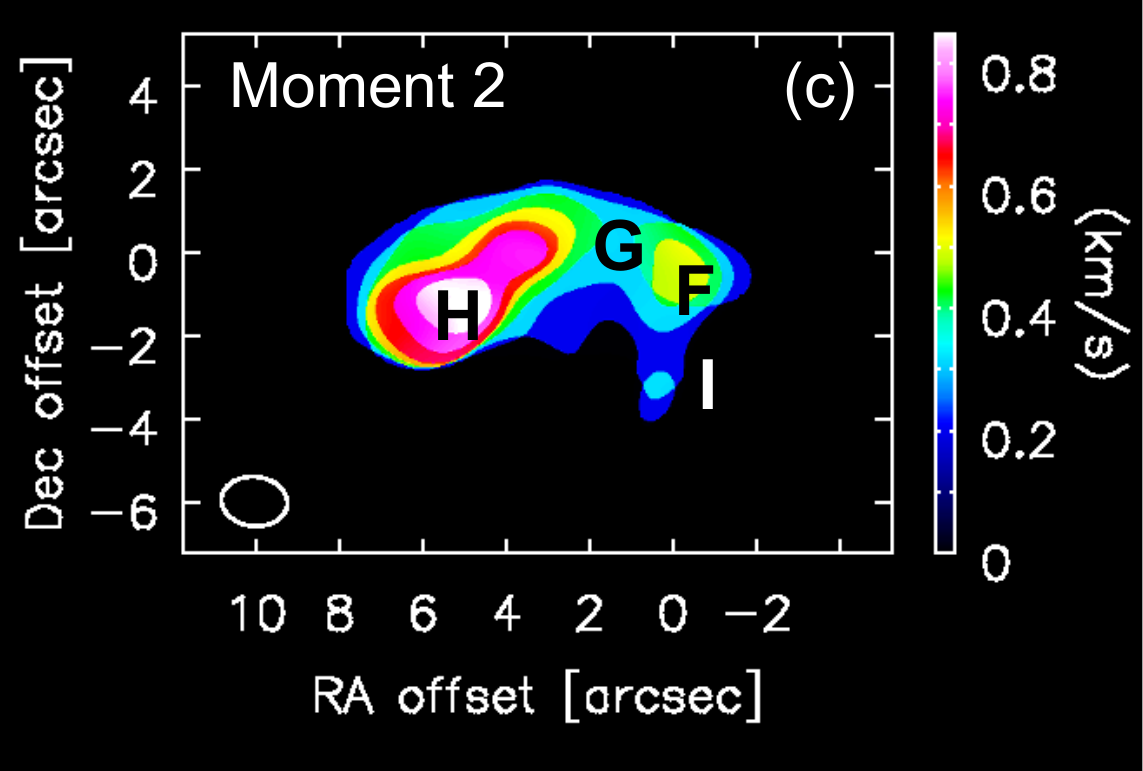}            
     \caption{The moment maps in the $^{12}$CO (2-1) line for ISO-OPH 200. `{\bf F G H I}' are the different structures seen in channel maps. {\bf (a)} Moment 0 map. Dashed lines are the PAs measured for the {\bf G H} structures. The colour bar shows the flux scale in units of Jy beam$^{-1}$.km s$^{-1}$. The 1-$\sigma$ rms is 1.0 Jy beam$^{-1}$.km s$^{-1}$. {\bf (b)} Moment 1 map and {\bf (c)} moment 2 map. The xy-axes show the position offset relative to the position of ISO-OPH 200. North is up, east is to the left. }
     \label{CO-moment}
  \end{figure*}

Figure~\ref{CO-moment} shows the moment maps in the $^{12}$CO (2-1) line. The individual velocity channel maps are discussed in Appendix~\ref{channel}. The brightest integrated CO emission in the moment 0 map (Fig.~\ref{CO-moment}a) is seen at the southern edge of the extended south-east lobe, labelled {\bf H}, at a position offset of approximately (+6, -2). A weaker peak is seen as a compact structure, labelled {\bf F}, at the (0,0) source position. An arc-like extended structure, labelled {\bf G}, is seen between {\bf F} and {\bf H}. The emission in {\bf G} is weaker than {\bf F} or {\bf H}. A much fainter blob-like feature is seen at a position offset of approximately (0, -4). Note that the PA changes at short position offsets from structure {\bf G} to {\bf H}, and these structures cannot be clearly separated from each other. We have measured a PA of 80$\degr \pm$5$\degr$ for {\bf G} and 127$\degr \pm$7$\degr$ for {\bf H}. A faint tail-like structure, labelled {\bf I}, is seen in the CO channel maps at velocities $\geq$6.4 km s$^{-1}$ (Fig.~\ref{CO-chmaps}). The emission in {\bf I} is detected at $<$3-$\sigma$ level, due to which this structure appears as diffuse emission or a tiny dot in the moment maps (Fig.~\ref{CO-moment}). 

The $^{12}$CO moment 1 map (Figs.~\ref{CO-moment}b) shows the peak CO emission at velocities of $\sim$6.5-7 km s$^{-1}$ in the bright extended lobe {\bf H}. This structure also shows the highest velocity dispersion ($\sim$1 km s$^{-1}$) compared to other structures, as seen in the moment 2 map (Figs.~\ref{CO-moment}c), and the velocity gradient increases from the structure {\bf F} along the eastward extension {\bf G} towards the emission peak {\bf H}.

The $^{12}$CO spectrum in Fig.~\ref{spectra} shows that all of the CO emission is red-shifted compared to the source V$_{LSR} \sim$4.0 km s$^{-1}$, as also seen in the moment 1 map (Fig.~\ref{CO-moment}b). The CO spectrum was extracted over a beam size of 10$\arcsec \times$6$\arcsec$ covering the full spatial scale of the emission observed in the velocity channel maps (Fig.~\ref{CO-chmaps}). The spectrum shows an extended broad wing at velocities of $\geq$8 km s$^{-1}$. As can be seen in Fig.~\ref{CO-chmaps}, the emission in the $\sim$8-10 km s$^{-1}$ channels mainly arises from the structure {\bf H} with faint emission arising from {\bf I}, while no definitive structure is seen in the channels at $v >$ 9.6 km s$^{-1}$. A sharp blue cut-off is seen in the CO spectrum (Fig.~\ref{spectra}) at around 5 km s$^{-1}$, such that the emission drops from the peak flux of $\sim$20 K to zero over just two velocity channels. The possible reasons for this sharp asymmetry in the observed CO emission are discussed in Section~\ref{3D-morph}.

\section{Origin of emission: Comparison with core collapse model}
\label{modelling}

To interpret the molecular line observations and understand the origin of emission in the various structures, we have conducted line radiative transfer modelling using an improvised version of the core collapse model for brown dwarf formation presented in Machida et al. (2009). These are 3D magneto-hydro-dynamic (MHD) simulations of brown dwarf formation via gravitational collapse of a very low-mass core. The evolution of the cloud in these simulations is calculated from the pre-stellar cloud core stage until about 0.06 Myr after the proto-brown dwarf formation. Without any artificial setting, four different components or zones appear naturally with the cloud evolution: (i) an infalling envelope that evolves into a circumstellar pseudo-disc; (ii) an inner Keplerian disc; (iii) a wide-angled low-velocity outflow; (iv) a collimated high-velocity jet.

\begin{figure*}
  \centering              
     \includegraphics[width=5in]{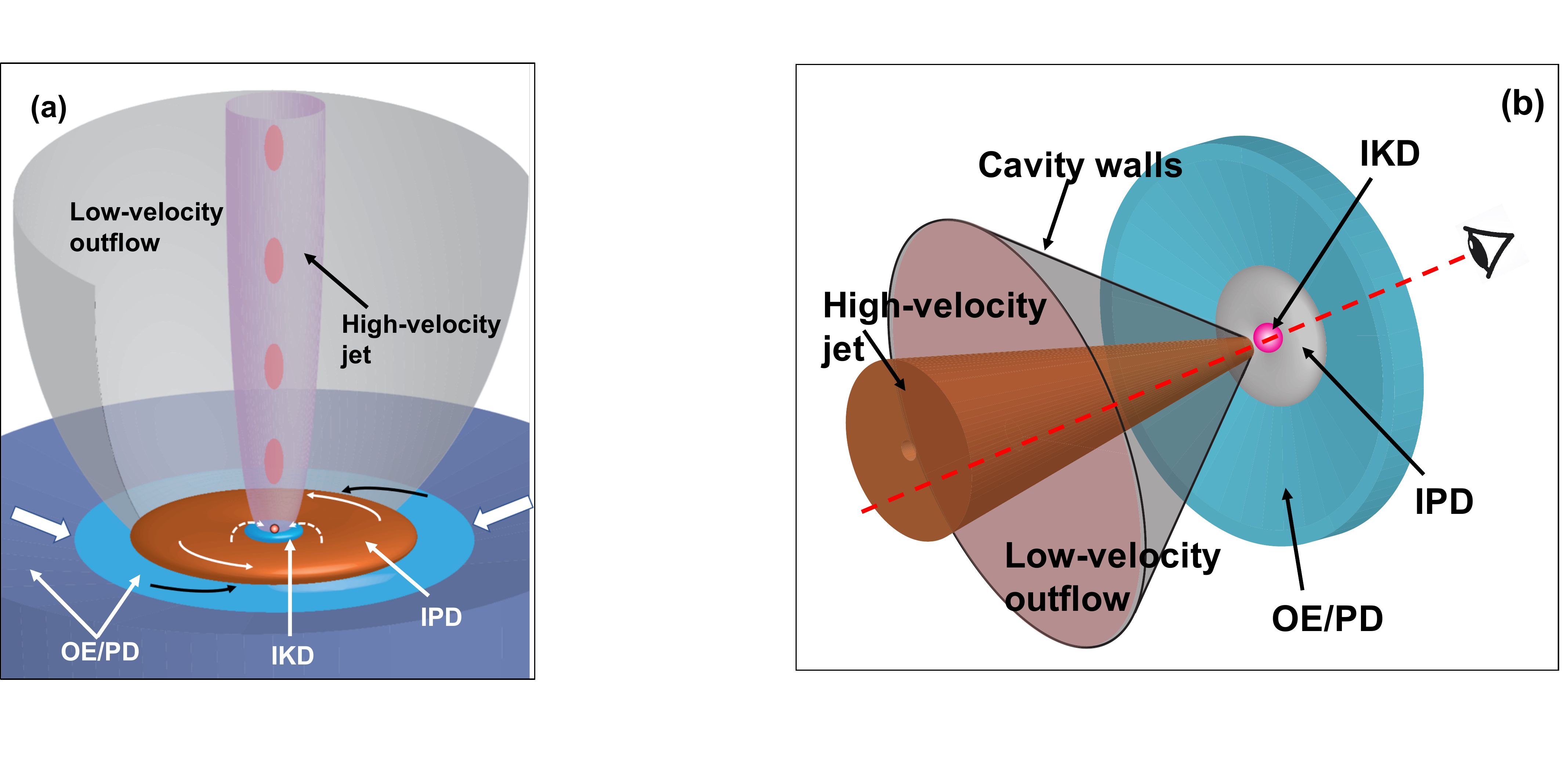}           
     \caption{Schematic view of the physical structure of the brown dwarf formation via core collapse model (Machida et al. 2009). The labels `OE/PD', 'IPD', and 'IKD' imply the outer envelope/pseudo-disc, inner pseudo-disc, and inner Keplerian disc regions, respectively. Panel {\bf (a)} shows the view at an edge-on (90$\degr$) inclination while panel {\bf (b)} shows the view at the best-fit 30$\degr$ inclination. The line of sight is marked by a red dashed line. }
     \label{model}
  \end{figure*}

Figure~\ref{model} shows a schematic diagram of the basic model structure at an edge-on (90$\degr$) and at a 30$\degr$ inclination that provides a best-fit to all molecular line data for this proto-BD system. The edge-on map (Fig.~\ref{model}a) shows the view perpendicular to the (pseudo-)disc axis and provides a clear view of the jet/outflow region, while in the 30$\degr$ map (Fig.~\ref{model}b), we are looking at the system partially through the outflow cavity, and have a direct view of a part of the jet. Figure~\ref{model} is useful to comprehensively understand observational features. However, it should be noted that it is difficult to correctly visualize the orientation (line of sight) of the observer with respect to the system and completely reproduce the observational characteristics with a simple 2D schematic view due to the complex nature of the structures seen in both observations and simulations. A better understanding of the morphology of the system can be obtained with the 3D schematic view presented in Sect.~\ref{3D-morph}.

\begin{figure*}
  \centering     
     \includegraphics[width=2.5in]{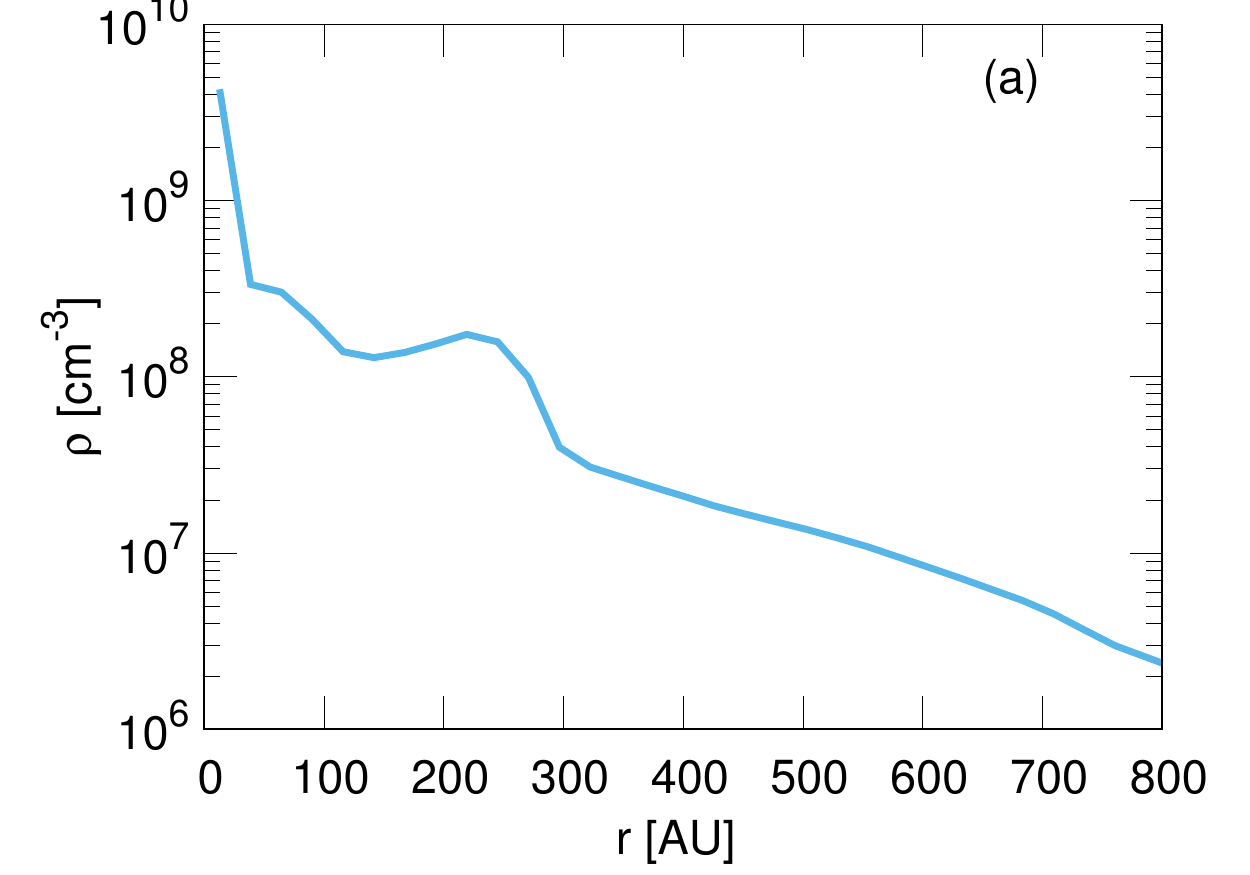}  
     \includegraphics[width=2.5in]{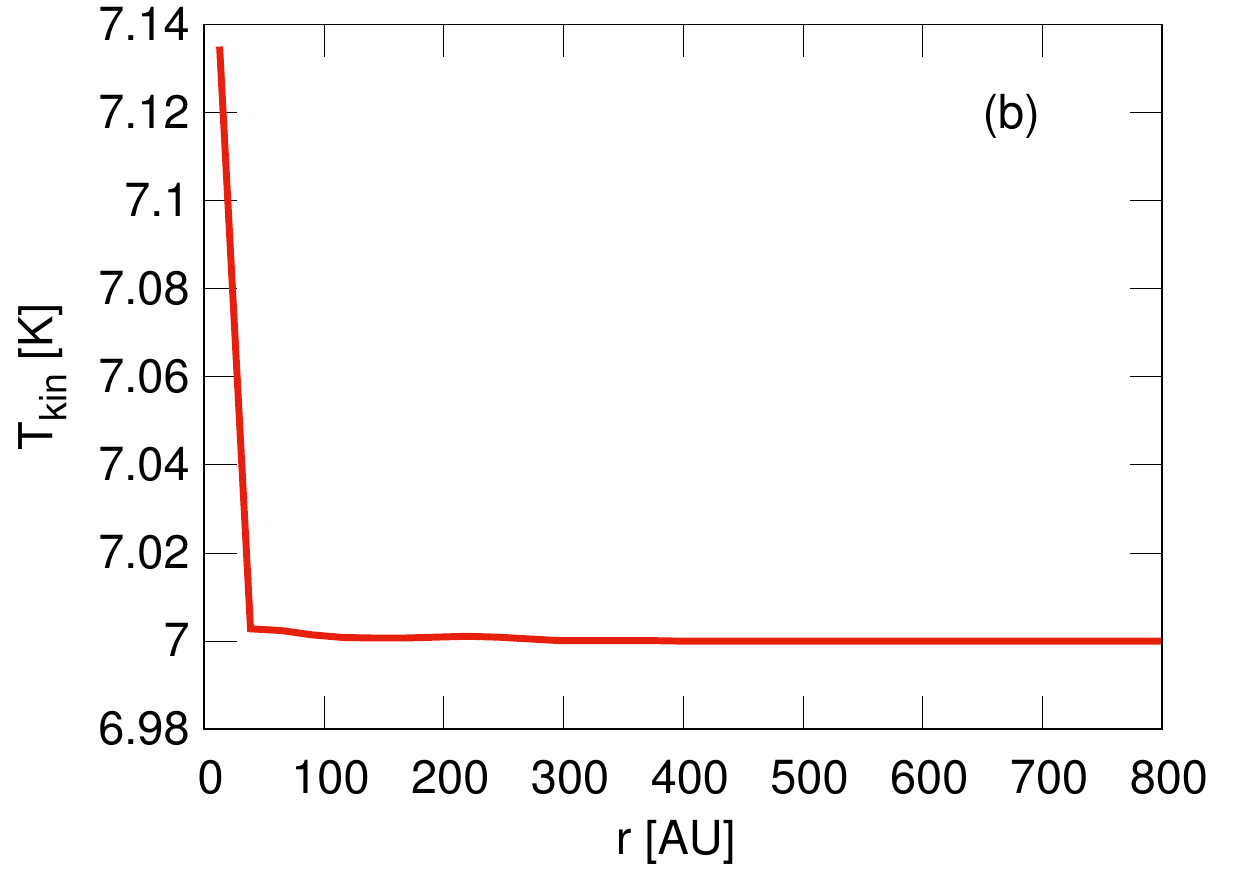}
     \includegraphics[width=2.5in]{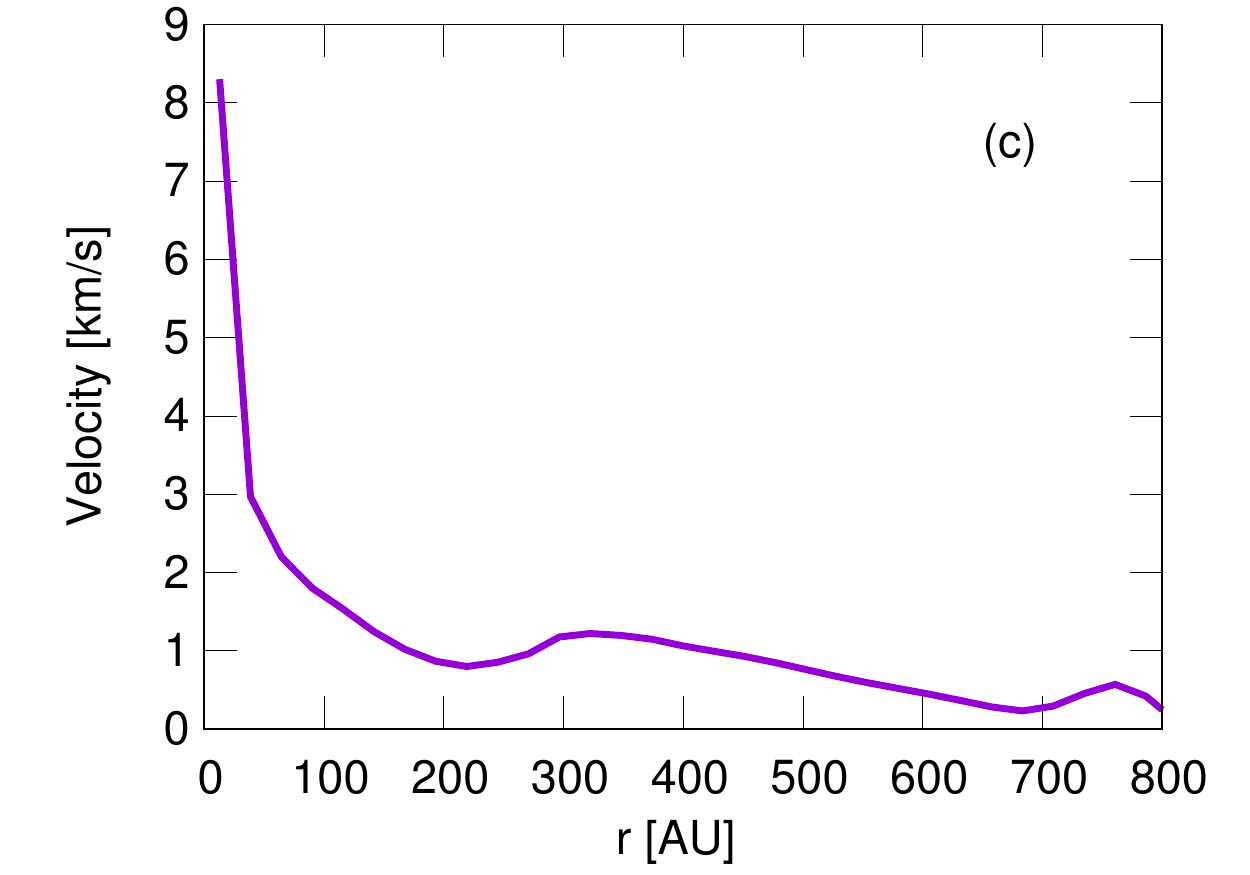}     
     \includegraphics[width=2.5in]{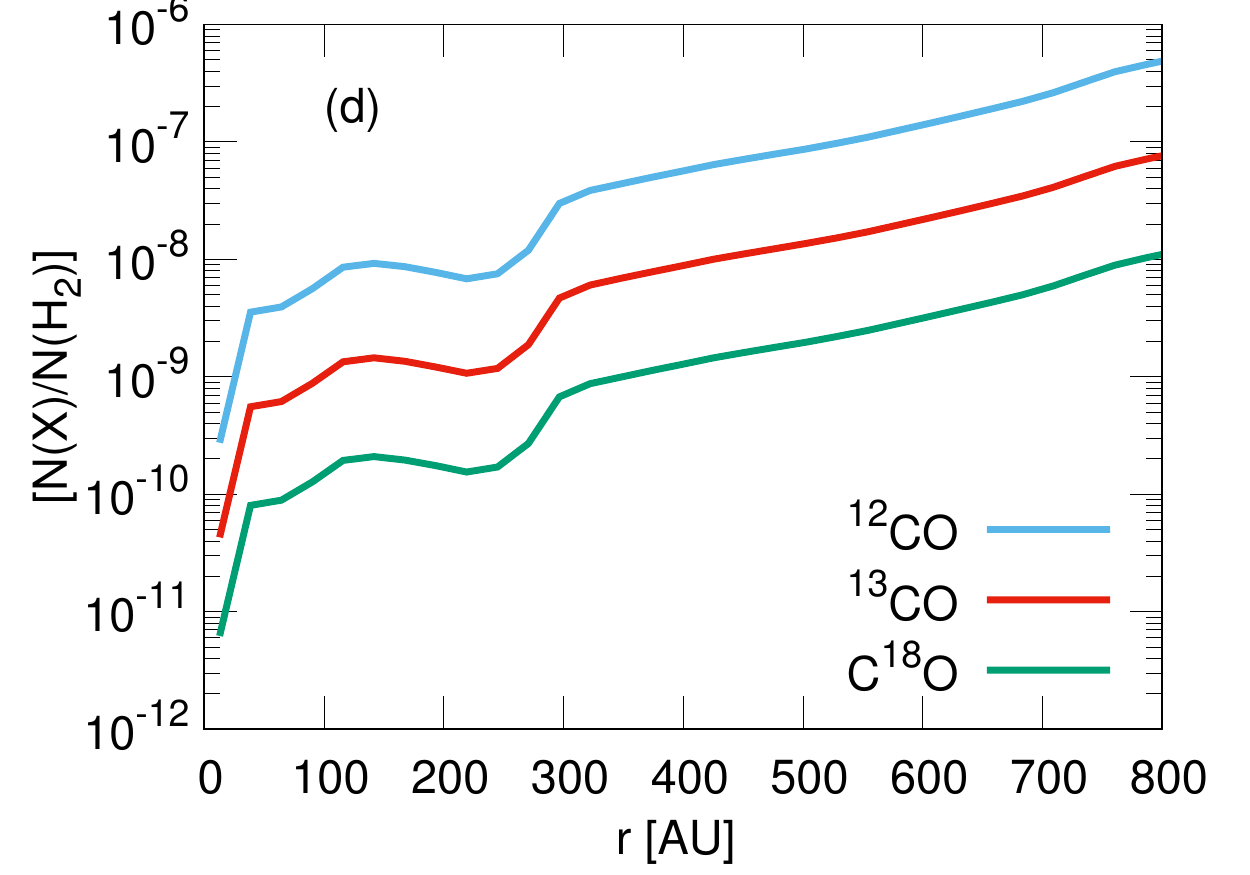}
     \caption{({\bf a,b,c}) The density, temperature, and velocity profiles at an evolutionary stage of 6000 yr in the core collapse model. ({\bf d}) The abundance profiles that provides the best-fit to the observed spectra. The radial profiles are the azimuthal average along the line of sight.  }
     \label{CC-model}
  \end{figure*}

Figures~\ref{CC-model}abc show the azimuthally averaged radial surface-brightness profiles for density, temperature, and velocity structure in the core collapse model. The value at each radius is defined as the azimuthal average of the pixel values along a suitably defined annulus (ring) of that radius centered on the object. The annuli are concentric circles and the average is taken along the line of sight at a 30$\degr$ inclination angle. There is a jump in the density profile (Fig.~\ref{CC-model}a) near $\sim$200 au, which corresponds to the boundary between the pseudo-disc and the infalling envelope. The outer edge of the pseudo-disc, or the centrifugal radius, is where material from the collapsing envelope falls onto the disc and cause an increase in the density. The sudden rise in the temperature (Fig.~\ref{CC-model}b) for $r <$20 au can be attributed to the presence of the high-velocity jet as well as some contribution from the stellar irradiation. The velocity profile (Fig.~\ref{CC-model}c) also shows a sudden rise as we approach the launching region of the high-velocity jet at $r <$20 au. The velocity is quite low at $<$ 1 km s$^{-1}$ for $r >$300 au. On or near the equatorial plane, the infalling gas is supported by the magnetic field (Lorentz force) and pressure gradient force. Thus, the radial velocity is slower than the infall velocity in the outer envelope/pseudo-disc regions. Figure~\ref{CC-model}d shows the radial profile of the molecular abundance that provides the best-fit to the observed spectra. The abundance profile is basically the inverse of the density structure and shows severe depletion in the abundance towards the central densest regions of the proto-brown dwarf. The profiles are similar for CO and isotopologues but the volume averaged abundances are different.

The physical structure, i.e., the density, temperature, and velocity radial profiles at different evolutionary stages from this model is coupled to the 3D non-LTE line radiative transfer code MOLLIE (Keto et al. 2004). We generate synthetic spectra for different radial profiles of the molecular abundance, which are then convolved by the ALMA 1.4$^{\prime\prime}$ beam size and compared with the observed spectra. A reasonable fit to the strength and width of the observed profile is reached from varying the molecular abundance and the line width. For each synthetic spectrum, a reduced-$\chi^{2}$ value is computed to determine the goodness of fit. We note that the core collapse simulation data is an appropriate model to use for a case such as ISO-OPH 200 where the proto-BD is known to drive a jet. Other models, such as, disc fragmentation or single or binary disc models presently available cannot reproduce a jet/outflow, as discussed in detail in Riaz et al. (2019a).

\subsection{C$^{18}$O and Continuum Emission -- Origin from pseudo-disc}
\label{c18o-model}

\begin{figure*}
  \centering           
     \includegraphics[width=2.25in]{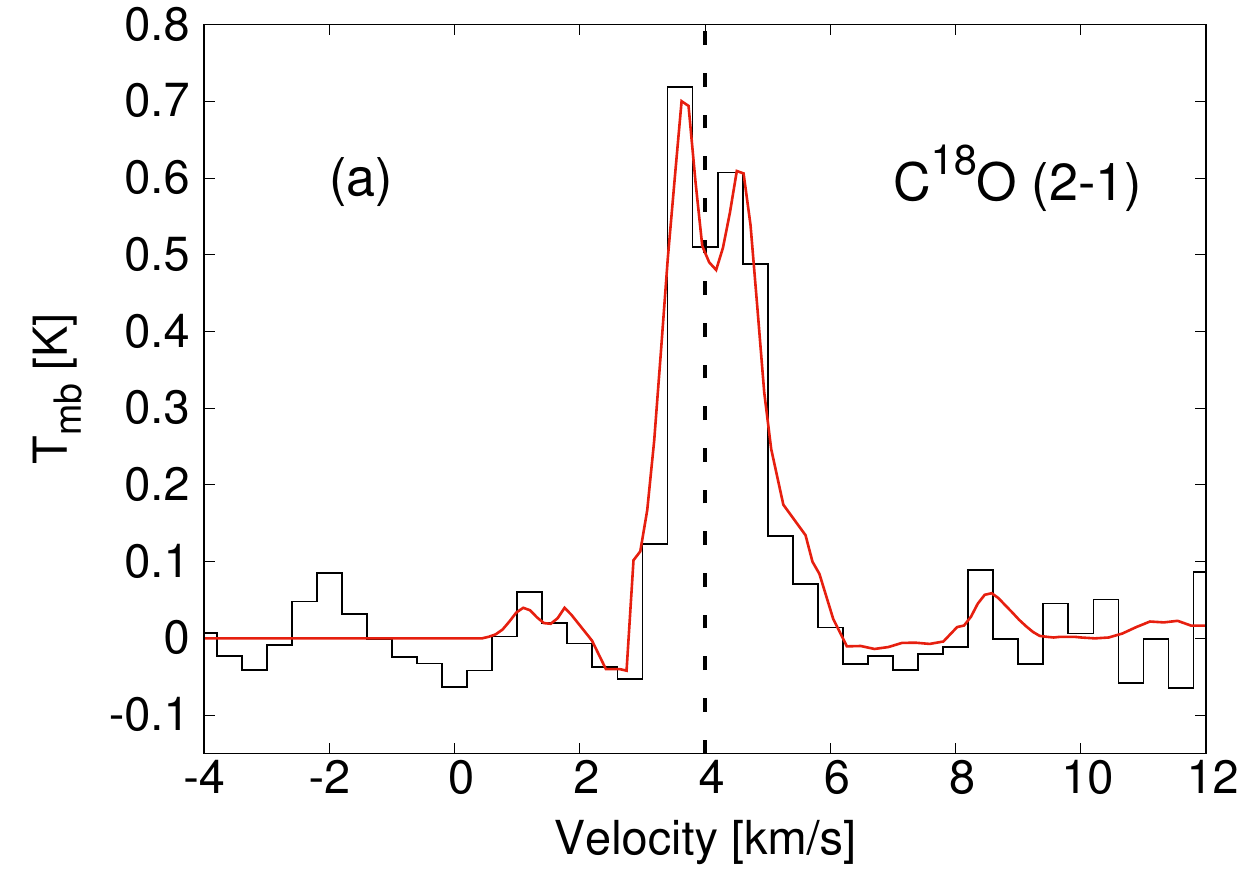}      
     \includegraphics[width=2.25in]{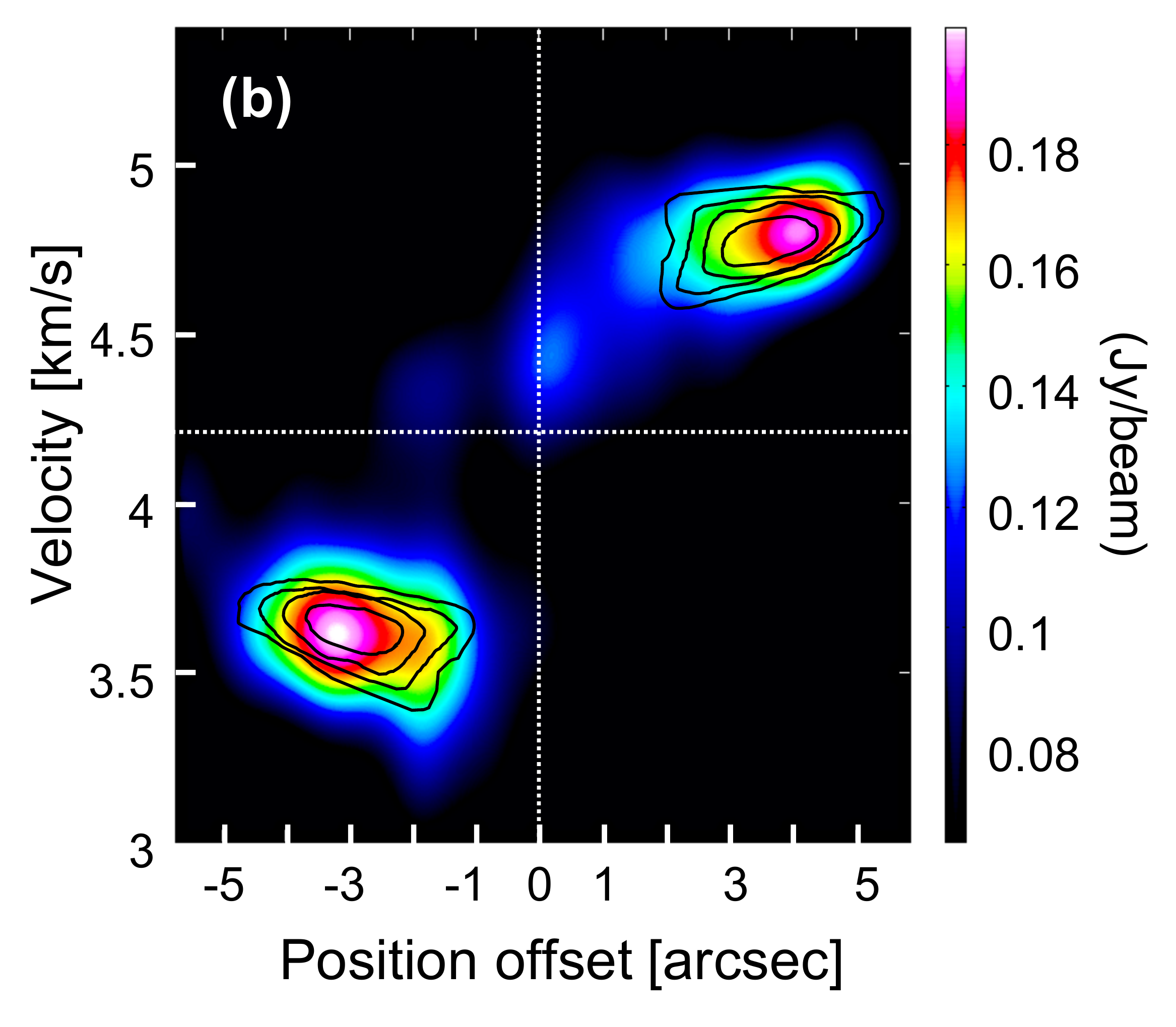}    
     \includegraphics[width=2.25in]{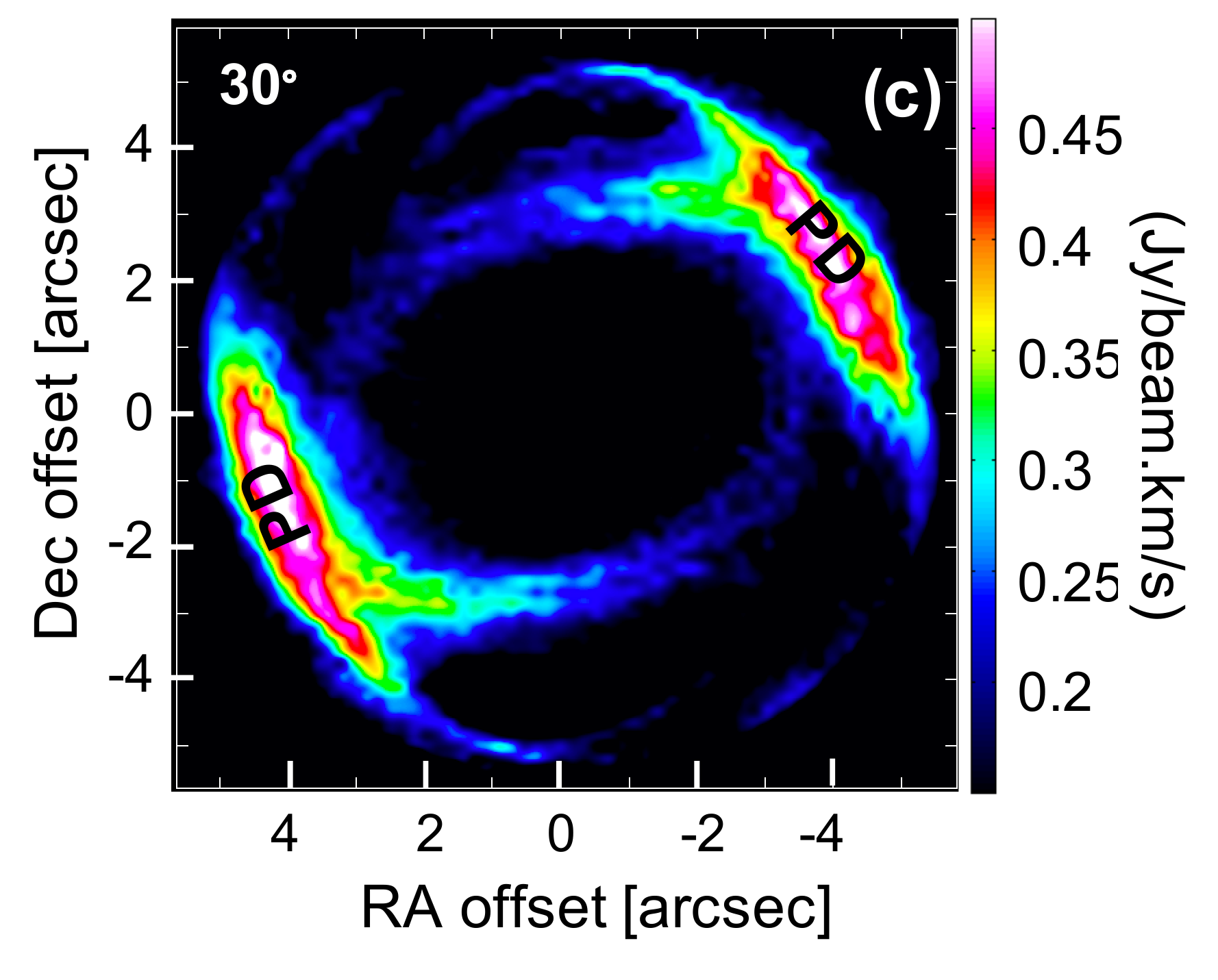}      
     \caption{ ({\bf a}) The core collapse model fit (red) to the observed C$^{18}$O spectrum (black). Dashed line marks the source V$_{LSR}$. ({\bf b}) Raster map shows the observed C$^{18}$O PVD. Overplotted is the model C$^{18}$O PVD in black contours. The contour levels are from 2-$\sigma$ to 10-$\sigma$ in steps of 2-$\sigma$. The 1-$\sigma$ rms is $\sim$0.05 Jy beam$^{-1}$. Dashed lines mark the source position and V$_{LSR}$. ({\bf c}) The C$^{18}$O integrated intensity map produced from the best model fit to the spectrum and PVD. The map is oriented at a 30$\degr$ inclination to the line of sight. The label `PD' implies the pseudo-disc.  }
     \label{C18O-modelfit}
  \end{figure*}

Figure~\ref{C18O-modelfit}ab shows the results from modelling of the C$^{18}$O spectrum and the PVD. The PVD was constructed by making a cut along the major axis PA = 37$\degr$ that shows a velocity spread in the moment 1 map (Fig.~\ref{pvd}; Section~\ref{C18O}). We simultaneously checked for a fit to the observed spectrum and the position and velocity offsets in the PVD using the simulation data at different evolutionary stages. As explained in Riaz et al. (2019b), the velocity spread in the model increases as the system evolves in time and the contribution from the various physical components becomes more prominent. From the best model fit to both the spectrum and the PVD, we can constrain the kinematical age of the ISO-OPH 200 system to be approximately 6000$\pm$1000 yr. The best model fit was obtained for an inclination angle of $\sim$30$\degr$. The reduced-$\chi^{2}$ value of the best-fit is 1.1.

\begin{figure*}
  \centering              
     \includegraphics[width=5in]{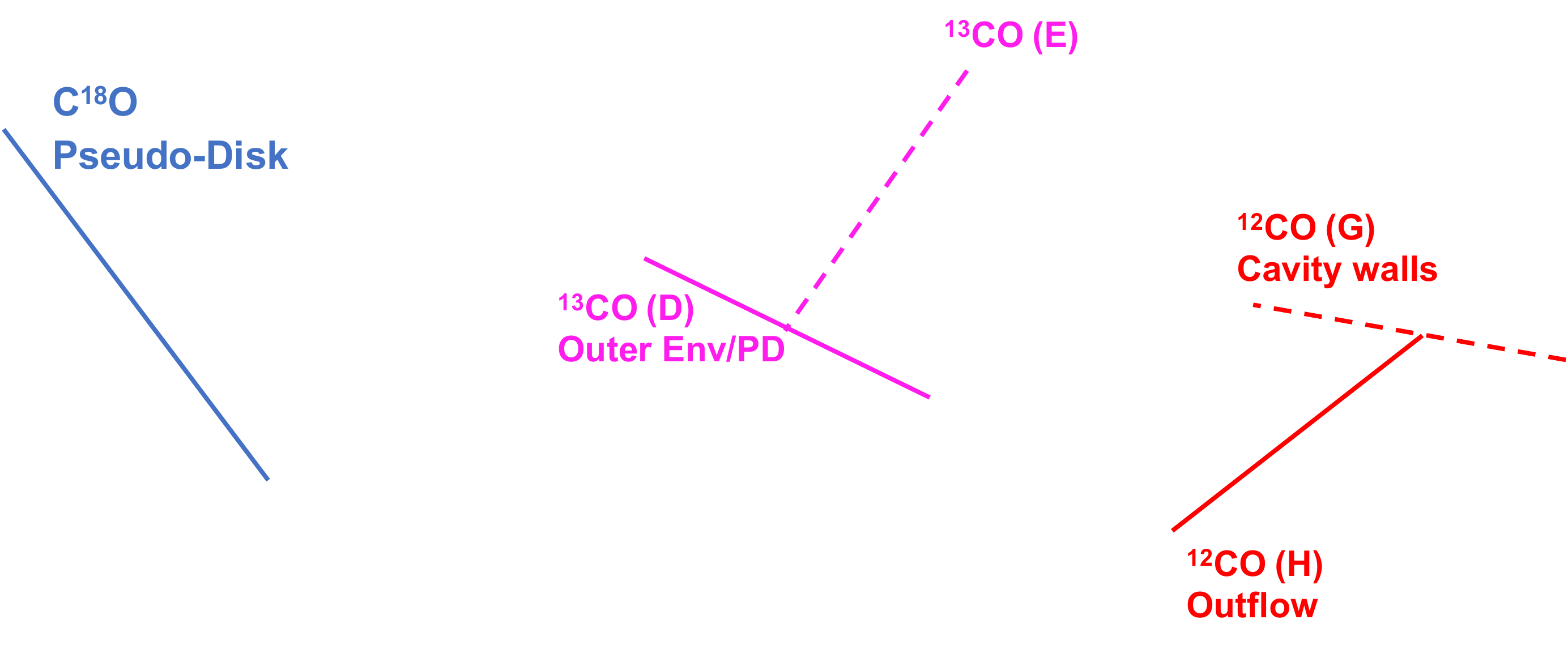} 
     \caption{A comparison of the position angles measured for the different structures seen in the moment maps.  }
     \label{PAs-all}
  \end{figure*}

From this best-fit (Fig.~\ref{C18O-modelfit}ab), we produced a model integrated intensity map at an inclination angle of 30$\degr$ to the line of sight (Fig.~\ref{C18O-modelfit}c). The peak C$^{18}$O emission, as predicted by the model map, arises from the pseudo-disc region at approximately $\pm$500 au from the central proto-BD (Fig.~\ref{C18O-modelfit}c). The pseudo-disc is the inner regions of the infalling envelope and accretes onto the central proto-brown dwarf and the Keplerian disc that is embedded in it. It is a dynamically collapsing structure expected to exhibit both infall and rotational kinematics, unlike a pure rotational Keplerian disc (Riaz et al. 2019a). The model map also shows signs of faint curved trails possibly tracing the infalling material. Based on this analysis, we can set the major axis PA = 37$\degr \pm$10$\degr$ as the pseudo-disc PA (Fig.~\ref{PAs-all}).

The overlap in the C$^{18}$O moment 0 and the 1.3 mm continuum images (Fig.~\ref{cont-c18o}) suggests that the continuum emission also has an origin from the pseudo-disc region. 
As predicted by the core collapse model (Machida et al. 2009), the inner Keplerian disc that is embedded in the densest, innermost region of the pseudo-disc has a spatial scale of $\sim$10 au or smaller. Interestingly, this predicted size is comparable to the $\sim$8.6 au source size measured from the 873 $\mu$m image and suggests that the continuum emission at a high angular resolution ($\sim$0.1$\arcsec$) may be tracing the inner Keplerian disc. The difference in the masses, however, is small; $\sim$4 M$_{Jup}$ measured from 873 $\mu$m compared to $\sim$6--7 M$_{Jup}$ measured in the larger beam size at 1.3 mm. This suggests that a large fraction of the disc mass reservoir is concentrated in the inner $\sim$12 au.

Note that the C$^{18}$O moment 1 map shows a clear velocity gradient due to which we can measure the PA and construct a PVD. This velocity information has allowed us to model both the spectrum and the position and velocity offsets. Due to this, the model map (Fig.~\ref{C18O-modelfit}c) will not match the moment 0 map in Fig.~\ref{C18O-moment}a but is expected to show the same position offsets for the peak emission as seen in the observed PVD (Riaz et al. 2019a).

We have also tested modelling using a pure Keplerian model but the results are poor (reduced-$\chi^{2}>$ 2) and degenerate. A poor fit of pure Keplerian kinematics is expected because at the very young kinematical age of the system ($\sim$6000 yr), infall activity is still very prominent, and it is expected to probe infall+rotational kinematics rather than pure Keplerian motions. The small dimensions of the inner Keplerian disc ($\leq$10 au) in a proto-brown dwarf system would require an angular resolution of $<$0.1$\arcsec$ to clearly resolve and separate the infall and rotational kinematics. At present, it is impossible to resolve and detect pure Keplerian motions in molecular line emission for a proto-brown dwarf at a high significance level above the noise even with ALMA.

\subsection{$^{13}$CO Emission -- Origin from Cavity walls, pseudo-disc, and Envelope}
\label{13co-model}

\begin{figure*}
  \centering           
     \includegraphics[width=2.3in]{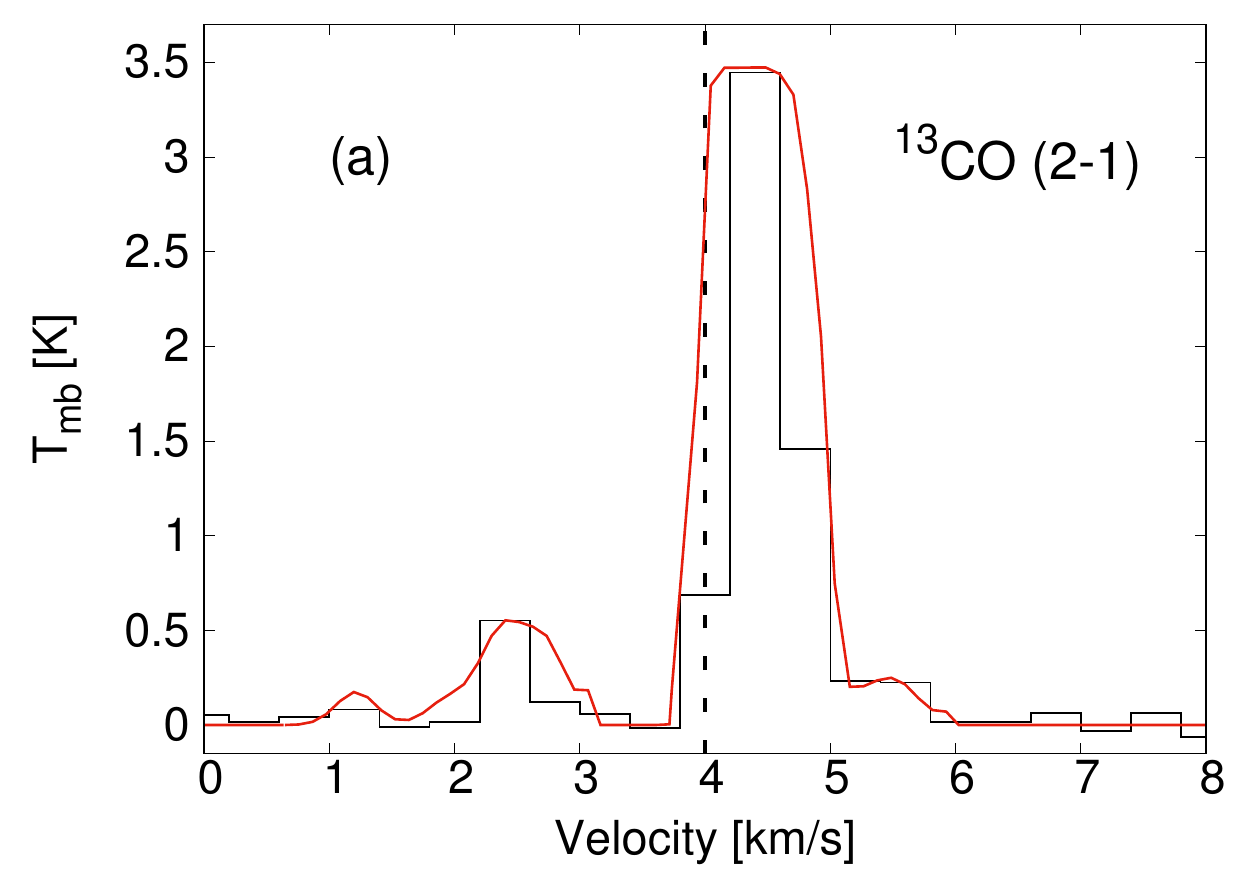}  \\
     \includegraphics[width=2.6in]{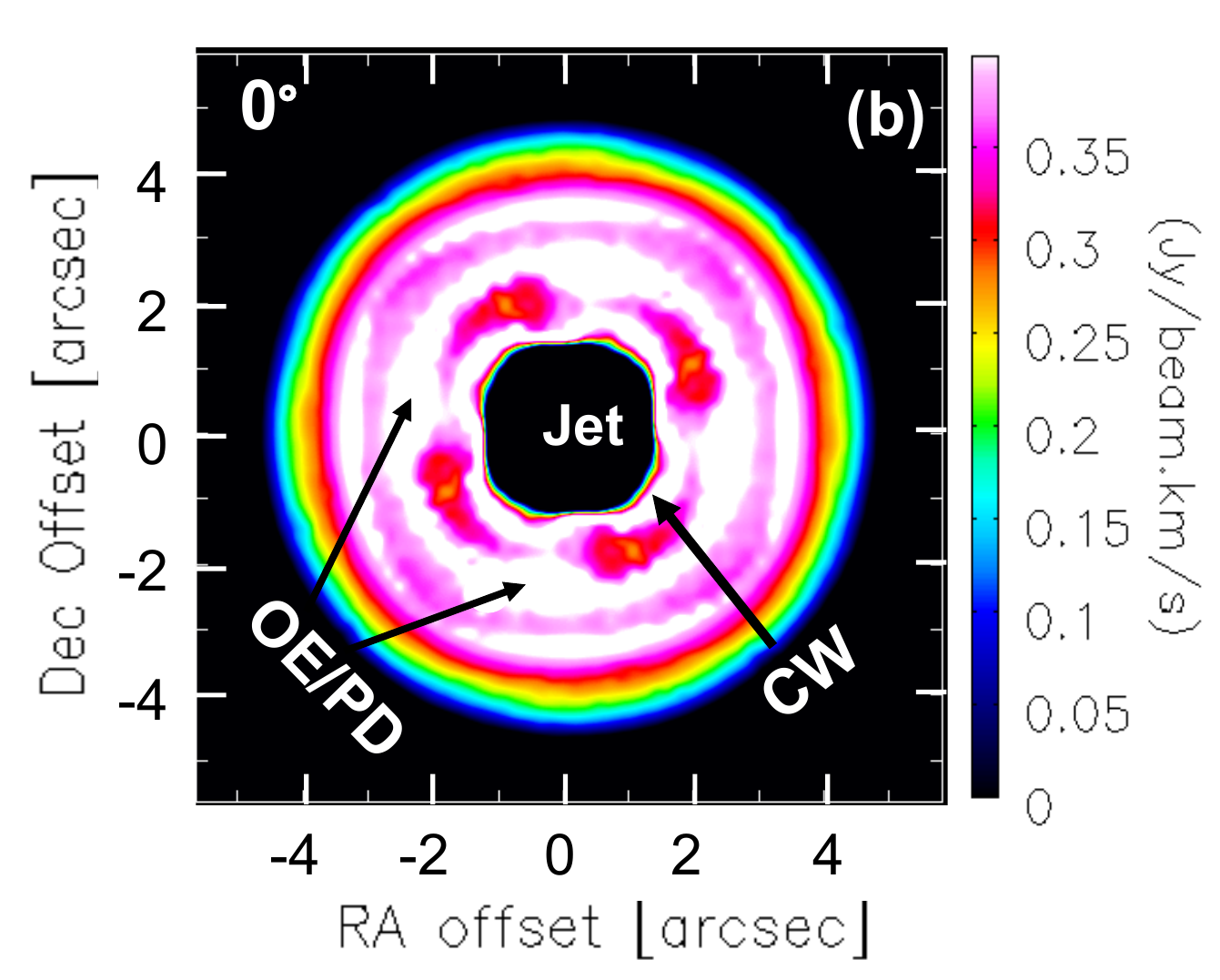} 
     \includegraphics[width=2.7in]{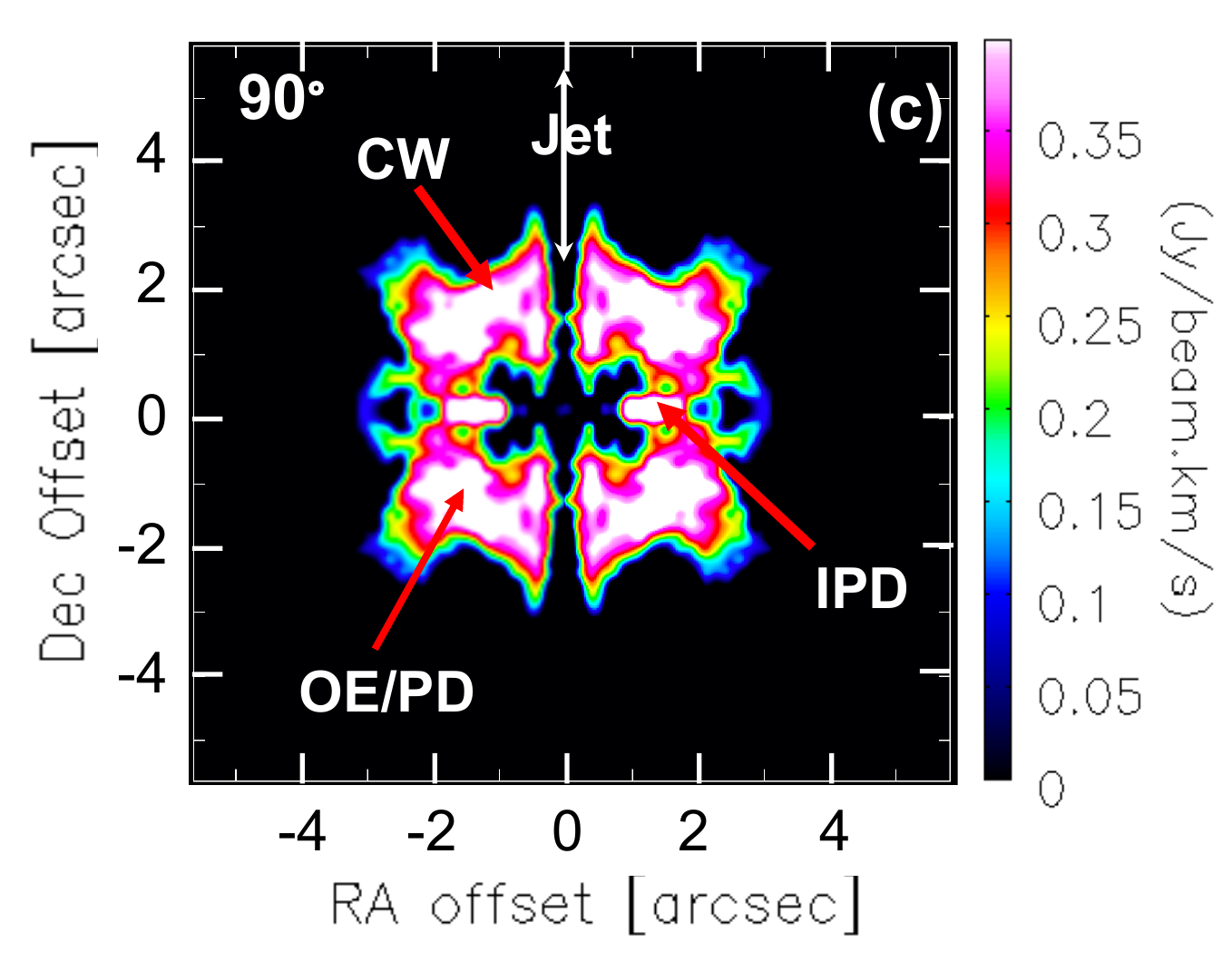}  \\
     \includegraphics[width=2.7in]{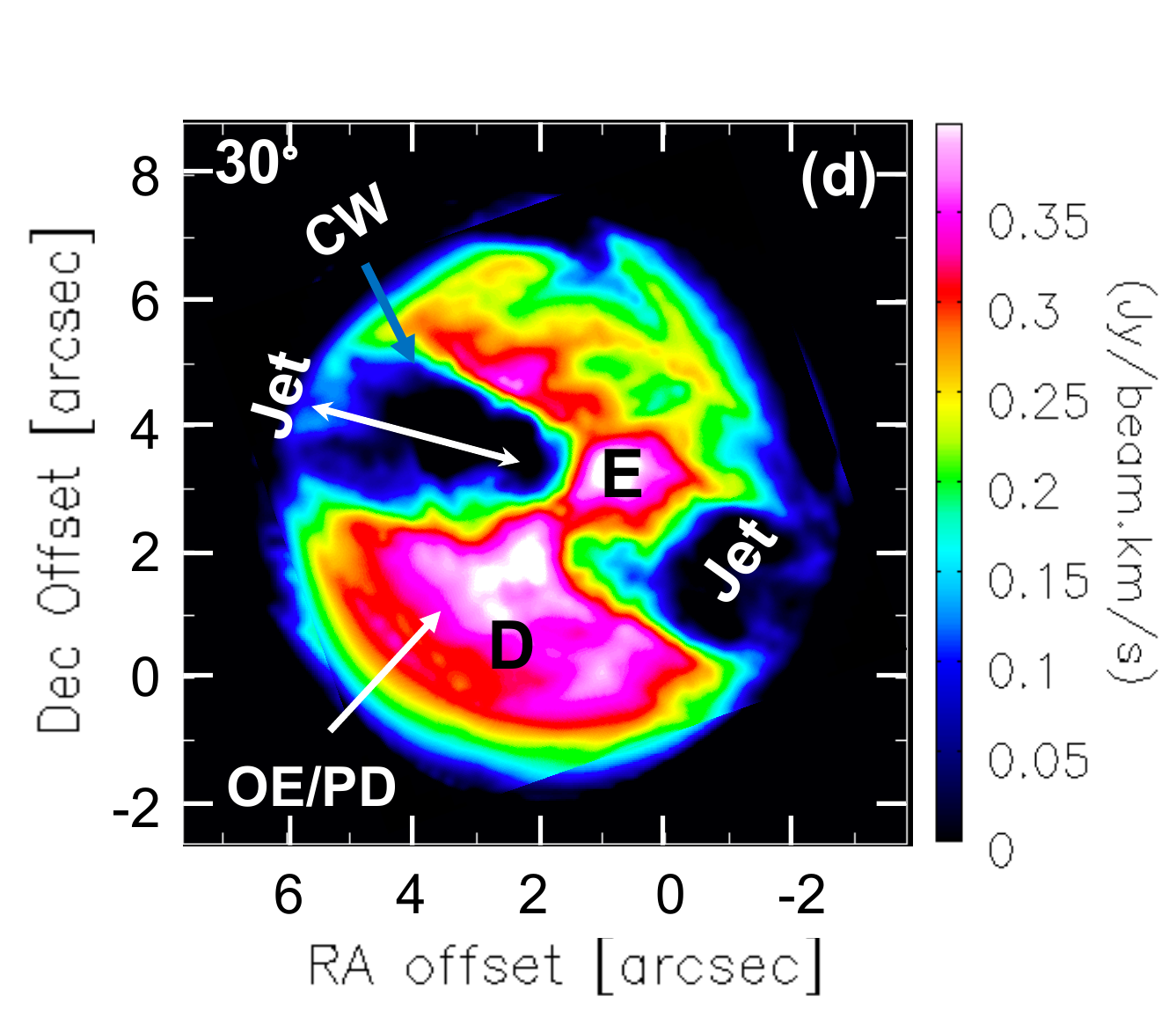}
     \includegraphics[width=2.65in]{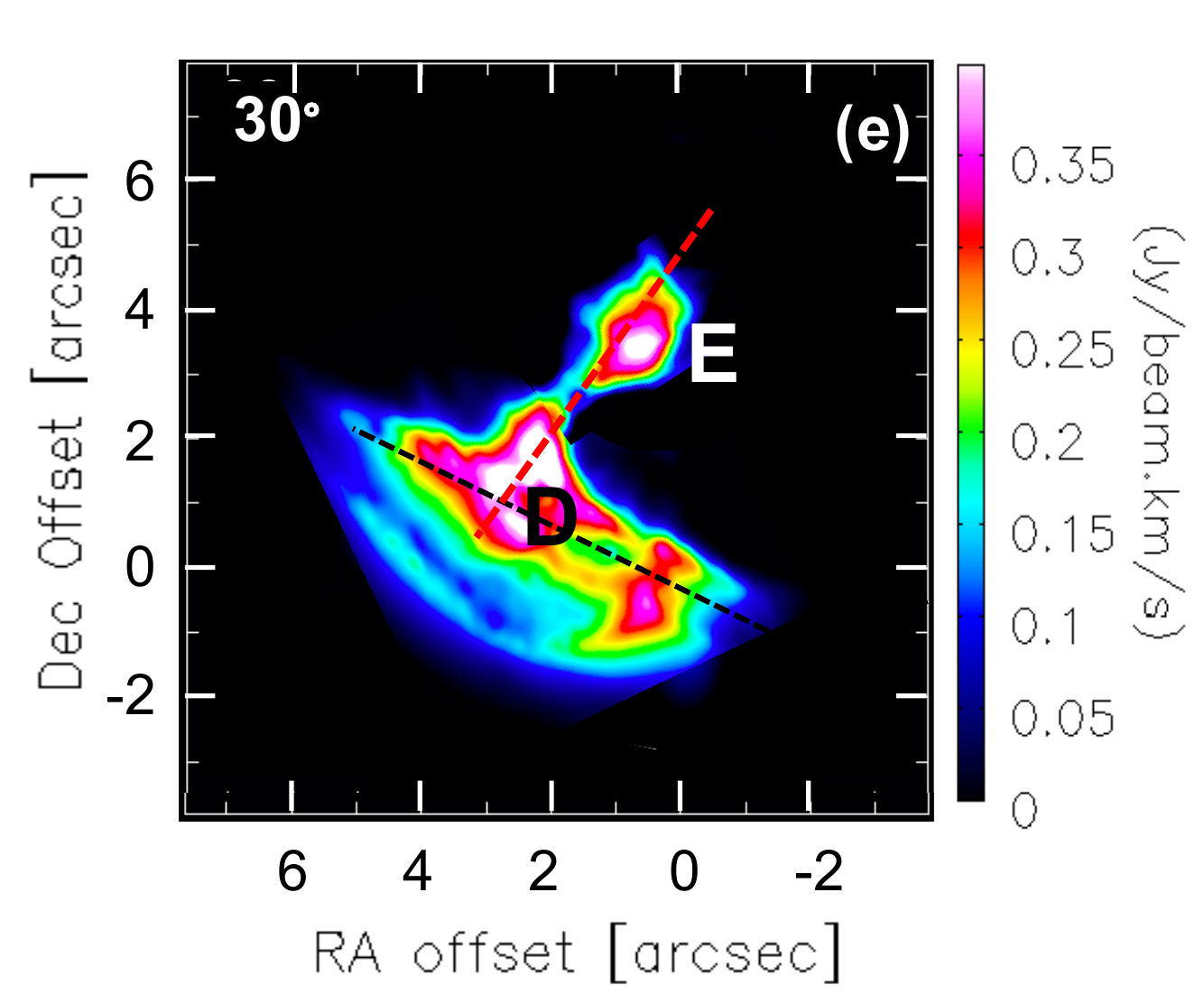}    
     \caption{ ({\bf a}) The core collapse model fit (red) to the observed $^{13}$CO spectrum (black). Dashed line marks the source V$_{LSR}$. Middle and bottom panels ({\bf b}), ({\bf c}), and ({\bf d}) show the $^{13}$CO integrated intensity model map at pole-on (0$\degr$), edge-on (90$\degr$), and the best-fit 30$\degr$ inclination to the line of sight, respectively. The peak emission is seen arising from the outer envelope/pseudo-disc regions (labelled `OE/PD') and the outflow cavity walls (labelled 'CW'), respectively. Panel ({\bf e}) shows the same 30$\degr$ model map as shown in ({\bf d}) but at a higher colour contrast, making it possible to see only the brightest emission. The labels {\bf D} and {\bf E} in these model maps mark the bright, complex structures seen in the observed $^{13}$CO moment 0 map (Fig.~\ref{13CO-moment})a, and the dashed lines are the PAs for these structures. }
     \label{13CO-modelFit}
  \end{figure*}

The observed $^{13}$CO emission (Fig.~\ref{13CO-moment}a) shows a much more complex structure compared to C$^{18}$O. We have used the same model at a stage of 6000 yr and at an inclination of $\sim$30$\degr$ that provided a good fit to the C$^{18}$O emission. Figure~\ref{13CO-modelFit}a shows the results from modelling of the $^{13}$CO spectrum. The reduced-$\chi^{2}$ value of the best-fit is 1.5. This model is able to fit the bright, narrow peak as well as the weak blue-shifted peak in the observed $^{13}$CO spectrum.

As discussed in Section~\ref{13CO}, the observed $^{13}$CO emission shows a broad, bright, extended structure labelled {\bf D}. The $^{13}$CO moment 1 map (Fig.~\ref{13CO-moment}b) does not show any clear velocity gradient along a single PA in the complex {\bf D} structure, which makes it difficult to measure the PA and construct a PVD. We have instead produced a model moment 0 map or an integrated intensity map from the best line model fit to the observed $^{13}$CO spectrum, and compared this with the observed $^{13}$CO moment 0 map (Fig.~\ref{13CO-moment}a). Figure~\ref{13CO-modelFit} shows these model maps at the best-fit 30$\degr$ inclination to the line of sight, as well as a model map at a pole-on (0$\degr$) and an edge-on (90$\degr$) inclination for a comparison. Note that the maps show symmetric structures about the (RA, Dec) = (0,0) position when viewing exactly at a pole-on or edge-on inclination, whereas asymmetries are seen at intermediate inclinations. 

As seen in the pole-on and edge-on model maps (Fig.~\ref{13CO-modelFit}bc), the bulk of the $^{13}$CO emission arises from the dense envelope and pseudo-disc regions, as well as near the dense walls of the outflow cavities. Comparing these maps with the model map at a $\sim$30$\degr$ orientation (Fig.~\ref{13CO-modelFit}d) makes it easier to identify the physical components. As seen in the 30$\degr$ map, the orientation of the system is such that the line of sight is partially brazing through the jet and the cavity walls and we have a partial view of the envelope/pseudo-disc region towards the south of the proto-BD position, while the emission towards the north and west is largely obstructed from the line of sight. No $^{13}$CO emission is seen along the jet/outflow regions in any of these maps. In Fig.~\ref{13CO-modelFit}e, we have shown the same 30$\degr$ model moment 0 map but with the intensity scaled in such a way that only the brightest emission with integrated flux of $>$0.3 Jy beam$^{-1}$.km s$^{-1}$ is seen. Enhancing the intensity contrast leaves behind only two prominent structures.

Comparing the model maps in Figs.~\ref{13CO-modelFit}de with the observed $^{13}$CO moment 0 map (Fig.~\ref{13CO-moment}a) then indicates that the broad {\bf D} structure is tracing the envelope/pseudo-disc regions as well as the walls of the outflow cavities, while the {\bf E} structure is likely arising from the far side of the envelope/pseudo-disc north of the source which is largely obstructed from the line of sight. The overall shape and orientation of the structures in these model maps are similar to the observed case, and lie along similar PAs as measured for the {\bf D} and {\bf E} structures, with a slight difference in the intensity of the structures. The fact that enhancing the intensity contrast in the model map produces a better match to the observed moment 0 map suggests that the observations are only tracing the densest, brightest regions and therefore the details of the outflow cavity walls carved out by the jet as seen in the 30$\degr$ model map in Fig.~\ref{13CO-modelFit}d cannot be identified in the observed moment 0 map. We cannot clearly distinguish the boundary between these regions at the present $\sim$1.4$\arcsec$ angular resolution of the observations, due to which these model fits and maps are likely not unique but can provide the first interpretation on the origin of the observed complex structure in the $^{13}$CO emission.

\subsection{$^{12}$CO Emission -- Origin from Jet/Outflow regions and Cavity walls}
\label{outflow}

\begin{figure*}
  \centering           
     \includegraphics[width=2.4in]{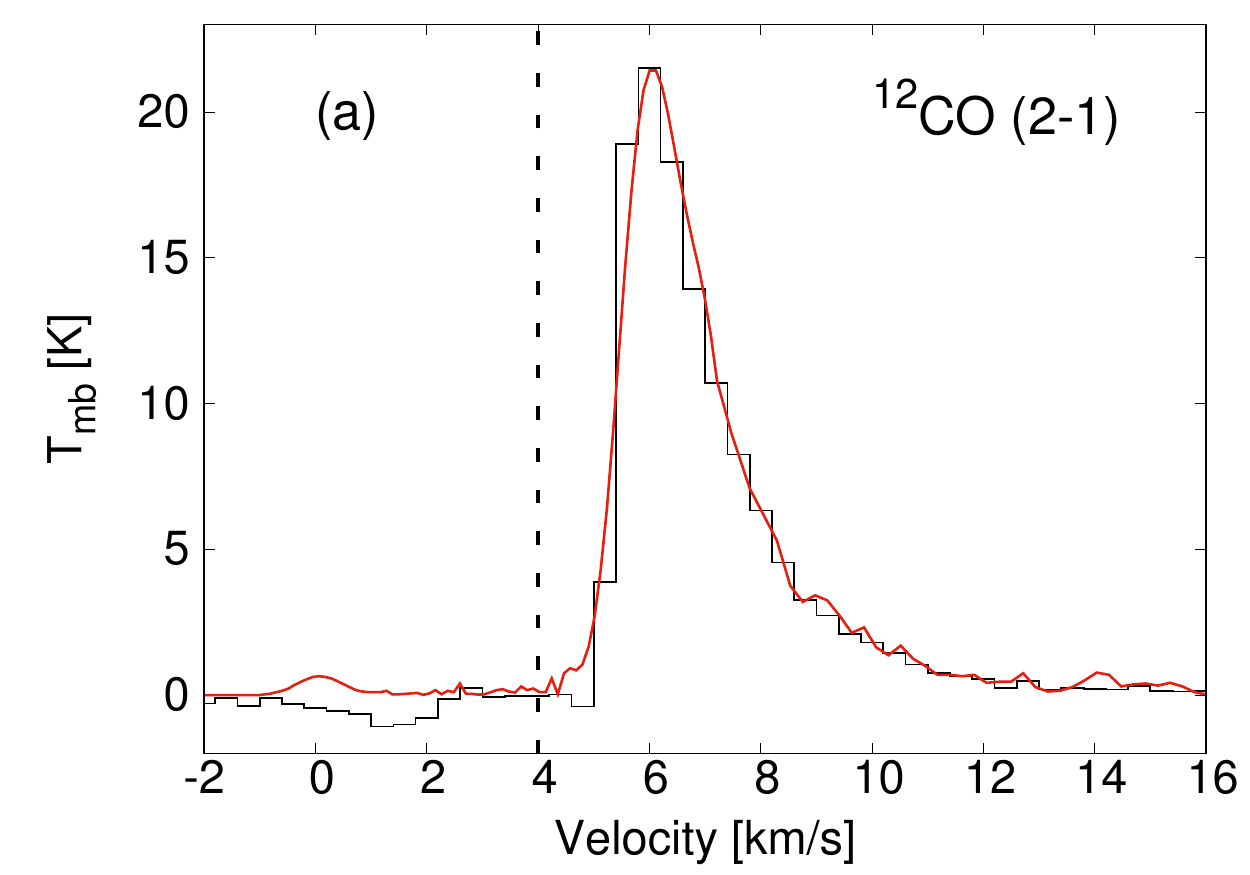}  \\
     \includegraphics[width=2.6in]{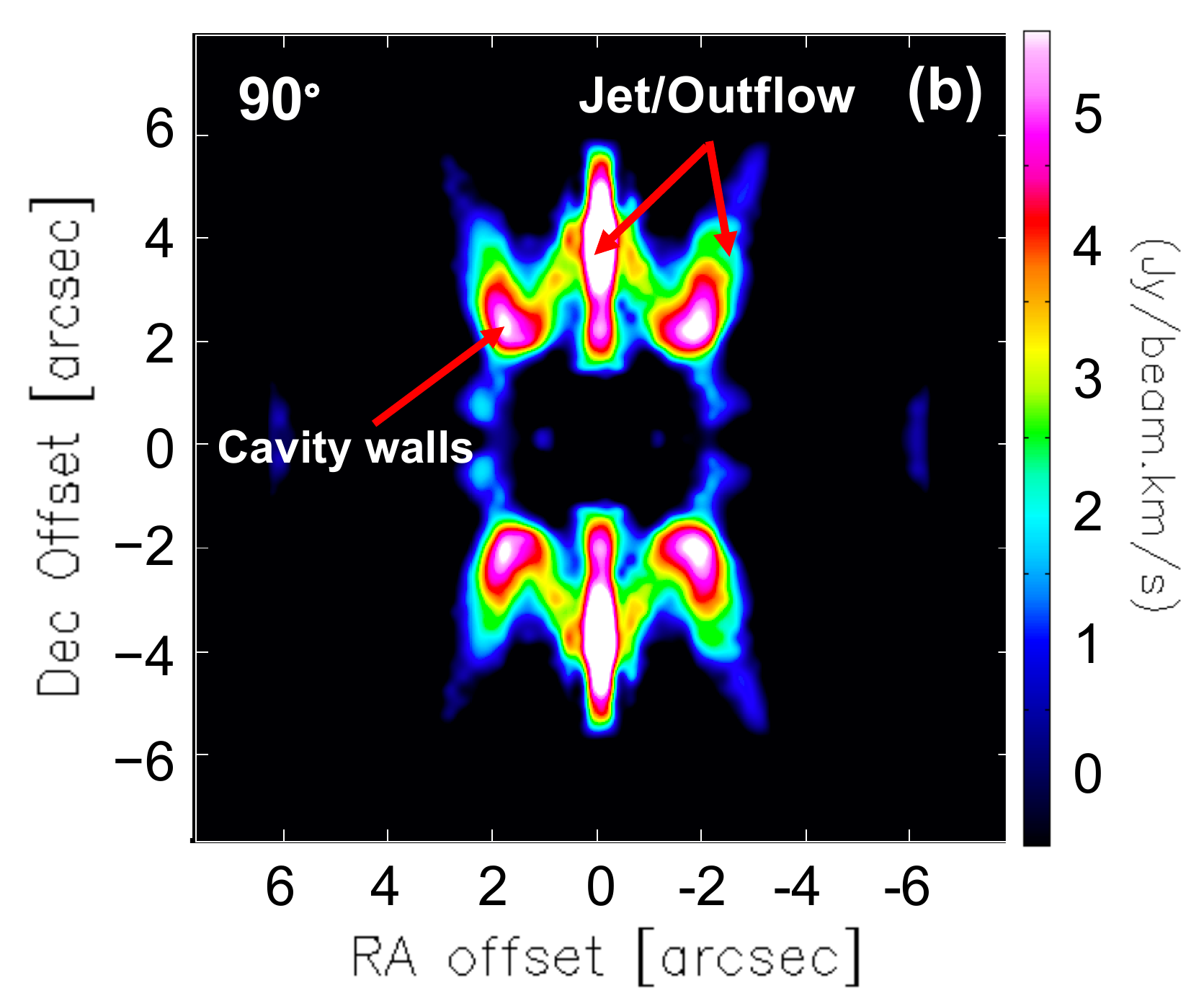}    \hspace{0.2in}
     \includegraphics[width=2.6in]{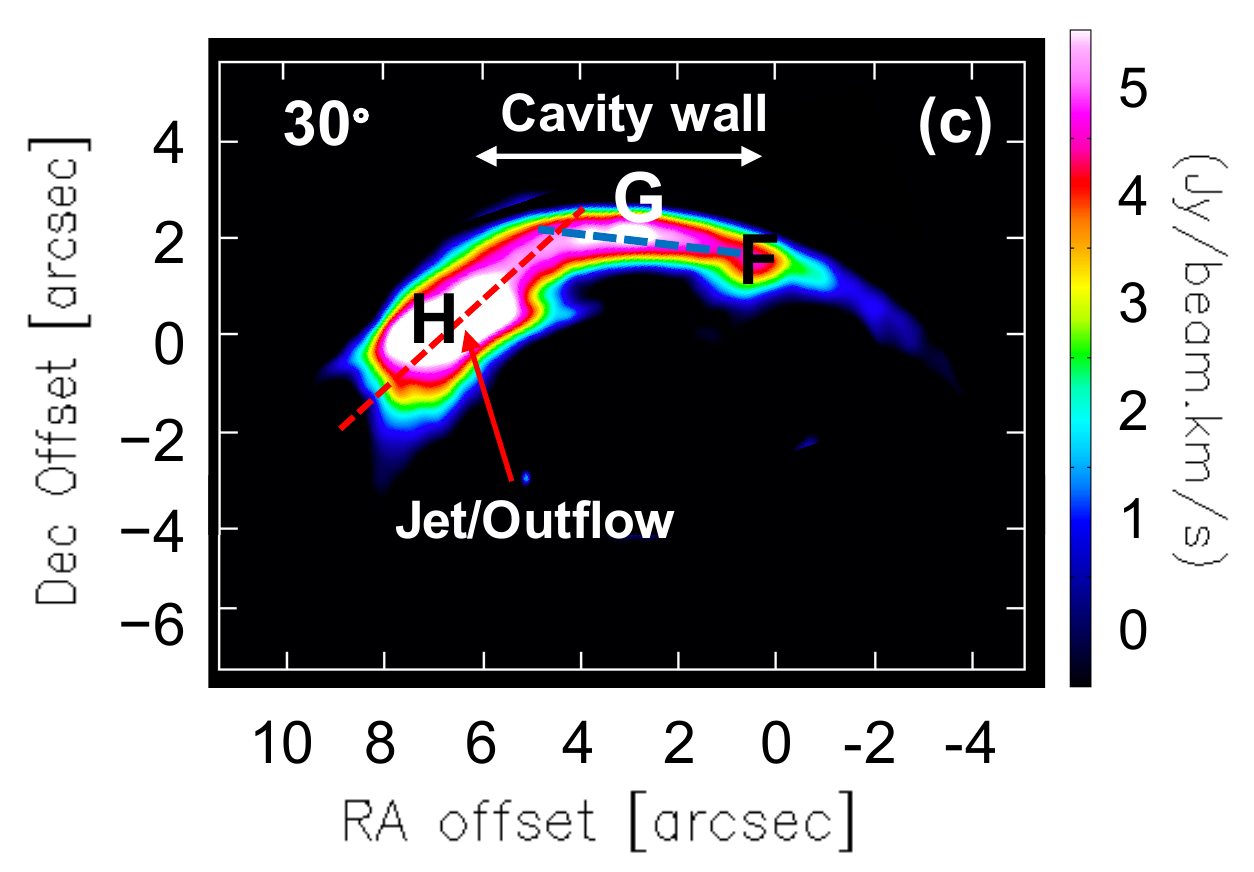}
     \caption{ ({\bf a}) The core collapse model fit (red) to the observed $^{12}$CO spectrum (black). Dashed line marks the source V$_{LSR}$. Bottom panels ({\bf b}) and ({\bf c}) show the $^{12}$CO integrated intensity model map at an edge-on (90$\degr$) inclination and the best-fit 30$\degr$ inclination to the line of sight, respectively. The peak emission is seen from the jet/outflow regions and the cavity walls (labelled `CW'). }
     \label{CO-modelFit}
  \end{figure*}

The $^{12}$CO emission also shows a complex, multi-component structure. We have used the same model at a stage of 6000 yr and at an inclination of $\sim$30$\degr$ that provided a good fit to the C$^{18}$O and $^{13}$CO observations. Figure~\ref{CO-modelFit}a shows the best-fit to the observed spectrum. The reduced-$\chi^{2}$ value of the best-fit is 1.1. This model is able to fit the bright peak and the broad, extended, red-shifted wing. As in the case of $^{13}$CO emission, the $^{12}$CO emission also shows a complicated shape with a different PA for the different structures, which cannot be clearly separated from each other (Fig.~\ref{CO-moment}a). There is no clear velocity gradient along a single PA seen in the moment 1 map (Fig.~\ref{CO-moment}b), which makes it difficult to construct a PVD and directly compare the position-velocity offsets with the model, as conducted in the case of C$^{18}$O.

To investigate the physical components from where the observed $^{12}$CO emission originates, we have produced CO integrated intensity model maps at an edge-on (90$\degr$) and at the best-fit 30$\degr$ inclination to the line of sight from the best model fit to the $^{12}$CO spectrum. The model maps show symmetric emission about the (RA, Dec) = (0,0) position, unlike the observed case. An edge-on map can provide a more clear view of the jet/outflow region compared to the pole-on (0$\degr$) case. The edge-on map (Fig.~\ref{CO-modelFit}b) shows strong CO emission arising from the jet/outflow region and the dense walls of the outflow cavities, whereas no emission is seen from the envelope/pseudo-disc regions. 

Comparing this edge-on map with the 30$\degr$ model map (Fig.~\ref{CO-modelFit}c) indicates that the bright, curved structure from where the bulk of the CO emission originates is tracing the jet/outflow region and part of the wide outflow cavity walls. Therefore, in the observed CO moment 0 map (Fig.~\ref{CO-moment}a), the structure {\bf H} is likely tracing the jet/outflow emission. Note that the PA of the outflow-associated {\bf H} structure (127$\degr \pm$7$\degr$) is perpendicular to the pseudo-disc axis (37$\degr \pm$10$\degr$) measured from the C$^{18}$O moment 1 map, as shown in Fig.~\ref{PAs-all}. The bright clump-like structure seen at the tip of the extended lobe {\bf H} could be a bright shock emission knot at the apparent end of the outflow (Fig.~\ref{CO-moment}a). The structure {\bf G} is likely tracing the cavity walls (Fig.~\ref{CO-modelFit}c;~\ref{CO-moment}a). The model cannot clearly distinguish between the {\bf F} and {\bf G} structures; it is likely that {\bf F} is also tracing the cavity walls as well as an unresolved molecular outflow originating close to the source position.

Strong outflow activity has been previously observed for ISO-Oph 200 in VLT near-infrared observations (Whelan et al. 2018). The near-infrared spectra show several [Fe~II] and H$_{2}$ emission lines. However, the [Fe~II] spectro-images do not show any extended jet emission. The peak in the [Fe II] emission is at $<$0.5$\arcsec$ ($<$ 72 au) position offset from the proto-BD driving source. This small position offset is within the position uncertainty. Thus, the PA estimated through a 2D fitting of the [Fe~II] peak emission region is uncertain as the [Fe~II] emission is unresolved. The near-infrared spectro-images are seeing-limited data due to which any emission at $<$1$\arcsec$ cannot be clearly resolved from the source.

The near-infrared observations also revealed slower H$_{2}$ emission that is extended to $\sim$4$\arcsec$ north-east from the source at a PA$\sim$67$\degr$--80$\degr$. As argued in Whelan et al. (2018), this H$_{2}$ emission is tracing the outflow cavity walls and not a jet. The PA of this extended H$_{2}$ emission is similar to the PA measured for the {\bf G} structure (80$\degr \pm$5$\degr$) in $^{12}$CO, which is tracing the cavity walls, as argued above based on modelling (Fig.~\ref{PAs-all}).

\section{Discussion}
\label{discuss}

\subsection{3D morphology}
\label{3D-morph}

Based on the modelling results, we can interpret that the $^{12}$CO emission is tracing the jet/outflow regions, $^{13}$CO is tracing the envelope/pseudo-disc regions, and C$^{18}$O and 1.3 mm continuum emission are tracing the pseudo-disc. Some contribution is also seen in the $^{12}$CO and $^{13}$CO emission from the walls of the outflow cavities. The spatial scales of the observed emission in these lines are different from each other, as can be seen from the moment 0 maps. While C$^{18}$O shows more compact emission with a projected size of $\sim$500 au, the emission in $^{12}$CO and $^{13}$CO is about twice more extended with a projected size of $\sim$1000--1300 au. While we cannot clearly distinguish the boundary between the infalling envelope and the pseudo-disc regions, comparing the dimensions with those predicted by the model suggests that C$^{18}$O is tracing the inner pseudo-disc while $^{13}$CO is tracing the outer envelope/pseudo-disc regions (Fig.~\ref{model}). The same physical model at the $\sim$6000 yr evolutionary stage and at a $\sim$30$\degr$ inclination can provide a good fit to all of the spectra and reproduce the main structures seen in the moment 0 maps.

\begin{figure}
  \centering       
       \includegraphics[width=2.6in]{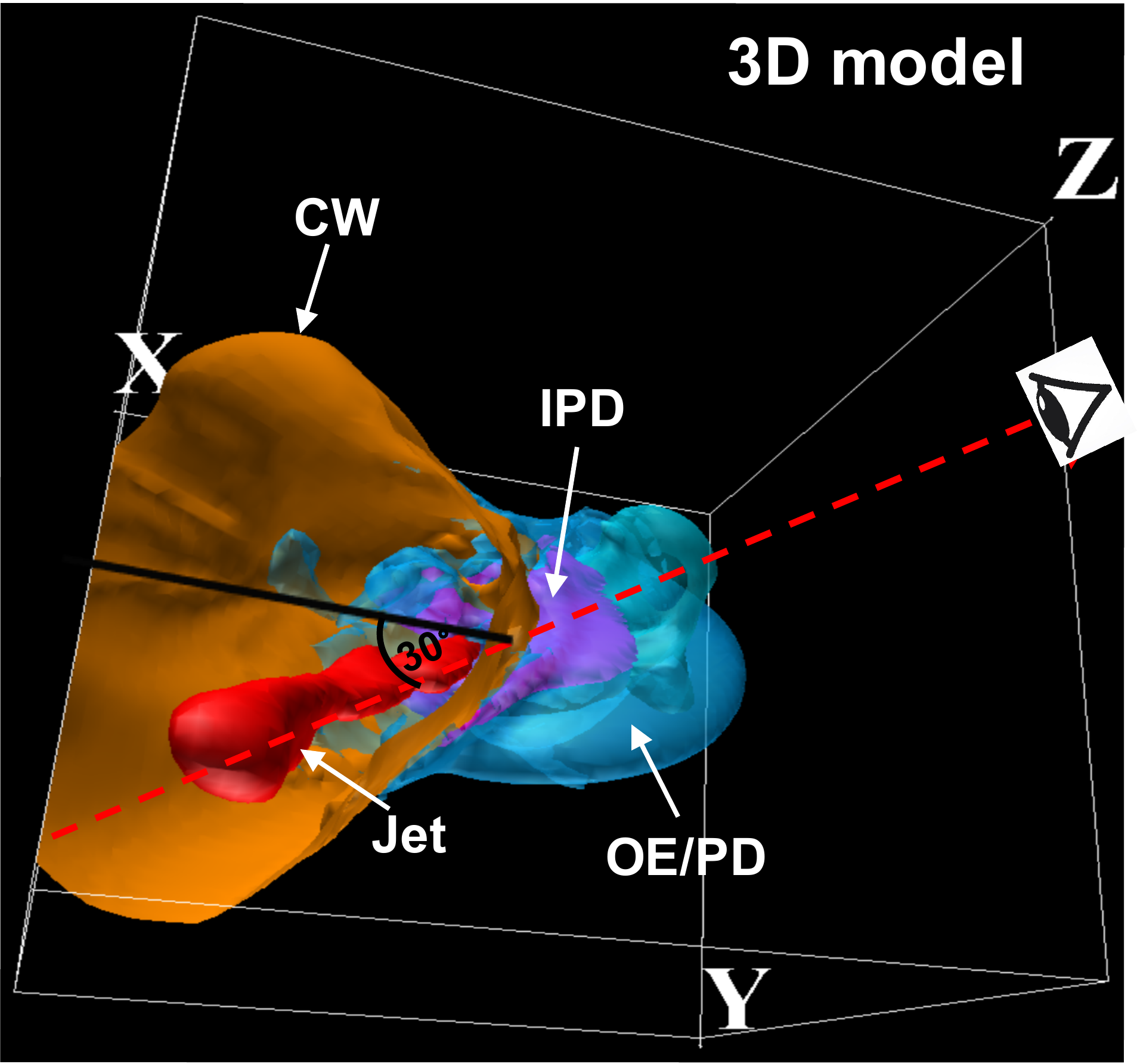}      
     \caption{A 3D model to interpret the structure of the ISO-OPH 200 system based on the morphology seen in the $^{12}$CO, $^{13}$CO, and C$^{18}$O model maps. The labels `IPD', `OE/PD', and 'CW' imply the inner pseudo-disc, the outer envelope/pseudo-disc regions and the outflow cavity walls, respectively. The line-of-sight (red dashed line) and the viewer are marked. The simulation boundary of the cube is drawn in thin white line. }
     \label{3D-model}
  \end{figure}

To understand the structure and morphology of the ISO-OPH 200 system more clearly, we have built a 3D view (Fig.~\ref{3D-model}) to replicate the morphology seen in the large-scale model maps produced from the $^{12}$CO and $^{13}$CO best model fits (Figs.~\ref{13CO-modelFit};~\ref{CO-modelFit}). The 3D view is directly made from the core collapse simulations in which we only changed the orientation of the system. The coordinate system and the way the inclination angle is set in MOLLIE to produce the synthetic observations is different from that in the 3D MHD simulations from Machida et al. (2009). The 3D view is similar to the 2D model schematic at 30$\degr$ inclination shown in Fig.~\ref{model}b. In MOLLIE, the orientation is set in terms of latitude and longitude along the Z-axis, whereas in the 3D MHD simulations, it is set with respect to the X-axis. The best model fit is obtained for a latitude of 110$\degr$ and longitude of 0$\degr$; this is the orientation with respect to the Z-axis. This translates into a 30$\degr$ angle with respect to the X-axis. We have marked the line-of-sight in the 3D model in Fig.~\ref{3D-model}.

The ISO-OPH 200 system is viewed partially through a wide outflow cavity resulting in a more direct view of the jet/outflow and a partial view of the envelope/pseudo-disc regions. A large part of the envelope/pseudo-disc/cavities/jet towards the northern and western side of the proto-BD are blocked from the view. It is possible that the faint tail-like structure {\bf I} seen in the $^{12}$CO channel maps (Fig.~\ref{CO-chmaps}) arises from the south-west part of the outflow hidden from the line of sight. We also do not see any blue-shifted emission in the $^{12}$CO and $^{13}$CO spectra (Fig.~\ref{spectra}). This suggests that the blue-shifted emission is towards the western side of the source, which is obstructed from the line of sight, and may cause the sharp cutoff in the observed spectra. However, there can be other reasons, such as an asymmetric outflow that are commonly seen in protostars, or the blue-shifted emission being too embedded resulting in a non-detection.

The 3D model can be considered as a good representative of the morphology of the system based on modelling of the present set of ALMA observations. However, this may not be a unique model fit to the complex structure of this system, in particular, to the complicated $^{13}$CO emission, and the interpretation of the structure seen in $^{13}$CO and $^{12}$CO is mainly based on a comparison between the observed and model integrated intensity maps. Modelling of higher angular resolution interferometric observations can further disintegrate the envelope/pseudo-disc/outflow/cavities emission and provide a better interpretation of the structure of this system.

\subsection{Physical parameters of the outflow}
\label{pars}

As shown in Section~\ref{outflow}, the bulk of the $^{12}$CO emission originates from the jet/outflow regions, indicating that CO is mainly tracing a molecular outflow driven by the proto-BD. Therefore, we have calculated the physical properties of the molecular outflow by integrating over the total CO emission observed over a velocity range of $\sim$4-12 km s$^{-1}$ (Fig.~\ref{spectra}c). While there is some contribution from the cavity walls to the CO emission, it is difficult to separate it in the velocity channel maps or the spectrum.

We have followed the standard methods described in Beuther et al. (2002), Palau et al. (2007), and Curtis et al. (2010). We have derived the molecular column density from the integrated and continuum subtracted $^{12}$CO emission, assuming an excitation temperature $T_{ex}$ of 30 K. The resulting CO column density is 1.3$\times$10$^{16}$ cm$^{-2}$. For a lower $T_{ex}$ = 10 K, the CO column density is 1.1$\times$10$^{16}$ cm$^{-2}$. Most outflow studies of low-mass protostars assume a $T_{ex}$ of 10 to 30 K and estimate an uncertainty of 20\% -- 30\% in the outflow properties if $T_{ex}$ was as high as 50-100 K (e.g., van Kempen et al. 2009; Dunham et al. 2014; Yildiz et al. 2013). In reality, the gas in the molecular outflow is not expected to be at the same excitation temperature and may vary both spatially and kinematically, particularly in the shocked emission knots or warm clumps.

To calculate the mass of the outflow, $M_{out}$, from the derived column density, we measured the size of the outflow from the CO moment 0 map and assumed a molecular abundance of 10$^{-4}$ for $^{12}$CO relative to H$_{2}$ (e.g., Frerking et al. 1982). The resulting outflow mass is (6$\pm$2)$\times$10$^{-5}$ M$_{\sun}$. The outflow momentum, $P_{out}$, and the kinetic energy of the outflow, $E_{kin}$, are calculated from $M_{out}$ and the range in the velocity over which CO emission is observed, and are estimated to be $P$ = (4$\pm$2)$\times$10$^{-4}$ M$_{\sun}$.km s$^{-1}$ and $E_{kin}$ = (2$\pm$1)$\times$10$^{43}$ erg.

The age or dynamical timescale, $t_{dyn}$, was derived by dividing the size of the outflow by the maximum velocity reached in the outflow with respect to the systemic velocity. The resulting dynamical age is 616$\pm$100 yr. The dynamical age is likely to be the lower limit to the true age of the flow and to the time over which the proto-BD driving the outflow has been accreting from its surroundings (e.g., Beuther et al. 2002). The very young $t_{dyn}$ compared to the $\sim$6000 yr kinematical age of the proto-BD system also suggests that this may be a recent episodic jet event and the outflow could be variable. Using $t_{dyn}$, the mass outflow rate, $\dot{M}_{out}$, the momentum rate or the mechanical force, $\dot{P}$, and the mechanical luminosity (mechanical power), $L_{mech}$, are estimated to be $\dot{M}_{out}$ = (9.7$\pm$8)$\times$10$^{-8}$ M$_{\sun}$ yr$^{-1}$, $\dot{P}$ = (7$\pm$3)$\times$10$^{-7}$ M$_{\sun}$.km s$^{-1}$ yr$^{-1}$, and $L_{mech}$ = (0.3$\pm$0.1) L$_{\sun}$.

The uncertainties on these measurements have been propagated from the opacity correction, the CO flux errors, the uncertainty on the size of the outflow, and the velocity range of CO emission. The line wings of outflows in low-$J$ transitions of $^{12}$CO are typically optically thick, due to which an opacity correction has to be applied to the outflow mass and dynamical properties. Following the method described in Curtis et al. (2010) and Dunham et al. (2014), we have estimated the opacity $\tau\sim$2 in the $^{12}$CO (2-1) line. We find that the opacity correction can lead to an increase in the outflow mass, momentum, luminosity, force, and energy by a factor of $\sim$4-6. This is consistent with previous studies where opacity correction factors ranging from 2 to 14 have been estimated for CO outflows (e.g., Curtis et al. 2010; Dunham et al. 2014). The correction factors depend on the distribution of mass moving at different velocities. In addition, there remain uncertainties due to the accuracy in the determination of the inclination angle (which will propagate to the space velocities).

The outflow mass, momentum, luminosity, force, and energy for ISO-OPH 200 are higher than the CO outflows driven by any other known proto-BD or VeLLO with L$_{bol}<$0.1 L$_{\sun}$ (e.g., Palau et al. 2014). Most notably, the kinetic energy and thus the mechanical luminosity for ISO-OPH 200 is an order of magnitude higher than previously known proto-BD/VeLLO outflows. In comparison with the more massive protostars with L$_{bol}\sim$0.5--5 L$_{\sun}$ (e.g., Beuther et al. 2004; 2006; Curtis et al. 2010; Dunham et al. 2014), the outflow mass and momentum are 2-3 orders of magnitude lower for ISO-OPH 200, whereas the kinetic energy, mechanical luminosity, and outflow force are within the range measured for the protostars. This proto-BD system follows the general trend of a decline in the outflow mass and force with decreasing bolometric luminosity (e.g., Palau et al. 2014). The comparatively higher dynamical properties of the outflow than proto-BDs/VeLLOs is consistent with ISO-OPH 200 being extremely young (Section~\ref{youth}), with a very fast moving and more massive outflow than other known proto-BD/VeLLO cases.

We note that the very low outflow mass of the order of 10$^{-5}$ M$_{\sun}$ and the very low disc mass of $\sim$4 M$_{Jup}$ in contrast to the high kinetic energy and mechanical luminosity of the outflow suggest the presence of a strong magnetic field in this proto-BD system. With a strong magnetic field, the angular momentum is excessively transported and a very small disc and weak outflow appears, as shown in the simulations by Machida et al. (2018). In a recent dust polarization study with ALMA, Sadavoy et al. (2019) have reported a non-detection in the dust polarized observations for ISO-OPH 200, based on which they report a polarization fraction of $<$1\% for this proto-BD. However, a low polarization fraction does not directly imply that there is a weak magnetic field. Instead, the dust grains may be too small to produce polarization signatures. We already see signs of sub-micron sized ISM-like dust grains from the silicate spectrum for ISO-OPH 200 (Section~\ref{youth}). It may also be the case that the dust grains are not aligned with the magnetic field. The polarization fraction is low when strong toroidal fields exist, because opposite direction of polarization vectors are canceled out. A low-polarization fraction indicates the complexity of magnetic field and that the magnetic configuration is not simple. 



\subsection{Collective source properties of ISO-OPH 200}
\label{youth}

\subsubsection{Sub-stellar nature}
\label{mass}

We can set constraints on the {\it present} mass of the central object for ISO-OPH 200 using the measured bolometric luminosity, L$_{bol}$, and numerical simulations of stellar evolution described in the accretion models of Baraffe et al. (2017) and Vorobyov et al. (2017). These authors calculated stellar properties starting from a protostellar seed of 1.0 M$_{Jup}$ and using the realistic mass-accretion rates derived from numerical hydrodynamics simulations of disc evolution. In both cases, the Lyon stellar evolution code was used to calculate the stellar properties. Note that the bolometric luminosity should be considered as an upper limit on the true internal luminosity, L$_{int}$, of the system. The L$_{int}$ is the luminosity of the central proto-BD that does not include any contribution from the external heating of the circumstellar envelope by the interstellar radiation field. The L$_{int}$ values are estimated to be $\sim$70\%--80\% of L$_{bol}$ (e.g., Riaz et al. 2016). The L$_{bol}$ and L$_{int}$ for ISO-OPH 200 are 0.08$\pm$0.02 L$_{\sun}$ and 0.06$\pm$0.02 L$_{\sun}$, respectively. These are the values derived after applying an extinction correction (A$_{v}\sim$8 mag). Using the L$_{int}$-M$_{obj}$ tracks for a Class 0 object from Vorobyov et al. (2017), the L$_{int}$ for ISO-OPH 200 implies the present central object mass of $\sim$10-20 M$_{Jup}$ in the current epoch. We can add to this the present circumstellar mass of 4-10 M$_{Jup}$, resulting in a total (central object + circumstellar) mass for ISO-OPH 200 of (14-30) M$_{Jup}$ in the present epoch.

To estimate the {\it final} central object mass for ISO-OPH 200, i.e. the mass at the end of the main accretion phase when the central object has reached the pre-main sequence, we have used the infrared photometry for this object and the evolutionary models by Baraffe et al. (2003). The $J$ band, in particular, is considered to be least affected by the potential effects of veiling at bluer wavelengths, and circumstellar disc emission further into the infrared (e.g. Hartigan et al. 1995). Assuming an age of 1 Myr, we can estimate a stellar mass of 20$\pm$5 M$_{Jup}$ and a stellar luminosity of L$_{*}$ = 0.0045 L$_{\sun}$ for this proto-BD. We have also checked the estimates using the DUSTY and BT-Settl models by Allard et al. (2003), and the values differ by at most 10\%. Note that none of these evolutionary models provide isochrones at an age of $<$1 Myr. Therefore, the model-derived estimates on the intrinsic stellar mass and luminosity using the 1 Myr isochrone should be considered as the possible final mass of the object.

Thus, both the present mass and the expected final mass of ISO-OPH 200 is within the sub-stellar limit, indicating a strong likelihood that this object will evolve into a brown dwarf, not a low-mass star.


A typically used method in protostellar studies is to estimate the mass of the central object by considering either the freefall velocity and radius of the infalling envelope, or the rotational velocity and radius of the Keplerian disc. In Sect.~\ref{c18o-model};~\ref{13co-model}, we have shown based on modeling results that the C$^{18}$O line is tracing the inner pseudo-disc region and $^{13}$CO emission is tracing the outer envelope/pseudo-disc regions. Some contribution is also seen in the $^{13}$CO emission from the walls of the outflow cavities. The pseudo-disc is expected to show both infall and Keplerian kinematics that cannot be disentangled at the present resolution. Also, since the pseudo-disc is produced by the magnetic effect, the Lorentz force cannot be ignored to determine the velocity. The Lorenz force can accelerate or decelerate the infalling gas (e.g. Aso et al. 2015). Likewise, we cannot directly estimate the central object mass from detecting Keplerian motions in the rotating disc due to the very small spatial scales ($\leq$10 au) of the inner Keplerian disc that cannot be resolved at the present angular resolution (Section~\ref{c18o-model}). Due to these uncertainties, it is difficult to directly measure the central object mass in an early-stage proto-brown dwarf system by assuming pure infall or Keplerian kinematics.

\subsubsection{Early-stage nature}

We now put together the various properties of ISO-OPH 200 to revisit its classification and justify the early-stage nature and the actual youth of this system. Results from modelling of the observed position and velocity offsets in the C$^{18}$O PVD indicate a very young kinematical age of $\sim$6000 yr for this system, and therefore ISO-OPH 200 must be at an early formation stage. The term `early formation stage' implies the Class 0 stage and needs to be understood in the context of brown dwarf formation. The duration of the Class 0 phase significantly differs between the brown dwarf formation and low-mass star formation cases. Simulations by Machida et al. (2009; 2012) have shown that the envelope dissipation timescale is very short ($<$10$^{4}$ yr) in brown dwarf formation. In contrast, observational signatures of infall motions from an envelope have been observed in low-mass protostars even at ages of $\sim$0.1 Myr (e.g., Davidson et al. 2011; Yen et al. 2011; Takakuwa et al. 2012). The very young kinematical age of $\sim$6000 yr for the ISO-OPH 200 system is consistent with being in the Class 0 stage (i.e., $<$10$^{4}$ yr).

The various observational signatures of the ISO-OPH 200 system also provide direct evidence of the early Class 0 stage. An important indicator is the detection in several high-density tracers of HCO$^{+}$, CS, H$_{2}$CO, CN, HNC in millimeter molecular line observations, none of which are detected in the more evolved Class Flat/Class II brown dwarfs (Riaz et al. 2018; 2019b). The integrated line intensity in the HCO$^{+}$ (3-2) line measured for ISO-OPH 200 satisfies the Class 0/Stage 0 criteria. The molecular abundances for this system in the high-density tracers are within the range of other Class 0/Stage 0 proto-BDs (Riaz et al. 2018; 2019).

The outflow rate of the order of 10$^{-7}$ M$_{\sun}$ yr$^{-1}$ measured for ISO-OPH 200 is within the range measured for Class 0 low-mass protostars and proto-BDs (e.g., Dunham et al. 2014; Antoinucci et al. 2017; Riaz \& Bally 2020). Such strong outflow activity and the very young $\sim$616 yr dynamical age of the CO outflow is realised only in the main accretion phase. The extended projected sizes of the outflow and envelope of $\sim$1000 au are also an indication of the early Class 0 stage (10$^{3}$--10$^{4}$ yr) when the outflow quickly extends to $>$1000 au spatial scales and reaches a projected length that is comparable to the size of the circumstellar envelope (Machida et al. 2009; 2019). 

\begin{figure}
  \centering       
       \includegraphics[width=2.6in]{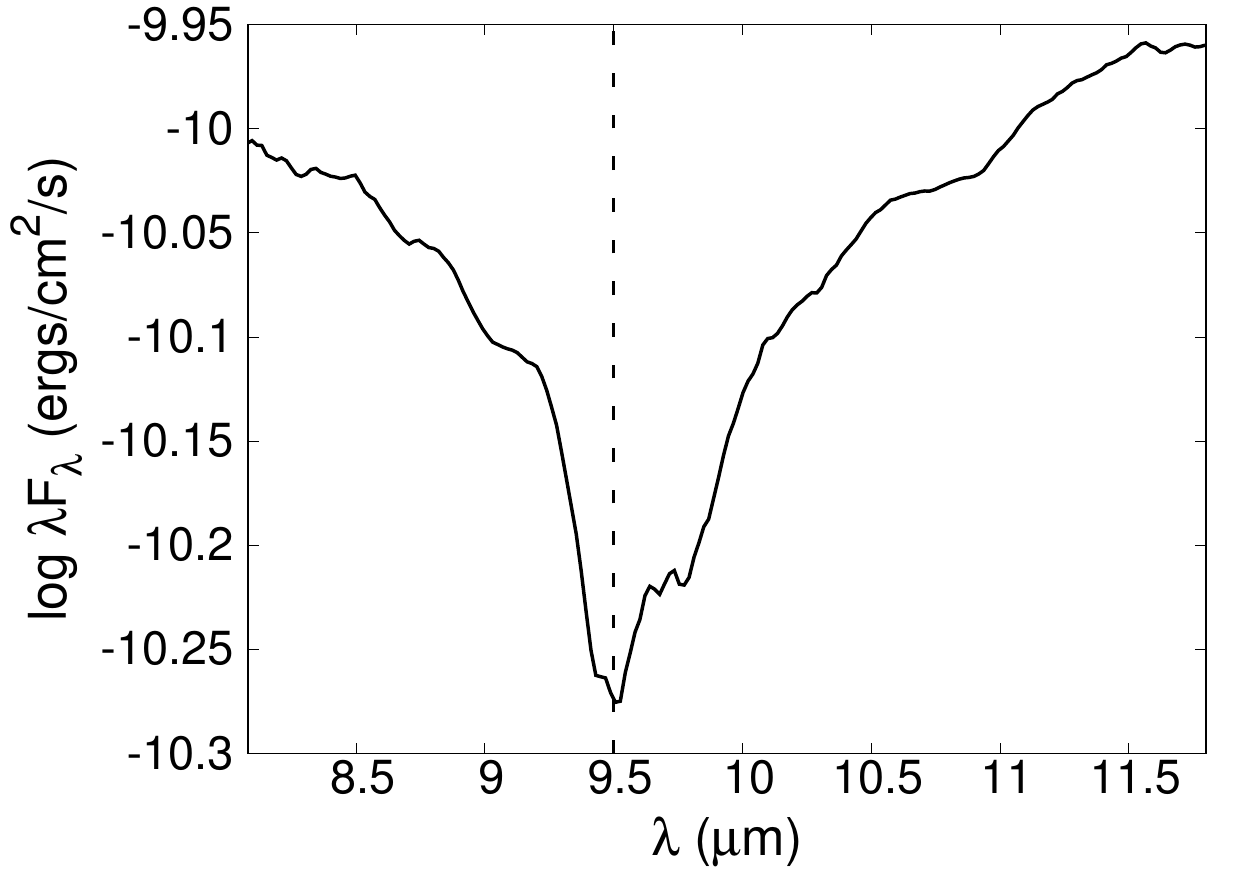}          
     \caption{The 10$\mu$m silicate absorption spectrum. Dashed line marks the peak absorption wavelength of $\sim$9.5 $\mu$m. }
     \label{target}
  \end{figure}

Another evidence of the youth and early stage of the ISO-OPH 200 system is the silicate absorption feature. Figure~\ref{target} shows the mid-infrared 10 $\mu$m silicate spectrum for ISO-OPH 200. The data reduction and analysis for the silicate spectrum is presented in Riaz et al. (2016). The silicate spectrum shows a deep absorption feature, as typically observed in Class 0 protostars (Kessler-Silacci et al. 2005) and proto-BDs (Riaz et al. 2016), and is caused by the optically thick circumstellar envelope surrounding the system. Considering the $\sim$30$\degr$ inclination of the system to the line of sight, the cause of the silicate absorption is not due to an occultation by an edge-on disc. We also note that the CanariCam observations were taken at a high resolution of $\sim$0.2$\arcsec$, and ISO-Oph 200 is not located in a confused region, as also seen in the ALMA and SCUBA-2 continuum maps, and lies $\sim$1.5$\arcmin$ away from the L1689S cloud. Therefore, the absorption is not caused by the L1689S cloud in the line of sight.

The silicate spectrum for ISO-Oph 200 is narrower than the typical absorption silicate features observed in low-mass protostars; the FWHM is $\sim$0.6 for ISO-OPH 200 compared to $\sim$2-3 in protostars. The narrow feature may be related to the partial occultation of the envelope/pseudo-disc due to the inclination of this proto-BD system. The peak in the silicate feature for ISO-OPH 200 is at $\sim$9.5$\mu$m, with a `shoulder' or lower optical depth seen near 9.8$\mu$m (Fig.~\ref{target}). Silicate spectrum of ISM-like dust shows a peak at $\sim$9.8$\mu$m due to the presence of amorphous olivine dust, whereas in more evolved systems such as comets, the dust is more processed and is dominated by crystalline enstatite, with peaks near $\sim$9.3$\mu$m, $\sim$9.5$\mu$m, and 11.3$\mu$m (e.g., Kessler-Silacci et al. 2005). The `shoulder' near 9.8$\mu$m in Fig.~\ref{target} could be due to the presence of crystalline forsterite (e.g., Riaz et al. 2009). The presence of a heating mechanism such as strong jet/outflow activity could raise the local temperature resulting in thermal annealing of amorphous dust, thus expediting silicate crystallization (e.g., Riaz et al. 2012b). Such a localized process can explain the presence of (amorphous) crystalline silicates in the circumstellar material of ISO-OPH 200.

\begin{figure}
  \centering       
       \includegraphics[width=2.6in]{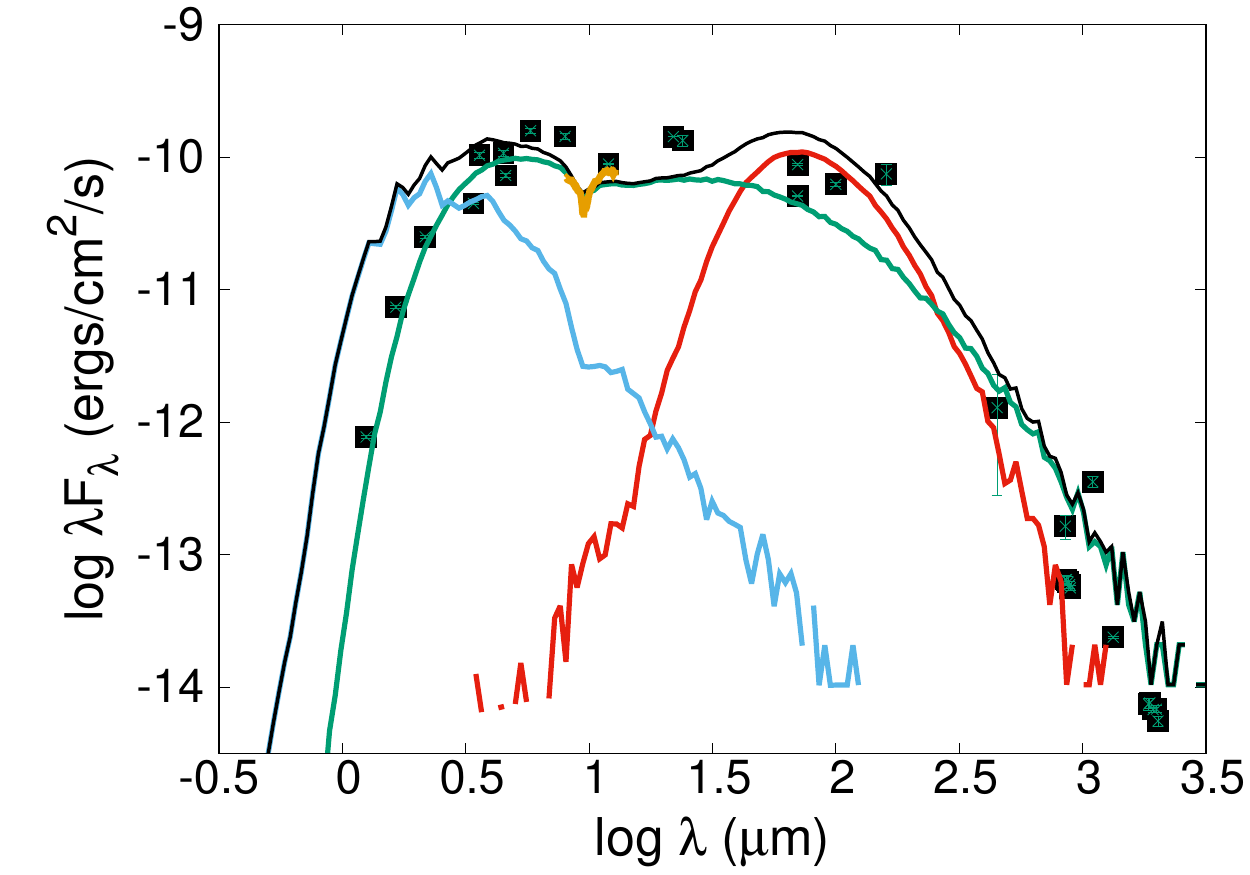} 
     \caption{The SED with model fit (solid black line) for ISO-OPH 200. The red, green, and blue lines indicate the contribution from the envelope, pseudo-disc, and stellar components, respectively. The 10$\mu$m silicate feature is plotted in yellow colour. The photometric data points are plotted as black filled squares, with error bars that are smaller than the point size for most data points.  }
     \label{obs-sed}
  \end{figure}

In contrast to the various signatures of an early-stage system, the 2-24 $\mu$m slope of the observed spectral energy distribution (SED) for ISO-OPH 200 is 0.1, which classifies it as a Class Flat object, i.e. a much more evolved system than Class 0/I object. Figure~\ref{obs-sed} shows the SED for ISO-OPH 200 built using near-infrared to millimeter photometry. The shape of the SED is nearly flat in the mid- and far-infrared. We have conducted radiative transfer modelling of the SED, following the methods described in Riaz et al. (2016). The best model fit indicates that the mid-infrared to millimeter emission has contribution from both the pseudo-disc and envelope components.

However, the flat shape of the SED can be explained by the inclination of the system. Since we are viewing this proto-BD system through a wide outflow cavity (Figs.~\ref{model}b;~\ref{3D-model}), nearly half of the system at the far side is hidden from the view, and only a partial view of the envelope/pseudo-disc is visible, while giving a rare direct view of the jet/outflow. The partial view of the envelope/pseudo-disc can produce a flat SED shape rather than a flared structure with a rising mid- to far-infrared slope expected from a typical Class 0 protostellar system, thereby leading to a mis-classification of the system based on the SED slope alone. Classification of a YSO based only on its SED slope can be erroneous, particularly for cases such as face-on embedded YSOs which may be mis-classified as Class II, or an edge-on flaring disc that could be mis-classified as a Class 0/I object (e.g., van Kempen et al. 2009; Heiderman \& Evans 2015; Riaz et al. 2016; Carney et al. 2016).  Instead, the diverse physical and chemical properties need to be taken into account for a robust classification of a YSO. Based on the various observational and model-derived characteristics detailed here, we conclude that ISO-OPH 200 has been mis-classified as a Class Flat object due to its unusual orientation to the line of sight. The properties of ISO-Oph 200 instead indicate a very young object in its early formation stages in line with the Class 0 phase of star formation.

\subsection{Formation mechanism of ISO-OPH 200}
\label{formation}

The main scientific result from this work is that no major differences are found in the mechanisms dominating the formation of brown dwarfs and low-mass stars. The various observational and model-derived signatures of ISO-OPH 200 are consistent with the predictions from the core collapse model for brown dwarf formation (Machida et al. 2009), making it an interesting benchmark case for star-like formation for brown dwarfs. This proto-BD system has a spatially extended envelope reaching out to $\sim$1000 au, which is comparable to the size of the extended outflow. The very young kinematical age of $\sim$6000 yr for ISO-OPH 200 is consistent with the $\sim$1000 au dimensions of the envelope/outflow, as predicted by the core collapse model, and indicates that the host cloud of proto-BD has retained the initial radius. This shows that there exists a large envelope even in the brown dwarf formation.

It is interesting to note that we see evidence of two kinds of flows for this proto-BD system, i.e. a high-velocity ($\sim$30--50 km s$^{-1}$) (unresolved) [Fe~II] atomic jet and a low-velocity ($\sim$5--10 km s$^{-1}$) CO molecular outflow, although we cannot resolve the launching region of the high-velocity jet. The high-velocity [Fe~II] jet emission is blue-shifted unlike the CO outflow emission that is red-shifted. There is also a possible relation between the velocity and dynamical age of the flows for ISO-OPH 200, such that the low-velocity CO molecular outflow is relatively older (dynamical age $\sim$616 yr) than the high-velocity [FeII] atomic jet (dynamical age $\sim$7-12 yr).

The appearance of two kinds of flows is consistent with the disc wind hypothesis and is also seen in the core collapse simulations for brown dwarf formation (Machida et al. 2009). A wide-angled low-velocity outflow ($\leq$10 km s$^{-1}$) in the core collapse simulations is driven near the pseudo-disc region, appears just after the first core formation ($\sim$100-1000 yr), and reaches a maximum projected length of about 2000 au. A collimated high-velocity jet is driven near the inner Keplerian disc, and emerges at $>$1000 yr. Thus, the low-velocity outflow precedes the high-velocity jet just after proto-BD formation (Tomisaka 2002; Banerjee \& Pudritz 2006; Tomida et al. 2013; Machida \& Basu 2019). The fact that similar signatures are observed for ISO-OPH 200 is another evidence of formation via core collapse for this proto-brown dwarf system.

An alternative to a core collapse formation mechanism for brown dwarfs is considered to be the disc fragmentation model, wherein the circumstellar disc around a massive star undergoes gravitational fragmentation resulting in the formation of very low-mass/sub-stellar cores that are then ejected from the system due to dynamical interactions (Stamatellos \& Whitworth 2009). The ejected brown dwarf possesses some circumstellar material in the form of a disc that shows an asymmetric structure and is truncated ($<$100 au in size) due to its violent dynamical history (Riaz et al. 2019a).

Other formation mechanisms proposed for brown dwarfs include formation due to ejection from larger cores (e.g. Reipurth \& Clarke 2001; Goodwin, Whitworth \& Ward-Thompson 2004), or being liberated from binaries (e.g., Goodwin \& Whitworth 2007), or in small cores in filaments (e.g. Bonnell, Clark \& Bate 2008; Clarke \& Bonnell 2008; Bate 2009). It is difficult for a brown dwarf to sustain an infalling envelope when formed by any of these alternative mechanisms. Since outflows/jets are powered by the release of the gravitational energy of the accreting matter, it is expected that these components are weak for brown dwarfs formed via fragmentation/ejection. All of these models thus predict a small, truncated disc ($<$100 au) at best, while an extended envelope and outflow of $\sim$1000 au sizes are not predicted by any of these alternative formation mechanisms. We refer to Riaz et al. (2012) for a detailed discussion on the predictions from different formation mechanisms for brown dwarfs.

In Riaz et al. (2019a), we had derived a kinematical age of 30,000- 40,000 yr for the Class I proto-BD Mayrit 1701117 (M1701117) in Orion, based on similar modelling of ALMA CO line data as conducted here. A comparison of the various properties of the much younger ISO-OPH 200 proto-BD with M1701117 indicates an evolutionary sequence with a few notable changes that can be traced with the kinematical age of the system. We had measured $\alpha$ = 2.1 and $\beta$ = 0.1 for M1701117, which indicates significant grain growth in the circumstellar material (Riaz et al. 2019a). In comparison, $\alpha$ = 3.2$\pm$0.3, and $\beta$ = 1.2$\pm$0.3 for ISO-Oph 200, which indicates weak signs of grain growth and the presence of pristine dust in the envelope/disc of the system.

Also notable is the complex structure of ISO-OPH 200 as seen in the ALMA observations. Such a complex structure where multiple physical components are dominant is expected at a very young age, as shown in the Machida et al. (2009; 2019) simulations. At later ages of $>$10,000 yr, we expect to see a more simplified structure dominated by only the pseudo-disc component, as seen for M1701117 (Riaz et al. 2019a). The size or spatial scale of the physical components is also dependent on the kinematical age; ISO-OPH 200 has a $\sim$1000 au envelope/outflow and $\sim$500 au pseudo-disc compared to the $\sim$200 au size of the pseudo-disc measured for M1701117.

There is a clear dependence of the spatial extent of the molecular outflow vs. atomic jet emission with the kinematical age. M1701117 is the driving source of the spatially extended ($\sim$0.26 pc $\sim$53,629 au) atomic jet HH~1165 (Riaz et al. 2017); however, only a compact (unresolved) CO molecular outflow was detected in ALMA observations, with peak emission originating from close to the central source and a spatial extent of $<$155 au (Riaz et al. 2019a). In contrast, ISO-OPH 200 drives an extended ($\sim$1000 au) CO outflow but a compact ($<$72 au) unresolved atomic jet. In Riaz \& Bally (2021), we had noted a possible evolutionary trend from a molecular to an ionic composition in the near-infrared jets/outflows driven by proto-BDs, such that the Class 0 proto-BDs show strong emission in the H$_{2}$ lines but the [Fe II] lines are undetected. There are also obvious differences in the environment surrounding these objects that can affect their evolution, most notably, the surroundings of M1701117 are devoid of dense gaseous material that has allowed the atomic jet to propagate to a large distance (Riaz et al. 2017).

These two objects form a potential evolutionary sequence, such that the extent of grain growth, the sizes of envelope/outflow/pseudo-disc components, accretion and outflow activity rates, complex  multiple component vs. simple single component structure, all show a dependence on the kinematical age and hence the evolutionary stage of the system. A detailed observational+modeling analysis of high-resolution observations for a number of proto-brown dwarfs can reveal more definitive trends.

\section{Conclusions}

We present ALMA $^{12}$CO (2-1), $^{13}$CO (2-1), C$^{18}$O (2-1) molecular line observations for a very young proto-BD system, ISO-OPH 200, located in the Ophiuchus region. We have conducted physical+chemical modelling of the complex internal structure of this proto-BD using the physical structure from the core collapse simulations for brown dwarf formation at different evolutionary stages. We have shown that the model at a stage of $\sim$6000 yr and a $\sim$30$\degr$ inclination can simultaneously provide a good fit to the C$^{18}$O spectrum and the observed position-velocity offsets. The same model can also provide a good fit to the $^{13}$CO and $^{12}$CO spectra, and the complex structures seen in the integrated intensity maps. 

Results from modelling indicate that $^{12}$CO emission is tracing an extended ($\sim$1000 au) molecular outflow, $^{13}$CO is tracing the outer ($\sim$1000 au) envelope/pseudo-disc regions, and C$^{18}$O is tracing the inner ($\sim$500 au) pseudo-disc. This is the most extended CO molecular outflow identified to date for a proto-BD, with a bright shock emission knot at the apparent end of the outflow. The presence of both a high-velocity jet and low-velocity outflow indicates that the disc wind hypothesis could be applicable in the case of ISO-OPH 200. The source size of $\sim$8.6 au measured in the 873$\mu$m continuum image is comparable to the inner Keplerian disc size predicted by the model.

We have constructed a 3D model of the structure of the proto-BD system. ISO-OPH 200 is viewed partially through a wide outflow cavity, due to which we have a partial view of the southern and south-eastern part of the envelope/pseudo-disc regions and a more direct view of the jet/outflow. We have argued that it is due to the unusual orientation to the line of sight that ISO-OPH 200 has been mis-classified as a Class Flat object based on its SED slope alone. The sub-mm to mm spectral slope of the SED $\beta\sim$1 for ISO-OPH 200, which indicates the presence of pristine ISM-like dust in the envelope/disc of this system. The 10$\mu$m silicate absorption spectrum is indicative of (amorphous) crystalline enstatite and forsterite silicates. Crystallization was likely expedited in this system due to strong jet/outflow activity.

The overall observational and model-derived signatures of ISO-OPH 200, in particular, the very young $\sim$6000 yr kinematical age estimated for the proto-BD system, the $\sim$616 yr dynamical age of the outflow, the high outflow rate ($\sim$1$\times$10$^{-7}$ M$_{\sun}$ yr$^{-1}$), the deep 10$\mu$m silicate absorption feature, the comparable sizes of the spatially extended envelope and outflow, indicate that ISO-OPH 200 is in the early Class 0 stage of formation when there exists a large envelope and the outflow begins to extend to large spatial scales. Multiple characteristics of ISO-OPH 200 are consistent with the predictions from the core collapse model for brown dwarf formation (Machida et al. 2009), making it an interesting benchmark case for star-like formation for brown dwarfs.

\section*{Acknowledgements}

We thank the referee Rafael Bachiller for his valuable comments on the paper. This paper makes use of the following ALMA data: ADS/JAO.ALMA\#2015.1.00741.S, 2016.1.00545.S, and 2017.1.00107.S. ALMA is a partnership of ESO (representing its member states), NSF (USA), and NINS (Japan), to- gether with NRC (Canada), MOST and ASIAA (Taiwan), and KASI (Republic of Korea), in cooperation with the Republic of Chile. The Joint ALMA Observatory is operated by ESO, AUI/NRAO, and NAOJ. BR acknowledges funding from the Deutsche Forschungsgemeinschaft (DFG) Projekt number RI 2919/2-1. The present research used the computational resources of the HPCI system provided by (Cyber Sciencecenter, Tohoku University; Cybermedia Center, Osaka University, Earth Simulator, JAMSTEC) through the HPCI System Research Project (Project ID:hp180001,hp190035). The present study was supported by JSPS KAKENHI Grant Numbers JP17K05387, JP17H02869, JP17H06360 and 17KK0096 (MNM). Simulations reported in this paper were also performed by 2018 and 2019 Koubo Kadai on Earth Simulator (NEC SX-ACE) at JAMSTEC.

\section{Data Availability}

The data underlying this article are available in the ALMA archives at https://almascience.nrao.edu/asax/.




\appendix

\section{Channel Maps}
\label{channel}

\subsection{C$^{18}$O (2-1) Emission}

\begin{figure*}
  \centering              
     \includegraphics[width=6.4in]{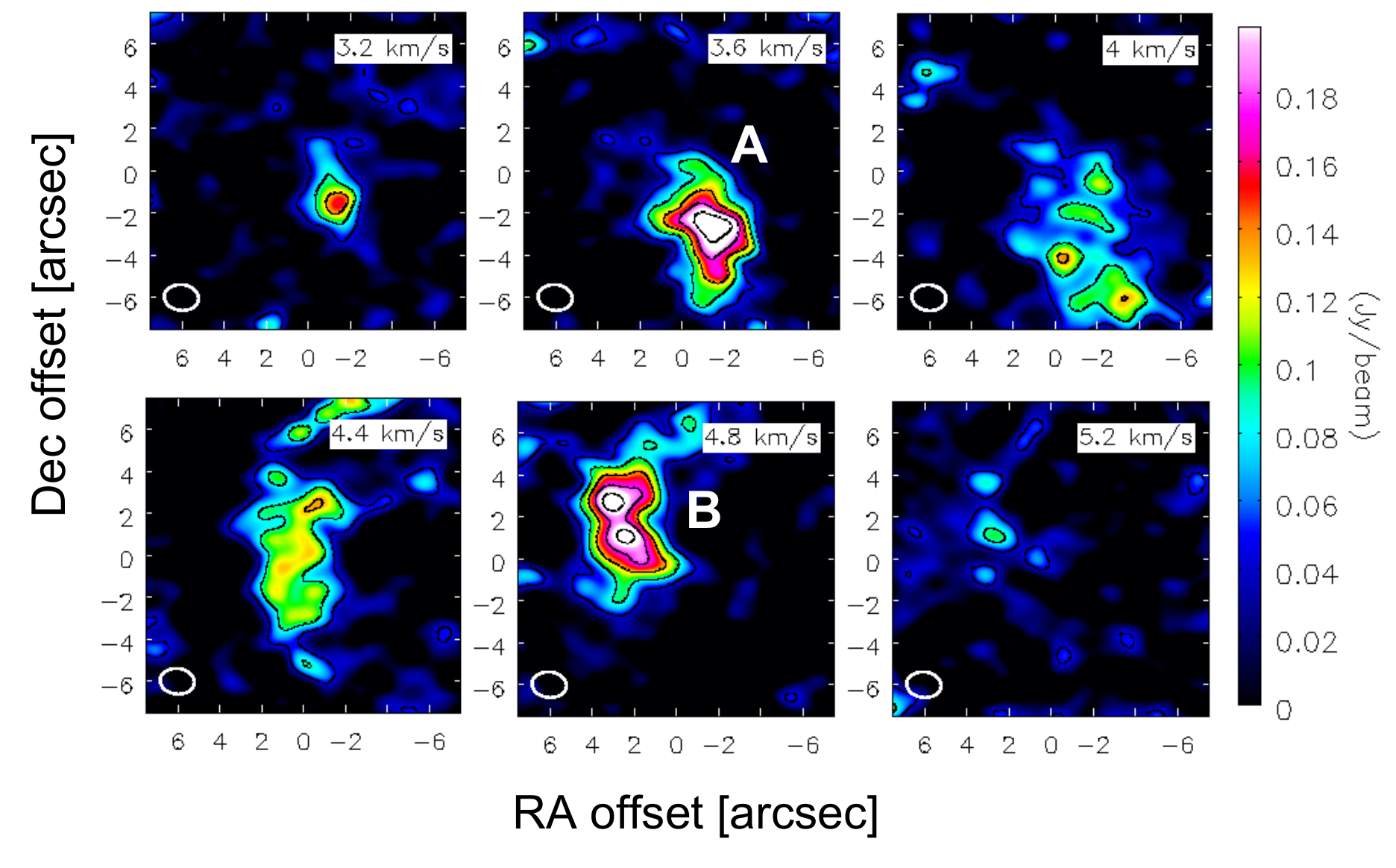}
     \caption{The velocity channel maps in the C$^{18}$O (2-1) line emission for ISO-OPH 200. `{\bf A B}' are the different structures seen in channel maps. The colour bar on the right shows the flux scale in units of Jy beam$^{-1}$. The contour levels are from 2-$\sigma$ to 10-$\sigma$ in steps of 2-$\sigma$; 1-$\sigma$ rms is 0.03 Jy beam$^{-1}$. The channel velocity is noted at the top, right corner, and the beam size is shown at the bottom, left corner. The xy-axes show the position offset relative to the target at (0,0) position. North is up, East is to the left.  }
     \label{C18O-chmaps}
  \end{figure*}

The velocity channel maps in the C$^{18}$O (2-1) line are shown in Fig.~\ref{C18O-chmaps}. Two bright structures, labelled {\bf A} and {\bf B}, are seen at position offset of about (-1,-2) and (2,2), respectively. The emission thus shifts from the south-west towards the north-east direction. The emission in {\bf A} peaks in the 3.6 km s$^{-1}$ velocity channel, and is therefore blue-shifted with respect to the source velocity V$_{LSR} \sim$ 4.2 km s$^{-1}$, while {\bf B} is red-shifted with a peak in emission in the 4.8 km s$^{-1}$ channel. The spatial extent of the structure {\bf A} is estimated to be $\sim$7.8$\arcsec$ or $\sim$1123 au, and for {\bf B} is $\sim$8.3$\arcsec$ or $\sim$1195 au. The sizes for {\bf A} and {\bf B} are about the same although both show a complex structure with multiple clumps or knot like features. The 4--4.4 km s$^{-1}$ channels show diffuse emission with no clear structure. No notable C$^{18}$O emission is detected in the higher velocity channels.

\subsection{$^{13}$CO (2-1) Emission}

\begin{figure*}
  \centering              
     \includegraphics[width=7.2in]{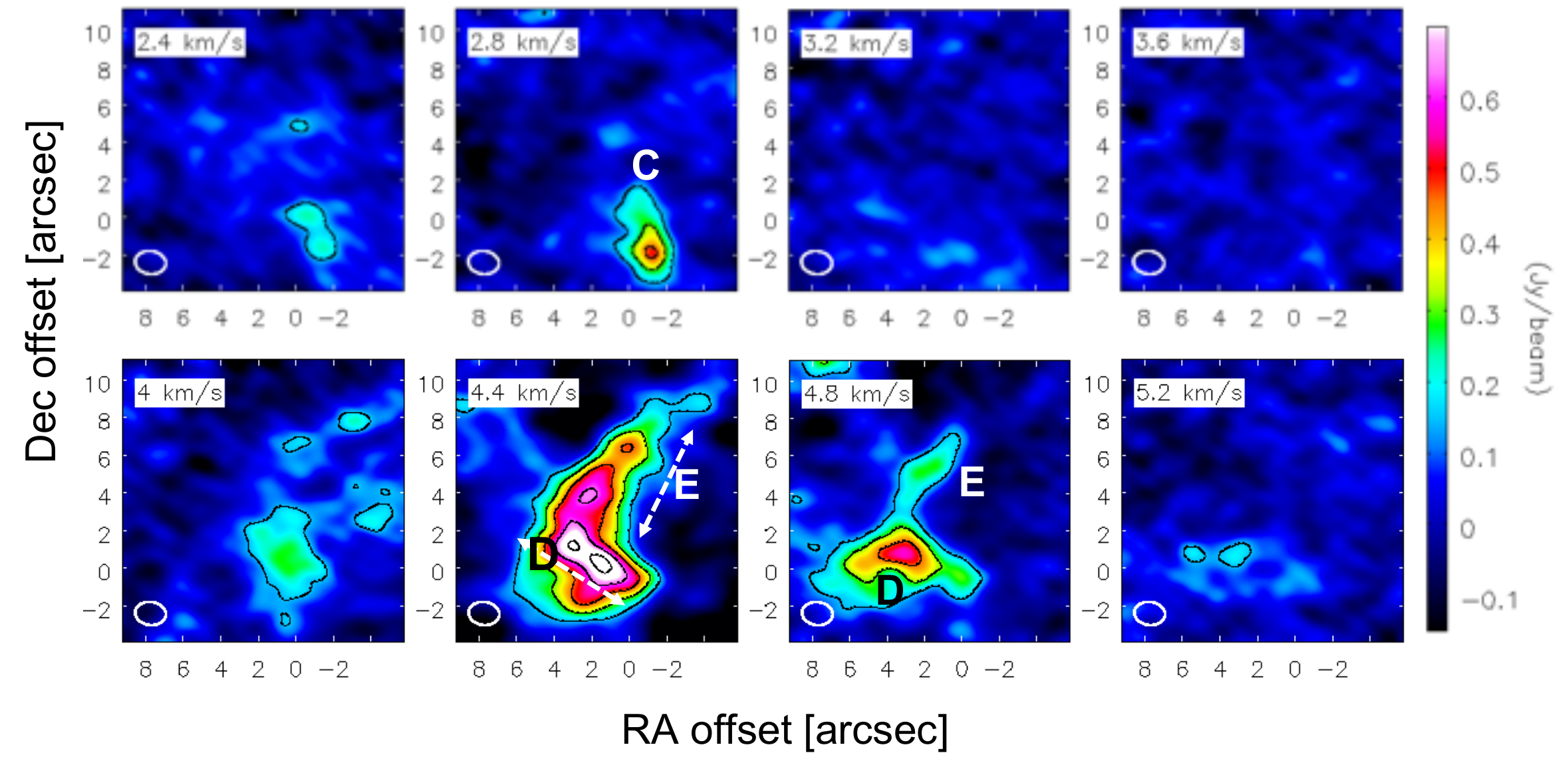}
     \caption{The velocity channel maps in the $^{13}$CO (2-1) line emission for ISO-OPH 200.  `{\bf C D E}' are the different structures seen in channel maps. The colour bar on the right shows the flux scale in units of Jy beam$^{-1}$. The contour levels are from 2-$\sigma$ to 10-$\sigma$ in steps of 2-$\sigma$; 1-$\sigma$ rms is 0.05 Jy beam$^{-1}$. The channel velocity is noted at the top, left corner, and the beam size is shown at the bottom, left corner. The xy-axes show the position offset relative to the target at (0,0) position. North is up, East is to the left.  }
     \label{13CO-chmaps}
  \end{figure*}

Figure~\ref{13CO-chmaps} shows the velocity channel maps for the $^{13}$CO (2-1) line emission. There are two complex structures that appear to be elongated in different orientation. One is the structure {\bf D} that extends north-east from the source position (0,0) up to approximately (+6,+2) position offset. The second is the structure {\bf E} that extends towards north and then curves towards the north-west direction extending up to a position offset of approximately (-3,+9). Also notable are multiple blobs or knots within both of these structures. The structure {\bf D} is much brighter than {\bf E}, and both are seen at red-shifted velocities of $\sim$4.4--4.8 km s$^{-1}$ (with respect to source V$_{LSR}\sim$4.2 km s$^{-1}$). A much fainter and compact structure labelled {\bf C} is seen at blue-shifted velocity of 2.8 km s$^{-1}$ and at position offset of approximately (-1,-2). The structure {\bf D} is merged with {\bf E}, which makes it difficult to correctly measure the spatial extent of these structures. We can roughly estimate it to be from RA $\sim$ +6$\arcsec$ to -2$\arcsec$ and Dec $\sim$ -2$\arcsec$ to +2$\arcsec$, implying a projected size of $\sim$1296 au for {\bf D}. The size of {\bf E} can be roughly estimated to be from RA $\sim$ +5$\arcsec$ to 0$\arcsec$ and Dec $\sim$ +2$\arcsec$ to +9$\arcsec$, implying a projected size of $\sim$1238 au.

\subsection{$^{12}$CO (2-1) Emission}

\begin{figure*}
  \centering              
     \includegraphics[width=7.2in]{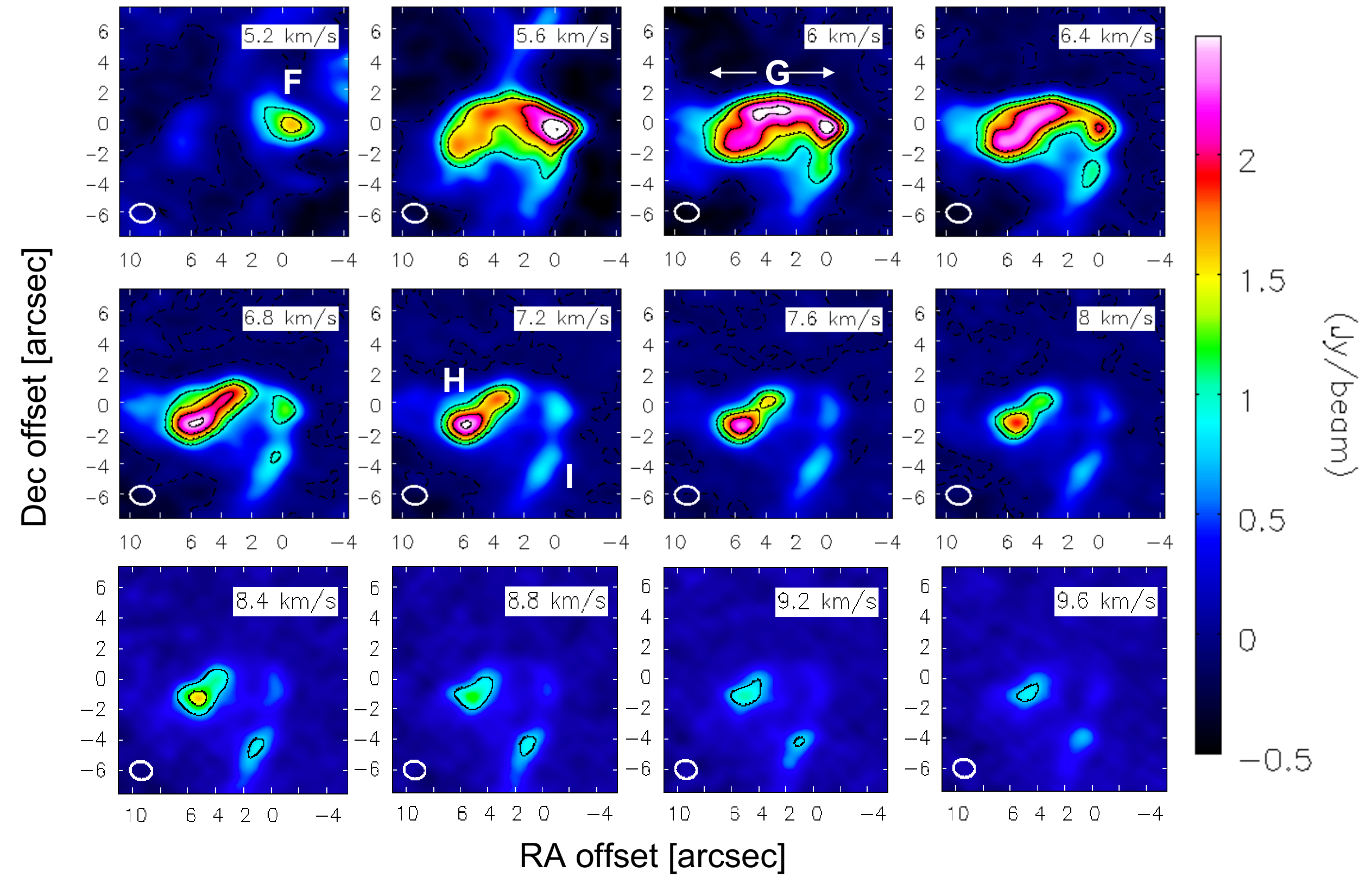}
     \caption{The velocity channel maps in the CO (2-1) line emission for ISO-OPH 200. `{\bf F G H I}' are the different structures seen in channel maps. The colour bar on the right shows the flux scale in units of Jy beam$^{-1}$. The contour levels are from 2-$\sigma$ to 10-$\sigma$ in steps of 2-$\sigma$; 1-$\sigma$ rms is 0.1 Jy beam$^{-1}$. The channel velocity is noted at the top, right corner, and the beam size is shown at the bottom, left corner. The xy-axes show the position offset relative to the target at (0,0) position. North is up, East is to the left.  }
     \label{CO-chmaps}
  \end{figure*}

Figure~\ref{CO-chmaps} shows the CO emission observed in different velocity channel maps. The observed CO emission in all of the channels is red-shifted compared to the source velocity (V$_{LSR} \sim$ 4.2 km s$^{-1}$; Section~\ref{C18O}). We have labelled the different structures seen in the channel maps, as explained below.

\noindent (i) At $\sim$5-5.4 km s$^{-1}$, the CO emission is compact and mainly concentrated at the source position (0,0), with a spatial extent of $\sim$2$\arcsec$ ($\sim$288 au). This compact structure is labelled as `{\bf F}'. 

\noindent (ii) Between $\sim$5.5--6.6 km s$^{-1}$, we see an arc-like elongated structure labelled `{\bf G}' that extends mainly horizontally towards the east of the source, and has a wide spatial extent from $\sim$ -2$\arcsec$ to $\sim$8$\arcsec$, or $\sim$10$\arcsec$ ($\sim$1,440 au). The structure {\bf G} is also quite wide and shows extended emission along north-south of the source from $\sim$ +2$\arcsec$ up to $\sim$ -3$\arcsec$; the width of this structure can thus be estimated as $\sim$5$\arcsec$ ($\sim$720 au). The peak in emission in the structure {\bf G} shifts from the source (0,0) position at $\sim$5.5 km s$^{-1}$ to a position offset of (+6, -2) at $\sim$6.5 km s$^{-1}$.  

\noindent (iii) At higher velocities, the emission in the $\sim$6.7--8 km s$^{-1}$ velocity channels appears as an extended lobe labelled `{\bf H}' that extends south-east from approximately (+2,+1) to (+7,-2) position offsets. The spatial extent of {\bf H} is about 7$\arcsec$ or 1,008 au. The lobe {\bf H} appears to be the south-east part of the elongated structure {\bf G} but is more prominently seen at higher red-shifted velocities. This high-velocity lobe {\bf H} also appears disjointed from the source position since we only see weak diffuse emission at the (0,0) offset.

\section{ALMA vs. single-dish observations}
\label{alma-iram}

\begin{figure}
  \centering              
     \includegraphics[width=3in]{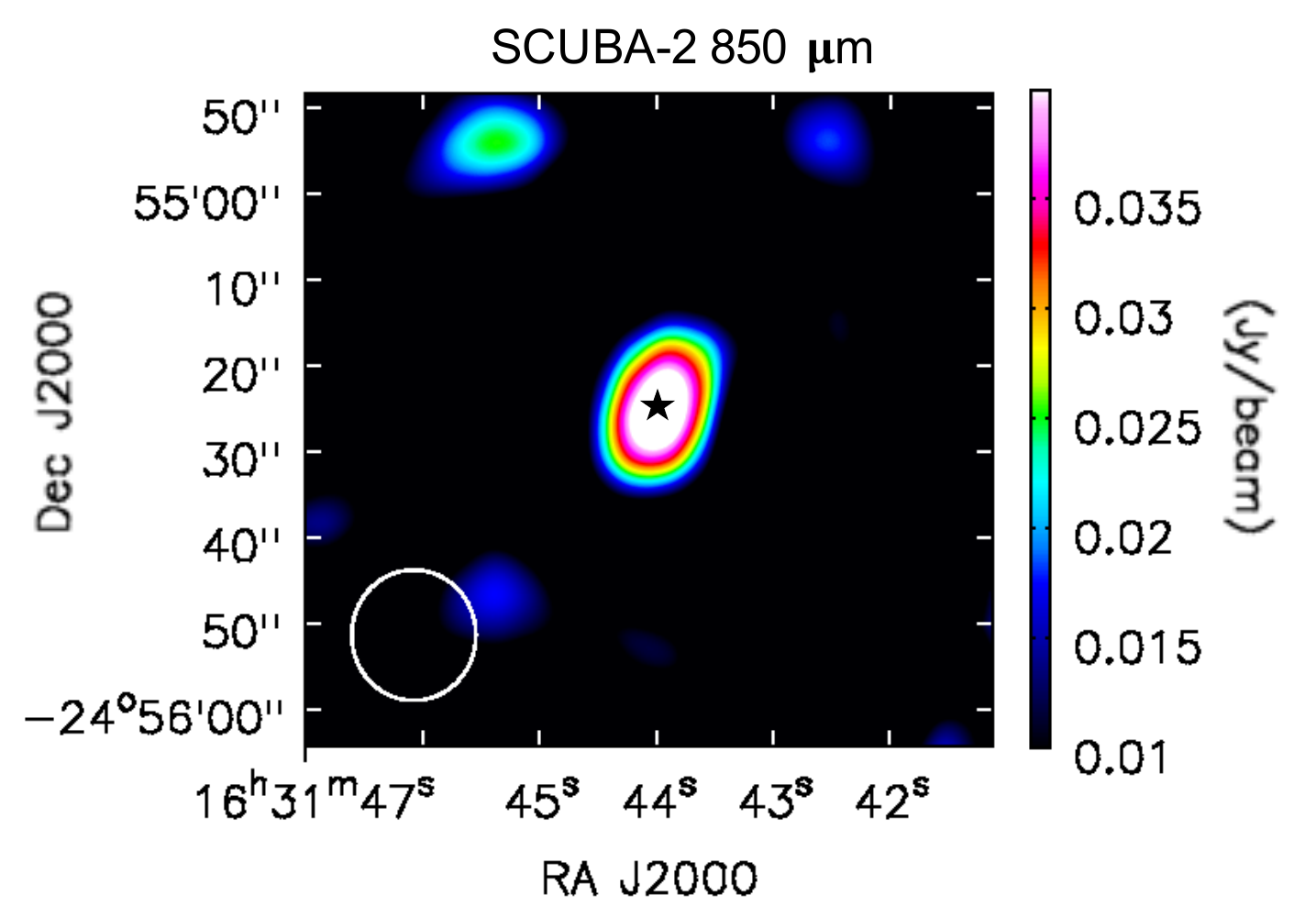}
     \caption{The JCMT SCUBA-2 continuum images for ISO-Oph 200. The target position is marked by a star. The colour bar shows the flux scale in units of Jy beam$^{-1}$. The 1-$\sigma$ rms is $\sim$4 mJy beam$^{-1}$. The beam size is shown at the bottom, left corner.   }
     \label{scuba-2}
  \end{figure}

Figure~\ref{scuba-2} shows the JCMT SCUBA-2 850 $\micron$ continuum image for ISO-OPH 200. The sub-millimeter continuum observations were obtained using the JCMT SCUBA-2 (Holland et al. 2013) instrument. The default map pixels are 4$\arcsec$ and the half- power beam width is 14.5$\arcsec$ in the 850$\micron$ band. The observations were obtained in July, 2012, in Grade 2 weather (225 GHz opacity of 0.06). The CV Daisy observing mode was used, reaching a 1-$\sigma$ rms of $\sim$3 mJy/beam at 850$\micron$. The data reduction is described in Riaz et al. (2016). The image size of ISO-OPH 200 in the 850$\micron$ continuum is approximately 14$\arcsec \times$ 17$\arcsec$ (Fig.~\ref{scuba-2}). The peak and integrated fluxes are 42 mJy beam$^{-1}$ and 46 mJy, respectively. The mass M$_{d+g}$ derived using the method described in Section~\ref{continuum} is 10$\pm$2 M$_{Jup}$.

\begin{figure*}
  \centering              
     \includegraphics[width=2in]{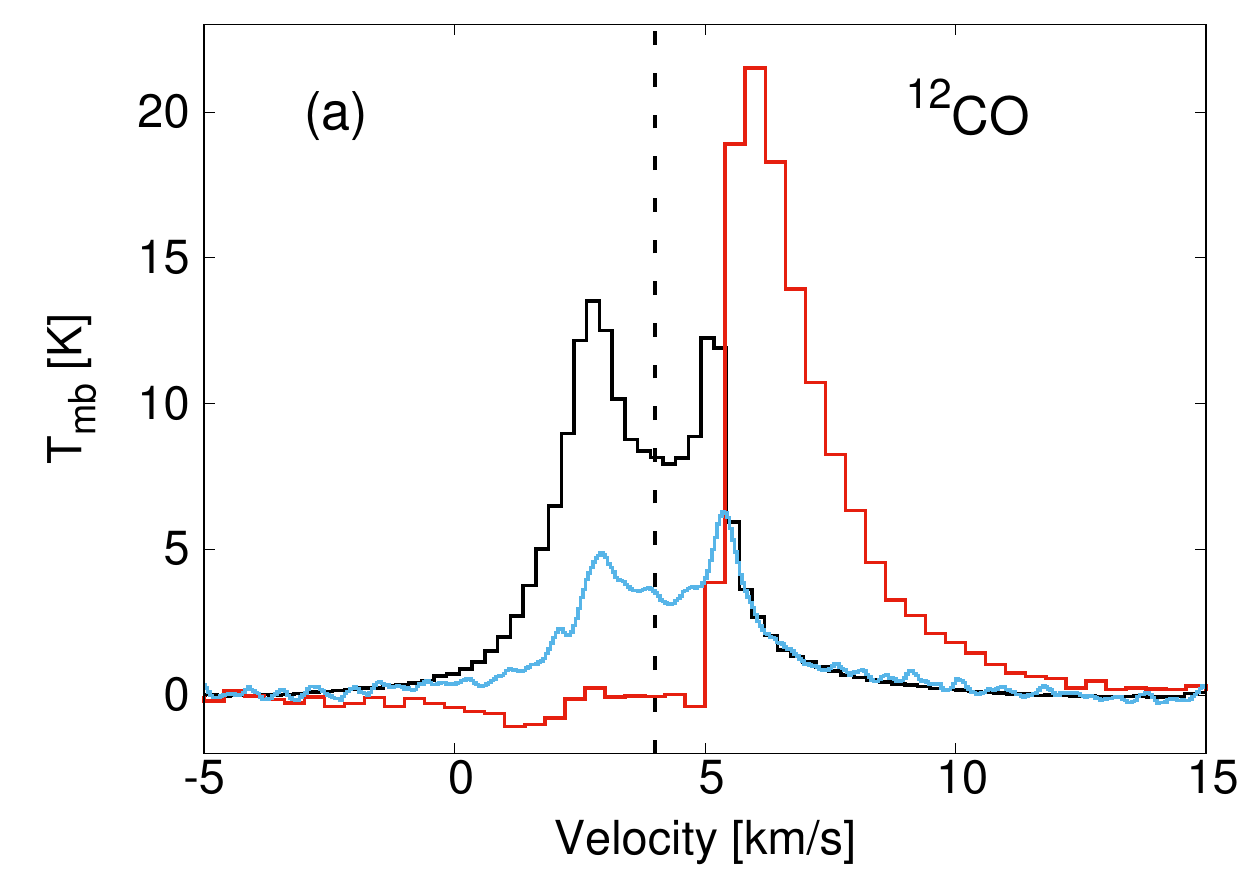}
     \includegraphics[width=2in]{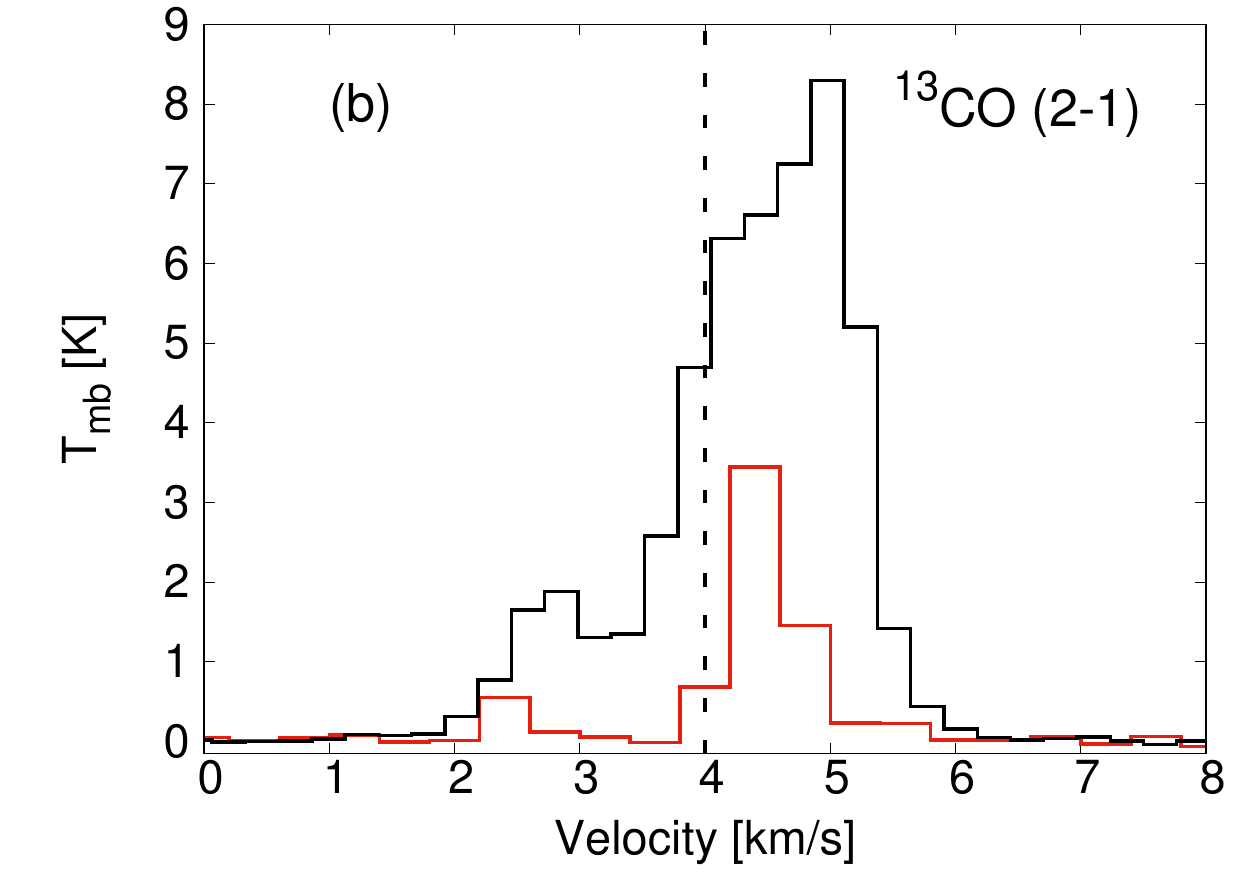}
     \includegraphics[width=2in]{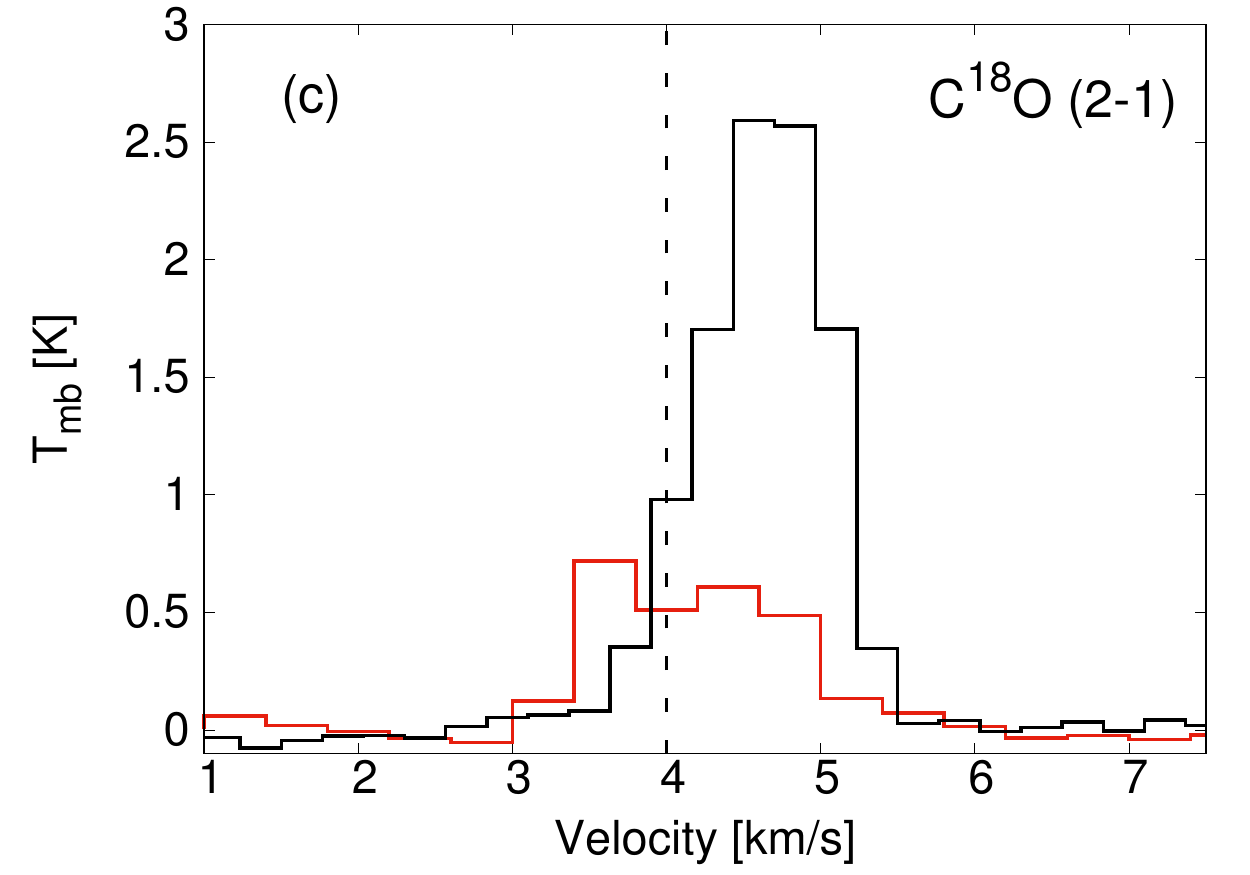}          
     \caption{A comparison of the ALMA (red), IRAM 30m (black), and APEX (blue) spectra for ISO-OPH 200. The IRAM data is from Riaz et al. (2019b). Dashed line marks the cloud systemic velocity.  }
     \label{iram}
  \end{figure*}

Figure~\ref{iram} shows a comparison of ALMA $^{12}$CO (2-1) spectrum with IRAM $^{12}$CO (2-1) and APEX $^{12}$CO (4-3) single-dish spectra. The IRAM observations are from Riaz et al. (2019b) and were obtained with the EMIR heterodyne. We observed $^{12}$CO (2-1), $^{13}$CO (2-1), and C$^{18}$O (2-1) lines. The beam size of the IRAM/EMIR spectra is $\sim$10$\arcsec$. The APEX observations were obtained using the FLASH$^{+}$ receiver. We only observed the $^{12}$CO (4-3) line with APEX. The beam size of the FLASH$^{+}$ spectrum is $\sim$17$\arcsec$.

The single-dish IRAM and APEX spectra of $^{12}$CO show two peaks at $\sim$2.7 and $\sim$5.2 km s$^{-1}$ (Fig.~\ref{iram}a). The APEX CO (4-3) spectrum is a factor of $\sim$2 weaker than the IRAM CO (2-1) spectrum. In contrast, the ALMA CO (2-1) spectrum shows a strong peak at $\sim$6 km s$^{-1}$. A red-dominated asymmetry in $^{12}$CO is typically indicative of outflowing material that can cause the line profiles to shift from a blue- to red-skewed profile (e.g., Evans 1999; Bally 2016). The overlap seen in the red-shifted broad wing of the ALMA, IRAM, and APEX spectra indicates a common origin from an outflow. The lack of absorption in the ALMA $^{12}$CO spectrum at the cloud systemic velocity of $\sim$4-4.4 km s$^{-1}$ indicates that the emission from any unrelated surrounding cloud material is resolved out in the high-resolution ALMA interferometric observations. Likewise, the significant contamination from the ambient molecular cloud seen in the IRAM $^{13}$CO and C$^{18}$O spectra is resolved out in the ALMA spectra, thus enhancing the emission from the small-scale proto-BD structures of envelope/pseudo-disc (Sect.~\ref{c18o-model};~\ref{13co-model}).

Based on these strong filtering effects, we have argued that the sharp blue cutoff seen in the ALMA $^{12}$CO spectrum is caused by an internal obstruction from the envelope/pseudo-disc within the proto-BD system, rather than foreground/background (external) cloud emission in the line-of-sight. Such obstruction from unrelated surrounding cloud material is expected to be seen in single-dish spectra in a ten times larger beamsize than ALMA. As discussed in Sect.~\ref{3D-morph}, we do not see any emission in the ALMA CO line images towards the northern and western sides of the source, and it is likely that the blue-shifted emission arises from these obstructed components.



\bsp	
\label{lastpage}
\end{document}